\documentclass[12pt]{article}
\usepackage[top=2cm,bottom=3cm,left=2cm,right=2cm]{geometry}
\usepackage{amsmath}
\usepackage{amssymb} 
\usepackage{hyperref}
\usepackage{graphicx}
\usepackage{amsthm}
\usepackage{bm} 
\usepackage{bbm}
 
\numberwithin{equation}{section}
\numberwithin{figure}{section}
\newtheorem{claim}{Claim}
\newtheorem{definition}{Definition}
\newtheorem*{SC}{Schwarz-Christoffel map with conical singularities}

\def\eq#1{(\ref{eq:#1})}
\def\lineup{\!\!\!\!\!\!\!\! &&}

\def\d{\partial}
\def\eps{\epsilon}

\def\SW{S_\mathrm{W}}
\def\SKZ{S_\mathrm{KZ}}
\def\Slc{S_\mathrm{lc}}

\def\PsiW{\Psi_\mathrm{W}}
\def\PsiKZ{\Psi_\mathrm{KZ}}
\def\Psilc{\Psi_\mathrm{lc}}
\def\Psicov{\Psi_\mathrm{cov}}
\def\Psilong{\Psi_\mathrm{long}}

\def\mW{m^\mathrm{W}}
\def\mlc{m^\mathrm{lc}}

\def\H{\mathcal{H}}
\def\Hcov{\mathcal{H}_\mathrm{cov}}
\def\Hlc{\mathcal{H}_\mathrm{lc}}
\def\Hperplc{\mathcal{H}_\mathrm{lc}^\perp}
\def\HDDF{\mathcal{H}_\mathrm{DDF}}
\def\Hlong{\mathcal{H}_\mathrm{long}}
\def\Hparallellc{\mathcal{H}_\mathrm{lc}^\parallel}

\def\QQ{{\bf Q}}
\def\Qlc{Q^\text{lc}}
\def\QQlc{{\bf Q}^\mathrm{lc}}
\def\mm{{\bf m}}
\def\mmeff{{\bf m}^\mathrm{eff}}
\def\mmW{{\bf m}^\mathrm{W}}
\def\mmlc{{\bf m}^\mathrm{lc}}
\def\TT{{\bf T}}
\def\II{{\bf I}}
\def\GG{{\bf G}}
\def\mmu{\bm{\mu}}

\def\VW{V^\mathrm{W}}
\def\Vlc{V^\mathrm{lc}}
\def\Veff{V^\mathrm{eff}}

\def\Sigmalcm{\Sigma^{\mathrm{lc},m}}

\def\Tlc{T^\mathrm{lc}}
\def\TW{T^\mathrm{W}}

\def\Alc{A^\mathrm{lc}}

\def\betalc{\beta^\mathrm{lc}}

\def\f#1{f_{(3,#1)}}
\def\fp#1{f_{(3,#1)}\ \!\!'}
\def\fW#1{f^\mathrm{W}_{(3,#1)}}
\def\flc#1{f^\mathrm{lc}_{(3,#1)}}
\def\flcp#1{f^\mathrm{lc}_{(3,#1)}\ \!\!'}
\def\flcm#1{f^{\mathrm{lc},m}_{(4,#1)}}
\def\flcmp#1{f^{\mathrm{lc},m}_{(4,#1)}\ \!'}
\def\fl#1{f^\ell_{(3,#1)}}
\def\flm#1{f^{\ell,m}_{(4,#1)}}

\def\L{L_\mathrm{long}}
\def\b{b_\mathrm{long}}

\begin{document}
\begin{titlepage}
\rightline\today

\begin{center}

\vskip 3.5cm

{\large \bf{Mapping between Witten \\ and Lightcone String Field Theories}}

\vskip 1.0cm

{\large Theodore Erler$^{(a,b)}$\footnote{tchovi@gmail.com}, Hiroaki Matsunaga$^{(c)}$\footnote{matsunaga@karlin.mff.cuni.cz}}

\vskip 1.0cm

$^{(a)}${\it CEICO, FZU - Institute of Physics of the Czech Academy of
Sciences}\\
{\it Na Slovance 2, 182 21 Prague 8, Czech Republic}\\
\vskip .5cm
$^{(b)}${\it Institute of Mathematics, the Czech Academy of Sciences}\\ {\it Zinta 25, 11567 Prague 1, Czech Republic}\\
\vskip .5cm
$^{(c)}${\it Mathematical Institute, Faculty of Mathematics and Physics,Charles University}\\ {\it Sokolovska 83, Prague 3, Czech Republic}

\vskip 2.0cm

{\bf Abstract}

\end{center}

We propose a transformation between the off-shell field variables of Witten's open bosonic string field theory and the traditional lightcone string field theory of Kaku and Kikkawa, based on Mandelstam's interacting string picture. This is accomplished by deforming the Witten vertex into lightcone cubic and quartic vertices,  followed by integrating out the ghost and lightcone oscillator excitations from the string field. Surprisingly, the last step does not alter the cubic and quartic interactions and does not generate effective vertices, and leads precisely to Kaku and Kikkawa's lightcone string field theory.

\end{titlepage}

\tableofcontents

\section{Introduction}

Since the earliest days of string theory, there have mainly been two ways to quantize the string~\cite{Thorn}. Either one fixes the worldsheet symmetries and quantizes in lightcone gauge, or one quantizes covariantly, implementing the symmetries as constraints at the quantum level. The later approach ultimately leads to the BRST invariant formulation of the worldsheet theory  \cite{KatoOgawa}. Understanding the relation between these descriptions is a classical problem. The celebrated no-ghost theorem \cite{Brower,Goddard_Thorn,KatoOgawa,Freeman,Thorn_ghost} establishes that covariant and lightcone approaches lead to the same spectrum of physical states. The equivalence of on-shell amplitudes was a topic of significant interest in the 1980's, and has more-or-less been established \cite{Wolpert,GiddingsDHoker,Giddings,DHokerPhong}. Ultimately, however, the most complete understanding of the relation between covariant and lightcone descriptions should be given in the language of string field theory. Mandelstam's picture of interacting strings in lightcone gauge \cite{Mandelstam1,Mandelstam2} lends itself very naturally to a field theoretic description \cite{Kaku1,Kaku2,Hopkinson}, and, following progress in the development of superstring field theories \cite{SenBV,KunitomoOkawa,WittenSS}, covariant string field theory has been argued to provide a complete defintion of string perturbation theory~\cite{SenErbin}. 

Nevertheless, it is unclear how covariant and lightcone string field theories are related.\footnote{Previous work in this direction appears in \cite{Siopsis}.} In this paper we wish to clarify this connection by constructing a transformation relating Witten's open bosonic string field theory \cite{Witten} and the traditional lightcone string field theory of Kaku and Kikkawa \cite{Kaku1}. This is accomplished following three steps: 
\begin{itemize}
\item The first is establishing a procedure for integrating out longitudinal degrees of freedom from a covariant string field theory. We approach this using a generalization of the no-ghost theorem due to Aisaka and Kazama \cite{AisakaKazama}.\footnote{A closely related construction was given before and corrected in \cite{Furuuchi}.} The resulting string field theory will be called a {\it lightcone effective field theory}. We further explain that a lightcone effective field theory can be equivalently interpreted as a covariant string field theory which has been fixed to lightcone gauge.
\item The second step is to find a covariant string field theory whose lightcone effective action represents Kaku and Kikkawa's lightcone string field theory. The relation between the form of a covariant string field theory and the form of its lightcone effective action is not necessarily straightforward. In fact, it turns out that lightcone effective field theories are generically singular. However, we find that Kaku and Kikkawa's theory emerges from fixing lightcone gauge in a covariant string field theory characterized by lightcone-style vertices. This is similar to the original covariant string field theory constructed by Hata, Itoh, Kugo, Kunitomo and Ogawa (HIKKO) \cite{HIKKO}, but the string length parameter is explictly equated with the minus component of the string momentum. The closed string version of this theory was considered by Kugo and Zwiebach~\cite{KugoZwiebach}, so we call it the {\it Kugo-Zwiebach string field theory}.
\item Finally, we need to relate the Kugo-Zwiebach theory to Witten's string field theory. This is achieved, following an idea of Kaku \cite{Kaku}, by gluing strips of string of fixed height $\ell$ to the open string boundaries in Mandelstam diagrams. The strips can be viewed as Chan-Paton indices carried by the endpoints of lightcone open strings. In the limit $\ell\to 0$ the strips disappear and we recover the conventional Mandelstam diagrams of the Kugo-Zwiebach theory. In the limit $\ell\to\infty$, with the appropriate rescaling of coordinates, we obtain the diagrams of Witten's open bosonic string field theory. In this way we have a 1-parameter family of string field theories which connect the Witten and Kugo-Zwiebach formulations. Following~\cite{HataZwiebach}, this implies the existence of a field redefinition connecting the theories which can be computed explicitly. 
\end{itemize}
After further integrating out ghost and lightcone oscillator excitations, we obtain a relation between the off-shell field variables of Witten's string field theory and Kaku and Kikkawa's lightcone string field theory. The particular relation we describe encounters difficulties related to the generically singular nature of lightcone effective actions. We expect that these complications can be avoided with a suitable improvement of our proposal, though we will not attempt such an improvement here. 

This paper is organized as follows. After some notational preliminaries in section \ref{sec:setup}, in section \ref{sec:effective} we develop the notion of ``lightcone effective field theory," defined by integrating out longitudinal states from a covariant string field theory. We characterize the longitudinal states using a similarity transformation derived by Aisaka and Kazama \cite{AisakaKazama}, which we review in subsection \ref{subsec:Kazama}.  In subsection \ref{subsec:decomposition} we note that the similarity transformation implements a diffeomorphism which maps to the lightcone gauge parameterization of the worldsheet, and use this result to characterize the longitudinal states in terms of a ``longitudinal dilatation operator" previously known from the work of Brink and Olive~\cite{BrinkOlive}. In subsection \ref{subsec:integrate} we explain the systematics of integrating out the longitudinal states with a judicious choice of gauge.  In subsection \ref{subsec:lightcone} we show that the resulting lightcone effective field theory can be interpreted as a covariant string field theory which has been fixed to lightcone gauge, where the lightcone gauge condition is
\begin{equation}
\left(b_0+i p_-\oint \frac{d\xi}{2\pi i}\frac{b(\xi)}{\d X^+(\xi)}\right)\Psi = 0 .
\end{equation}
The operator on the left hand side plays an important role in Freeman and Olive's proof of the no-ghost theorem \cite{Freeman}.

In section \ref{sec:KZ} we show that the traditional lightcone string field theory of Kaku and Kikkawa can be derived as the lightcone effective action of Kugo and Zwiebach's string field theory. The surprising fact is that integrating out longitudinal states does not change the form of the interactions, a phenomenon we refer to as {\it transfer invariance}.  In subsection \ref{subsec:transfer} we show that transfer invariance follows from a peculiar property of correlation functions on Mandelstam diagrams, that insertions of $\d X^+$ can be replaced by a $c$-number in the absence of contractions with $\d X^-$. In subsection \ref{subsec:genericcubic} we evaluate the cubic vertex of a generic lightcone effective field theory. We find that the vertex has the structure of a cubic lightcone vertex attached to ``stubs" whose purpose is to ensure that the local dilatation at the punctures is the same as that of the cubic vertex of the parent covariant string field theory. We note that if one state in the vertex has sufficiently low momentum relative to the others, the stub lengths will typically be {\it negative}, resulting in singular couplings between highly excited string states. We refer to this as the  {\it soft string problem} of lightcone gauge. 

In section \ref{sec:Kaku} we construct a field redefinition relating Witten and Kugo-Zwiebach string field theories based on Kaku's idea of attaching ``Chan-Paton strips" to lightcone vertices \cite{Kaku}. In subsection \ref{subsec:Kaku} we derive expressions for the local coordinate maps and the quartic vertex region of the moduli space which follow from  Kaku's deformation, and in subsection \ref{subsec:infinitesimal} we present the infinitesimal field redefinition which slightly alters the length of the Chan-Paton strips following~\cite{HataZwiebach}. Continuously iterating the infinitesimal field redefinition allows us to relate Witten and Kugo-Zwiebach string fields. 

In section \ref{sec:transformation} we describe a transformation between Witten and lightcone string field theories based on the results of previous sections. The transformation is given by mapping the lightcone string field into the Kugo-Zwiebach string field and then performing the field redefinition constructed in section \ref{sec:Kaku}. We find that the transformation encounters difficulty with highly excited string states related to the soft string problem discussed in subsection \ref{subsec:genericcubic}. As a sample calculation we evaluate the relation between the tachyon fields of the two theories explicitly up to quadratic order. We conclude the paper with a list of questions for future inquiry.

\subsubsection*{Conventions}

We assume $\alpha'$ and the string field coupling constant is equal to one. Ghost correlators are normalized as $\langle c\d c\d^2 c(u)\rangle^\text{gh}_\text{UHP} = -2$. Commutators are graded with respect to  ``degree,'' the degree of a string field is its Grassmann parity plus $1$ (mod $\mathbb{Z}_2$). We introduce the symbol
\begin{equation}\delta_K=\left\{\begin{matrix}1\ \ K\ \text{is true}\\
0\ \ K\ \text{is false}\end{matrix}\right..\end{equation}
Lightcone coordinates are denoted $x^+,x^-, x^1,...,x^{24}$. These are related to conventional Minkowski coordinates $x^0,...,x^{25}$ through
\begin{equation}x^+=\frac{1}{\sqrt{2}}(x^0+x^{25}),\ \ \ x^-=\frac{1}{\sqrt{2}}(-x^0+x^{25}).\end{equation}
Choosing the mostly plus metric in Minkowski space, the metric in lightcone coordinates is
\begin{equation}
ds^2  = 2dx^+dx^- + (dx^1)^2+...+(dx^{24})^2.
\end{equation}
In particular
\begin{equation}x^+ = x_-,\ \ \ \ \ x^-=x_+\ \ \ \ \  x^i=x_i,\ \ (i=1,...,24).\end{equation}
The spacetime index $i$ always ranges over transverse directions.

\section{Setup}
\label{sec:setup}

To formulate lightcone string field theory it is necessary to have a background with at least two flat and noncompact spacetime dimensions. To facilitate discussion we will consider a space-filling D-25 brane in flat space, though nothing essential depends on this setup. The worldsheet boundary conformal field theory (BCFT) consists of 25 spacelike and one timelike free boson all subject to Neumann boundary conditions, together with the $bc$ system of central charge $-26$. We take one spacelike and the timelike free boson to form a pair of lightlike free bosons, so the worldsheet fields are 
\begin{eqnarray} 
\lineup \ \ \ \ \ \ \ X^i(z,\overline{z})\ \ i=1,...,24;\nonumber\\
\lineup X^+(z,\overline{z}),\ \ X^-(z,\overline{z}),\ \ b(z),\ \ c(z).
\end{eqnarray}
The $X^i(z,\overline{z})$s define what will be called the {\it transverse} factor of the BCFT. The lightlike free bosons and the $bc$ system  constitute what will be called the {\it longitudinal} factor of the BCFT. The position zero mode of $X^+(z,\overline{z})$ will be identified with lightcone time, and, in lightcone diagrams, the momentum zero mode of $X^+(z,\overline{z})$ will be the string length. Our conventions for free bosons are written in appendix \ref{app:free}.

The actions for Witten's open bosonic string field theory, the Kugo-Zwiebach theory, and Kaku and Kikkawa's lightcone string field theory can be written respectively as 
\begin{eqnarray}
\SW\lineup = -\frac{1}{2}\omega(\PsiW,Q\PsiW)-\frac{1}{3}\omega(\PsiW,\mW_2(\PsiW,\PsiW)),\phantom{\Bigg)}\\
\SKZ\lineup = -\frac{1}{2}\omega(\PsiKZ,Q\PsiKZ)-\frac{1}{3}\omega(\PsiKZ,\mlc_2(\PsiKZ,\PsiKZ))-\frac{1}{4}\omega(\PsiKZ,\mlc_3(\PsiKZ,\PsiKZ,\PsiKZ)),\phantom{\Bigg)}\ \ \ \ \ \ \\
\Slc \lineup = -\frac{1}{2}\omega(\Psilc,c_0L_0\Psilc)-\frac{1}{3}\omega(\Psilc,\mlc_2(\Psilc,\Psilc))-\frac{1}{4}\omega(\Psilc,\mlc_3(\Psilc,\Psilc,\Psilc)).\phantom{\Bigg)}\label{eq:Slc}
\end{eqnarray}
The actions are expressed from the point of view of a shifted $\mathbb{Z}_2$ grading on the open string vector space which we call {\it degree}. The degree of a string field $A$ will be denoted as $|A|$, and is given by the Grassmann parity of the corresponding vertex operator plus 1. The shifted grading allows us to make use of the coalgebra description of $A_\infty$ algebras, which will be useful at a few points. An introduction to the coalgebra formalism can be found in \cite{WBsmall}, whose notation we follow. The lightcone string field theory action is not usually written as \eq{Slc}, but it will be helpful to have the theories expressed in a common language.  Let us explain the ingredients. The dynamical fields of the respective theories are $\PsiW,\PsiKZ$ and $\Psilc$. They are degree even and worldsheet ghost number 1 elements of the vector space $\H$ of the full matter+ghost BCFT.  $Q$ is the BRST operator, $c_0$ is the zero mode of the $c$ ghost and $L_0$ is the zero mode of the total energy-momentum tensor. The symplectic form $\omega:\H^{\otimes 2}\to\mathbb{C}$ is related to the BPZ inner product through
\begin{equation}\omega(A,B) = (-1)^{|A|}\langle A,B\rangle,\end{equation}
and 
\begin{eqnarray}
\lineup \mW_2:\H^{\otimes 2}\to \H,\\
\lineup \mlc_2:\H^{\otimes 2}\to \H,\\
\lineup \mlc_3:\H^{\otimes 3}\to \H
\end{eqnarray}
are string products which define the Witten vertex and the cubic and quartic lightcone vertices. We review the definition of these vertices in appendix \ref{app:vertex}. The product $\mW_2$ is related to Witten's open string star product by
\begin{equation}\mW_2(A,B) = (-1)^{|A|}A*B.\end{equation}
Both the Witten and Kugo-Zwiebach string field theories are characterized by cyclic $A_\infty$ algebras. In the coalgebra formalism, this means that the string products define coderivations on the tensor algebra of $\H$
\begin{eqnarray}
\mmW\lineup =\QQ + \mmW_2,\\
\mmlc\lineup = \QQ+\mmlc_2+\mmlc_3,
\end{eqnarray}
which satisfy
\begin{equation}
(\mmW)^2 = 0,\ \ \ \ \ (\mmlc)^2 = 0,
\end{equation}
and 
\begin{equation}
\langle \omega|\pi_2\mmW = 0,\ \ \ \ \ \ \langle\omega|\pi_2\mmlc = 0.\phantom{\Big)}
\end{equation}
Lightcone vertices define an $A_\infty$ algebra because open string lightcone diagrams provide a single cover of the moduli spaces of disks with boundary punctures. We will refer to the Kugo-Zwiebach string field theory as a ``covariant" string field theory, even though the interactions are not Lorentz invariant. A covariant string field theory for our purposes is one which possesses the usual BRST kinetic term and linearized gauge invariance.

Still more information is needed to explain our expression for the lightcone action. The dynamical fields  $\PsiW,\PsiKZ$ and $\Psilc$ are elements of the vector space $\H$ of the BCFT. But it will be helpful to imagine $\PsiW$ and $\PsiKZ$ as elements of a ``covariant" vector space $\Hcov$ while $\Psilc$ as an element of a ``lightcone" vector space $\Hlc$:
\begin{equation}\PsiW,\PsiKZ\in\Hcov,\ \ \ \ \Psilc\in\Hlc.\end{equation}
One way to understand the distinction is that $\Hcov$ and $\Hlc$ are defined by pairing $\H$ with a nilpotent operator, which in the first case is the usual BRST operator and in the second case is not. Namely,
\begin{equation}\Hcov = (\H,Q),\ \ \ \ \Hlc = (\H,\Qlc).\end{equation}
In other words, $\Hcov$ and $\Hlc$ should be viewed as chain complexes which use isomorphic graded vector spaces but distinct differentials.
The operator $\Qlc$ is given by $c_0L_0$ plus a nonminimal term which cancels the contribution from ghost and lightcone oscillators to the cohomology. The full expression will be given in subsection \ref{subsec:Kazama}, but only $c_0L_0$ appears in the lightcone string field theory action. It is important that the lightcone string field  is not an arbitrary element of $\Hlc$, but lives in a subspace defined by a linear constraint:
\begin{equation}L_0^\parallel A = 0,\ \leftrightarrow \ A\in\Hperplc\subset\Hlc.\end{equation}
The restricted vector space will be denoted $\Hperplc$. The operator $L_0^\parallel$ counts the level created by ghost and lightcone oscillators 
\begin{equation}
L_0^\parallel = \sum_{n=1}^\infty \Big(\alpha^+_{-n}\alpha^-_n +\alpha^-_{-n}\alpha^+_n\Big) + \sum_{n=1}^\infty n\Big(b_{-n}c_n+c_{-n}b_n\Big),\label{eq:L0parallel}
\end{equation}
with the normal ordering constant chosen so that the tachyon state has level zero. The effect of the constraint $L_0^\parallel =0$ is to forbid ghost and lightcone oscillator excitations from the string field. We may therefore form a basis for the restricted space $\Hperplc$ with the states
\begin{eqnarray}
\lineup \  \alpha_{-n_1}^{i_1}...\alpha_{-n_N}^{i_N} c_1 e^{ik\cdot X(0,0)}|0\rangle,\nonumber\\
\lineup \alpha_{-n_1}^{i_1}...\alpha_{-n_N}^{i_N} c_0 c_1 e^{ik\cdot X(0,0)}|0\rangle,
\end{eqnarray}
where $\alpha^i_n$ are oscillators for the transverse free bosons and $|0\rangle$ is the $SL(2,\mathbb{R})$ vacuum. Note that $\Hperplc$ only contains states at ghost number $1$ and ghost number $2$, and the dynamical field lives in the ghost number 1 part. In this way, the dynamical field is completely fixed in the longitudinal factor of the BCFT, so we can evaluate correlators in this factor to reduce the lightcone action \eq{Slc} to an expression which only involves correlators in the transverse component of the BCFT. This is the lightcone string field theory action as it would normally be written. Details are given in appendix \ref{app:long}. The operator $L_0^\parallel$ will later be interpreted as the zero mode of the appropriately defined ``longitudinal energy-momentum tensor," and the constraint $L_0^\parallel =0$ may be thought of as analogous to the level matching condition in closed string field theory. Incidentally, lightcone string field theory has a cyclic $A_\infty$ structure defined by the string products
\begin{eqnarray}
 c_0L_0:\lineup\Hperplc \to \Hperplc,\label{eq:KKprod1}\\
\delta(L_0^\parallel)\mlc_2:\lineup (\Hperplc)^{\otimes 2}\to \Hperplc,\\
 \delta(L_0^\parallel)\mlc_3:\lineup (\Hperplc)^{\otimes 3}\to \Hperplc,\label{eq:KKprod3}
\end{eqnarray}
where $\delta(L_0^\parallel)$ is the projector onto the kernel of $L_0^\parallel$. However, the $A_\infty$ relations follow trivially from ghost number counting. They do not contain information about covering of the moduli space. Consider for example the derivation property
\begin{equation}
c_0L_0\delta(L_0^\parallel)\mlc_2(A,B) + \delta(L_0^\parallel)\mlc_2(c_0L_0 A,B)+\delta(L_0^\parallel)\mlc_2(A,c_0L_0B)=0,\ \ \ A,B\in\Hperplc.
\end{equation}
All three terms separately vanish. The second and third terms vanish since $\delta(L_0^\parallel)$ acts on a state of ghost number not less than 3, and there are no such states in the kernel of $L_0^\parallel$. The first term vanishes since $c_0L_0$ acts on a state in $\Hperplc$ with ghost number not less than 2, and all such states are proportional to $c_0$. The triviality of the $A_\infty$ relations is related to the fact that lightcone string field theory does not possess gauge invariance, since there are no nonvanishing gauge parameters at ghost number 0.

\section{Lightcone effective field theory}
\label{sec:effective}

The first step in our analysis is understanding how to integrate out the longitudinal degrees of freedom from a covariant string field theory. The resulting ``physical" string field theory will be called a {\it lightcone effective field theory}. This is a slight abuse of terminology, since we are not integrating out high energy modes in a conventional sense. Still, the structure of lightcone effective field theory closely parallels that of low energy effective actions derived from string field theory, as described for example in \cite{SenEffective}. Part of the following discussion was previously reported in \cite{Matsunaga}.

Integrating out the longitudinal modes essentially means that we eliminate states outside the kernel of $L_0^\parallel$ using the equations of motion. However, this needs to be understood in the correct sense. The analogue of $L_0^\parallel$ in the covariant vector space is not simply \eq{L0parallel}, since the covariant and lightcone vector spaces are not identical. This is of course well-known. Transverse oscillators $\alpha^i_n$ of the string in lightcone gauge are embedded in the covariant vector space through DDF operators~\cite{DDF}
\begin{equation}A_n^i =  i\sqrt{2}\oint\frac{d\xi}{2\pi i}\d X^i e^{\frac{in}{p_-}X^+(\xi)}.\end{equation}
The analogue of $L_0^\parallel$ in the covariant vector space was given long ago by Brink and Olive \cite{BrinkOlive}. However, we approach this problem from the point of view of a transformation introduced by Aisaka and Kazama \cite{AisakaKazama}. Initially, the transformation is a linear map
\begin{equation}
S: \Hperplc \to \HDDF\subset\Hcov,
\end{equation}
which takes the restricted lightcone vector space into the subspace of off-shell DDF states. The transformation is explicitly
\begin{eqnarray}
S \Big(\alpha_{-n_1}^{i_1}...\alpha_{-n_N}^{i_N} c_1 e^{ik\cdot X(0,0)}|0\rangle\Big) \lineup = \Big(e^{\frac{in_1}{2k_-}x^+}A_{-n_1}^{i_1}\Big)...\Big(e^{\frac{in_N}{2k_-}x^+}A_{-n_N}^{i_N}\Big)c_1 e^{ik\cdot X(0,0)}|0\rangle,\phantom{\Bigg)}\nonumber\\
S \Big(\alpha_{-n_1}^{i_1}...\alpha_{-n_N}^{i_N} c_0 c_1 e^{ik\cdot X(0,0)}|0\rangle\Big) \lineup = \Big(e^{\frac{in_1}{2k_-}x^+}A_{-n_1}^{i_1}\Big)...\Big(e^{\frac{in_N}{2k_-}x^+}A_{-n_N}^{i_N}\Big)c_0 c_1 e^{ik\cdot X(0,0)}|0\rangle.\phantom{\Bigg)}\label{eq:SlcDDF}
\end{eqnarray}
The exponential factors cancel the $x^+$ dependence inside the DDF operators. This map preserves inner products, or equivalently, it is a symplectomorphism
\begin{equation}
\omega(SA,SB) = \omega(A,B),\ \ \ \ A,B\in\Hperplc.
\end{equation}
Aisaka and Kazama's construction upgrades $S$ so that it provides a symplectic isomorphism between the full vector spaces $\Hlc$ and $\Hcov$. The lightcone vector space can be decomposed into a direct sum 
\begin{equation}\Hlc = \Hperplc\oplus\Hparallellc,\end{equation}
where $\Hparallellc$ consists of linear combinations of states with nonzero $L_0^\parallel$ eigenvalue. This will then imply a decomposition of the covariant vector space
\begin{equation}
\Hcov = \HDDF\oplus\Hlong,
\end{equation}
where
\begin{equation}
\HDDF = S\Hperplc,\ \ \ \ \Hlong = S\Hparallellc.
\end{equation}
The longitudinal states of the covariant vector space live in $\Hlong$, and these are the states we need to integrate out. 

\subsection{Aisaka and Kazama's transformation}
\label{subsec:Kazama}

It is helpful to understand the meaning of ``longitudinal degrees of freedom" from the point of view of the string in lightcone gauge. Of course, the string in lightcone gauge only has the transverse free bosons $X^i(z,\overline{z})$ together with lightcone zero modes $p_+,p_-$. But we can imagine appending these variables with longitudinal degrees of freedom in such a way that the longitudinal degrees of freedom have been made ``invisible." From the point of view of Aisaka and Kazama's result, this turns out to mean that we should twist the $bc$ system so that the $b$ ghost carries weight 1 and the $c$ ghost carries weight $0$. Then  the $bc$ system together with the non-zero mode part of the lightcone scalars define a BCFT with vanishing central charge, and in this sense is ``invisible." This motivates the following definitions:
\begin{itemize}
\item The {\it transverse} energy-momentum tensor will be denoted $T^\perp(z)$. This is simply the energy-momentum tensor of the transverse scalars. It has central charge $c=24$. 
\item The {\it longitudinal} energy-momentum tensor will be denoted $T^\parallel(z)$. This is given by adding the energy-momentum tensor of the lightcone scalars evaluated at zero momentum to the energy-momentum tensor of the twisted $bc$ system. It has central charge $c=0$. The zero mode of the longitudinal energy-momentum tensor is the operator $L_0^\parallel$ we have already encountered. 
\end{itemize}
The sum of transverse and longitudinal energy-momentum tensors will be called the {\it lightcone} energy-momentum tensor,
\begin{equation}
T^\text{lc}(z) = T^\perp(z) + T^\parallel(z),\label{eq:Tlc}
\end{equation}
which has central charge $c=24$. Note that $p_+$ and $p_-$ do not appear in the lightcone energy-momentum tensor. We can interpret this to mean that the string in lightcone gauge sees $p_+$ and $p_-$ as external parameters, rather than worldsheet variables. In the conformal frame $w$ of the strip, the lightcone energy-momentum tensor is related to the total energy-momentum tensor by
\begin{equation}T^\text{lc}(w) = T(w)|_{p_\pm=0} - \d j_\text{gh}(w),\end{equation}
where $j_\text{gh}(z) = -:\!bc(z)\!:$ is the ghost current. The Virasoro modes are explicitly given by
\begin{eqnarray}
L_n^\perp \lineup = \frac{1}{2}\sum_{m\in\mathbb{Z}}:\!\alpha^i_{-m}\alpha^i_{m+n}\!: ,\\
L_n^\parallel \lineup = \sum_{m\in\mathbb{Z}}\delta_{m\neq 0}\delta_{m\neq n}:\!\alpha^+_{-m}\alpha^-_{m+n}\!:+\sum_{m\in \mathbb{Z}}(m+n):\! b_{-m}c_{m+n}\!:,\\
L^\text{lc}_n \lineup = L_n^\perp+L_n^\parallel.
\end{eqnarray}
It will be convenient to introduce the notation $\widetilde{X}^\pm(z)$ for the chiral lightcone scalars with the zero modes subtracted out, and $\widetilde{b}(z),\widetilde{c}(z)$ for the twisted $bc$ ghosts. These are related to the chiral lightcone scalars and the ordinary $bc$ ghosts by 
\begin{eqnarray}X^\pm(\xi) \lineup = \frac{1}{2}x^\pm -ip_\mp \ln(\xi)+\widetilde{X}^\pm(\xi),\\
\widetilde{b}(z) \lineup = \frac{b(z)}{f'(z)},\\
\widetilde{c}(z) \lineup = f'(z)c(z),
\end{eqnarray}
where $f(z)$ is a conformal transformation relating $z$ to the coordinate $w$ on the strip. In the conformal frame $\xi$ of radial quantization, we have explicitly $\widetilde{b}(\xi) = \xi b(\xi)$ and $\widetilde{c}(\xi) = c(\xi)/\xi$. The operators $\d \widetilde{X}^\pm(z)$, $\widetilde{c}(z)$ and $\widetilde{b}(z)$ are primaries with respect to the lightcone energy-momentum tensor,
\begin{eqnarray}
T^\text{lc}(z)\d \widetilde{X}^\pm(z') \lineup = \frac{1}{(z-z')^2}\d \widetilde{X}^\pm(z') + \frac{1}{z-z'}\d^2 \widetilde{X}^\pm(z')+\text{regular},\\
T^\text{lc}(z) \widetilde{b}(z') \lineup = \frac{1}{(z-z')^2}\widetilde{b}(z') + \frac{1}{z-z'}\d \widetilde{b}(z')+\text{regular},\\
T^\text{lc}(z)\widetilde{c}(z') \lineup = \frac{1}{z-z'}\d \widetilde{c}(z')+\text{regular},
\end{eqnarray}
with respective conformal weights of $1,1$ and $0$. 

We may expand the BRST operator so as to explicitly display its dependence on ghost and lightcone zero modes: 
\begin{equation}
Q = p_-\delta^- +c_0 L_0 + \widehat{Q} + b_0 M + p_+\delta^+,\label{eq:Qzero}
\end{equation}
where $L_0$ is the zero mode of the total energy-momentum tensor and
\begin{eqnarray}
\delta^\pm\lineup = 2i\oint\frac{d\xi}{2\pi i}\widetilde{c}\d \widetilde{X}^\pm(\xi)\ \  = \sqrt{2}\sum_{n\in \mathbb{Z}}\delta_{n\neq 0}c_{-n}\alpha_n^\pm,\label{eq:deltam}\\
M \lineup = \oint\frac{d\xi}{2\pi i}\widetilde{c}\d \widetilde{c}(\xi) \ \ \ \ \ \ \ = -\sum_{n\in\mathbb{Z}}nc_{-n}c_n,\\
\widehat{Q} \lineup = \sum_{n\in \mathbb{Z}}\delta_{n\neq 0} c_{-n}L^\text{matter}_n|_{p_\pm=0}-\frac{1}{2}\sum_{m,n\in\mathbb{Z}}\delta_{m\neq 0}\delta_{n\neq 0}\delta_{m+ n\neq 0}(m-n) c_{-m}c_{-n}b_{m+n}.\label{eq:hatQ}
\end{eqnarray}
It was originally realized by Kato and Ogawa \cite{KatoOgawa} that the first two terms in this expansion are together nilpotent, and the associated cohomology is precisely the vector space of physical states in lightcone gauge. Thus we define the {\it lightcone BRST operator},
\begin{equation}\Qlc = p_-\delta^- + c_0L_0.\end{equation}
The first term eliminates the ghost and lightcone oscillator states from the cohomology by the quartet mechanism, and the second term imposes the mass-shell condition. A simple way to see that the correct cohomology is reproduced is to observe that $L_0^\parallel$ is lightcone BRST exact:
\begin{equation}L_0^\parallel = [\Qlc,b_0^\parallel],\end{equation}
where the operator $b_0^\parallel$ is given by
\begin{eqnarray}
b_0^\parallel\lineup = \frac{i}{p_-}\oint \frac{d\xi}{2\pi i}\xi \widetilde{b}\d \widetilde{X}^+(\xi)\label{eq:b0parallel}\\
\lineup = \frac{1}{\sqrt{2}p_-}\sum_{m\in\mathbb{Z}}\delta_{m\neq 0}\alpha^+_{-n}b_n.
\end{eqnarray} 
This means that any $\Qlc$-closed state $\Psi\in\Hparallellc$ can be written
\begin{equation} \Psi = \Qlc\left(\frac{b_0^\parallel}{L_0^\parallel}\Psi\right),\end{equation}
and does not contribute to the cohomology. It should be mentioned that the above argument assumes $p_-\neq 0$, otherwise the operator $b_0^\parallel$ is not well-defined. In fact, $\Qlc$ does not reproduce the correct cohomology when $p_-=0$. The assumption $p_-\neq 0$ is always necessary in lightcone quantization, and we will take it as given.

Aisaka and Kazama's result is the construction of a similarity transformation $S$ satisfying
\begin{equation}Q = S\Qlc S^{-1}.\label{eq:QS}\end{equation}
We will give a derivation which is slightly different from the original one. First we introduce an auxiliary grading on the lightcone vector space called ``$N$-number." By definition, non-zero mode oscillators $\alpha^+_{n}$ and $c_{n}$ are assigned $N$-number $+1$, while non-zero mode oscillators $\alpha^-_n$ and $b_n$ are assigned $N$-number $-1$. Everything else has $N$-number $0$. The $N$-number counting operator is 
\begin{equation}
N = \sum_{n\in \mathbb{Z}} \delta_{n\neq 0}\left(\frac{1}{n}:\! \alpha_{-n}^+\alpha_n^-\!:+:\!c_{-n}b_n\!:\right),
\end{equation}
and is lightcone BRST exact,
\begin{equation}N = [\Qlc,K],\end{equation}
where
\begin{equation}
K = \frac{1}{\sqrt{2}p_-}\sum_{n\neq 0}\delta_{n\neq 0}\frac{1}{n}\alpha^+_{-n}b_n.\label{eq:K}
\end{equation}
The key observation is that the zero mode expansion of the BRST operator in \eq{Qzero} can be interpreted as an $N$-number expansion
\begin{equation}Q = Q_0+Q_1+Q_2,\end{equation}
where
\begin{eqnarray}Q_0 \lineup = \Qlc,\\
Q_1 \lineup = \widehat{Q},\\
Q_2 \lineup = b_0 M + p_+\delta^+.
\end{eqnarray}
The terms respectively have $N$-number $0$, $1$, and $2$. We have the relations 
\begin{eqnarray}
Q_0^2 \lineup = 0,\\
\ [Q_1,Q_0]\lineup = 0,\\
Q_1^2+[Q_0,Q_2]\lineup = 0,\\
\ [Q_1,Q_2]\lineup =0,\\
Q_2^2 \lineup = 0,
\end{eqnarray}
which follow from nilpotency of the BRST operator.

Next we introduce an $N$-number counting parameter $t$ and define a corresponding family of BRST operators
\begin{equation}Q(t) = t^N Q t^{-N} = Q_0+t Q_1+ t^2 Q_2.\end{equation}
The family connects the lightcone BRST operator at $t=0$ to the ordinary BRST operator at $t=1$. At the same time, if there is a similarity transformation satisfying \eq{QS} we must have 
\begin{equation}Q(t) = S(t) Q_0 S(t)^{-1},\end{equation}
where $S(t) = t^N S t^{-N}$. This implies the differential equation
\begin{equation}
\frac{d}{dt}Q(t) = [Q(t),r(t)],\label{eq:QRdot}
\end{equation}
where 
\begin{equation}r(t) = S(t) \frac{d}{dt}S(t)^{-1}.\label{eq:RdotS}\end{equation}
The strategy is to solve \eq{QRdot} to determine $r(t)$, after which \eq{RdotS} determines $S(t)$ as a path-ordered exponential. We know that $S(t)$ will have an $N$-number expansion of the form
\begin{equation}S(t) = \mathbb{I}+ t S_1 + t^2 S_2+\text{higher orders},\end{equation}
since the similarity transformation should reduce to the identity at $t=0$. This implies that $r(t)$ has an $N$-number expansion 
\begin{equation}r(t) = r_1 + t r_2+ t^2 r_3 + t^3 r_4+\text{higher orders}.\end{equation}
Substituting the $N$-number expansions into the differential equation \eq{QRdot} and equating like powers of $t$ gives 
\begin{eqnarray}
Q_1 \lineup = [Q_0,r_1],\\
2 Q_2 \lineup = [Q_0,r_2]+[Q_1,r_1], \\
0\lineup = [Q_0,r_3] + [Q_1,r_2]+[Q_2,r_1], \\
0\lineup = [Q_0,r_4]+[Q_1,r_3]+[Q_2,r_2], \\
\lineup\ \vdots\ . \nonumber
\end{eqnarray}
These relations are easy to solve since the operator $K$ implies that $Q_0$ has no cohomology when $N\neq 0$. We readily see that 
\begin{eqnarray}
r_1 \lineup = [K,Q_1],\label{eq:KR1}\\
r_2 \lineup = [K,Q_2]-\frac{1}{2}[K,[Q_1,r_1]],\\
r_3 \lineup = -\frac{1}{3}[K, [Q_1,r_2]+[Q_2,r_1]],\\
r_4 \lineup = -\frac{1}{4}[K,[Q_1,r_3]+[Q_2,r_2]],\\
\lineup \ \vdots \  .\nonumber
\end{eqnarray}
We obtain some simplification by plugging the solution from previous orders into subsequent orders. First, we note that the contribution from $r_1$ to $r_2$ drops out due to $K^2=0$, so we obtain simply
\begin{equation}r_2 = [K,Q_2].\label{eq:KR2}\end{equation}
Next we plug the solutions for $r_1$ and $r_2$ to determine $r_3$:
\begin{eqnarray}
r_3= -\frac{1}{3}\big[K, [Q_1,[K,Q_2]]+[Q_2,[K,Q_1]]\big]=0.
\end{eqnarray}
This vanishes due to the Jacobi identity and $[Q_1,Q_2]=0$.  Finally,
\begin{equation} r_4 = -\frac{1}{4}[K,[Q_2,[K,Q_2]]]=0,\end{equation}
which vanishes due to $Q_2^2=0$. It follows that all higher $r_n$s can be taken to vanish. Evaluating the commutators \eq{KR1} and \eq{KR2} gives
\begin{eqnarray}
r_1 \lineup = \frac{1}{\sqrt{2}p_-}\sum_{n\in\mathbb{Z}}\delta_{n\neq 0}\frac{1}{n}\alpha_{-n}^+\left(L_n^\text{matter}|_{p_\pm=0}-\sum_{m\in \mathbb{Z}}\delta_{m\neq 0}(m+n)b_{-m}c_{m+n}\right),\\
r_2 \lineup = \frac{2}{\sqrt{2}p_-}b_0\sum_{n\in\mathbb{Z}}\delta_{n\neq 0}\alpha_{-n}^+c_n,\\
r_n\lineup = 0,\ \ \ n\geq 3 .
\end{eqnarray}
Having found $r(t)$ we can derive $S(t)$ as a path-ordered exponential. However, one can verify that $r_1$ and $r_2$ commute,
\begin{equation}[r_1,r_2] = 0,\end{equation}
so path-ordering is not necessary. Then we readily find that 
\begin{equation}S = e^{-R},\end{equation}
where
\begin{equation}
R = r_1+\frac{1}{2}r_2 = \frac{1}{\sqrt{2}p_-}\sum_{n\in \mathbb{Z}}\delta_{n\neq 0}\frac{1}{n}\alpha^+_{-n}L_n^\text{lc}.
\end{equation}
The appearance of $L_n^\mathrm{lc}$ in this result is what motivates the definition of the lightcone energy-momentum tensor. One may verify that $R$ is BPZ odd. Therefore $S$ is a symplectic isomorphism between the full lightcone and covariant vector spaces, as desired. 

\subsection{Transverse/longitudinal decomposition}
\label{subsec:decomposition}

The goal is to have a concrete understanding of how to decompose the covariant string field  into transverse and longitudinal parts. This requires understanding what the transformation $S$ does to the lightcone vector space. For this we state a useful result:  
\begin{claim}
\label{claim:1}
Let $\mathcal{O}$ be an operator without contractions with $X^+(z,\overline{z})$. Then
\begin{equation}S\mathcal{O}S^{-1} = F^{-1}\circ_\mathrm{lc}\mathcal{O},\label{eq:SF}\end{equation}
where $F^{-1}$ is the inverse of the holomorphic function
\begin{equation}F(\xi) = e^{-\frac{ix^+}{2p_-}}\exp\left[\frac{i}{p_-}X^+(\xi)\right],\end{equation}
and $\circ_\mathrm{lc}$ denotes a conformal transformation computed with respect to the lightcone energy-momentum tensor \eq{Tlc}.
\end{claim}
\noindent The proof of this claim is technical so we leave it to appendix \ref{app:claim1}. Note that the conformal transformation explicitly depends on the operator $X^+(z)$. This is meaningful since, by assumption, $\mathcal{O}$ has no contractions with $X^+(z)$, so $X^+(z)$ effectively behaves as an ordinary function  (though its explicit form will depend on the correlation function where it appears). There is an interesting way to understand the meaning of the above result. In the coordinate $w = \tau+i\sigma$ on the strip, the conformal transformation takes the form
\begin{equation}\tau'+i\sigma' = \frac{i}{p_-}\left(X^+(e^{\tau+i\sigma})-\frac{1}{2}x^+\right).\label{eq:wF}\end{equation}
We can Wick rotate back to a Lorentzian worldsheet by writing $\tau = i t$, where $t$ is a timelike coordinate. The free boson on the Lorentzian worldsheet may be written
\begin{equation}X^\mu(t,\sigma) = X^\mu(z,\overline{z})|_{z=e^{i(t+\sigma)}}.\end{equation}
Then \eq{wF} is equivalent to 
\begin{eqnarray}
t' \lineup =\frac{1}{2p_-}(X^+(t,\sigma)-x^+),\\
\sigma' \lineup = \frac{1}{2p_-}\int_0^\sigma ds \frac{d}{dt}X^+(t,s).
\end{eqnarray}
This is precisely the coordinate transformation which maps the worldsheet parameterization of conformal gauge to the worldsheet parameterization of lightcone gauge. The inverse transformation appears in \eq{SF} since this relates an operator in the lightcone vector space to an equivalent operator in the covariant vector space.

Let us illustrate how the DDF operators emerge from this point of view. The transverse oscillators of the lightcone string map to
\begin{eqnarray}
S \alpha_n^i S^{-1}\lineup = F^{-1}\circ_\text{lc}\left(i\sqrt{2}\oint\frac{d\xi}{2\pi i}\xi^n \d X^i(\xi)\right)\nonumber\\
\lineup = i\sqrt{2}\oint\frac{d\xi}{2\pi i}\xi^n \frac{d F^{-1}(\xi)}{d\xi}\d X^i(F^{-1}(\xi)).
\end{eqnarray}
Here we used the fact that $\d X^i$ has no contractions with $X^+$ and transforms as a weight 1 primary with respect to the lightcone energy-momentum tensor.  Making a change of integration variable $\xi'=F^{-1}(\xi)$ and relabeling $\xi'$ as $\xi$, we obtain 
\begin{eqnarray}
S \alpha_n^i S^{-1}\lineup = i\sqrt{2}\oint\frac{d\xi}{2\pi i}F(\xi)^n \d X^i(\xi)\nonumber\\
\lineup = e^{-\frac{in}{2p_-}x^+}A_{n}^{i}.\label{eq:alphaDDF}
\end{eqnarray}
This is the DDF operator with the $x^+$ dependence subtracted out. One can further verify that 
\begin{equation}S c_1e^{ik\cdot X(0,0)}|0\rangle = c_1e^{ik\cdot X(0,0)}|0\rangle,\ \ \ \ \ Sc_0c_1e^{ik\cdot X(0,0)}|0\rangle=c_0c_1e^{ik\cdot X(0,0)}|0\rangle.
\end{equation}
It follows that $S$ maps $\Hperplc$ into $\H_\text{DDF}$ precisely as written in \eq{SlcDDF}. 

Continuing in a similar way we can derive the transformation of other mode operators. We write 
\begin{eqnarray}
\lineup\ \ \  \ \ \ \ S \alpha_n^i S^{-1}= e^{-\frac{in}{2p_-}x^+}A_n^i,\phantom{\bigg)}\\
S c_n S^{-1}\lineup = e^{-\frac{in}{2p_-}x^+}C_n,\ \ \ \ \ \ \ S b_n S^{-1} = e^{-\frac{in}{2p_-}x^+}B_n,\phantom{\bigg)}\\
S \alpha_n^+ S^{-1}\lineup = e^{-\frac{in}{2p_-}x^+}A_n^+,\ \ \ \ \ \ S \alpha_n^- S^{-1} = e^{-\frac{in}{2p_-}x^+}A_n^-,\phantom{\bigg)}
\end{eqnarray}
where we introduce analogues of DDF operators for the longitudinal modes. When $n\neq 0$ we find explicitly
\begin{eqnarray}
A_n^i \lineup =i\sqrt{2}\oint \frac{d\xi}{2\pi i}\d X^i  e^{\frac{in}{p_-}X^+(\xi)},\phantom{\bigg)} \\
C_n \lineup = \frac{i}{p_-}\oint \frac{d\xi}{2\pi i}\frac{1}{\xi} c\d X^+ e^{\frac{in}{p_-}X^+(\xi)},\phantom{\bigg)} \\
B_n \lineup = \oint \frac{d\xi}{2\pi i}\xi \,b e^{\frac{in}{p_-}X^+(\xi)},\phantom{\bigg)}\\
A_n^+\lineup = -\sqrt{2}p_-\oint\frac{ d\xi}{2\pi i}\frac{1}{\xi} e^{\frac{in}{p_-}X^+(\xi)},\phantom{\bigg)}\\
A_n^-\lineup = \frac{1}{\sqrt{2}p_-}\oint \frac{d\xi}{2\pi i}\xi\Big(\!:\!Te^{\frac{in}{p_-}X^+(\xi)}\!:+j_\text{gh}\d e^{\frac{in}{p_-}X^+(\xi)}\Big).\phantom{\bigg)}
\end{eqnarray}
The zero modes transform according to
\begin{equation}S p_i S^{-1} = p_i,\ \ \ S p_\pm S^{-1} = p_\pm,\ \ \ S b_0 S^{-1} = b_0,\ \ \ S c_0 S^{-1} =  \frac{i}{2p_-}[Q,x^+].\end{equation}
The form of $A_n^-$ does not immediately follow from claim \ref{claim:1} since $X^-$ has nontrivial contractions with $X^+$. But it can be inferred from the relation
\begin{equation}[\Qlc,b_n] = \sqrt{2}p_-\alpha^-_n-nc_0 b_n,\end{equation}
since we know that $\Qlc$ transforms into $Q$.

The image of $L_0^\parallel$ in the covariant vector space will be written
\begin{equation}\L = S L_0^\parallel S^{-1}.\end{equation}
To compute this we write 
\begin{equation}L_0^\parallel = L_0^\text{lc} - L_0^\perp.\end{equation}
The zero mode of the lightcone energy-momentum tensor is invariant because $S$ preserves mode number. The zero mode of the transverse energy-momentum tensor has no contractions with $X^+$, so the transformation can be computed with claim \ref{claim:1}. In this way we find 
\begin{equation}
\L = L_0^\text{lc} +i p_-\oint\frac{d\xi}{2\pi i}\frac{1}{\d X^+(\xi)}\bigg[T^\perp(\xi)-2\{X^+,\xi\}\bigg]-\mathbb{I},\phantom{\Bigg)}
\label{eq:LL0par}\end{equation}
where
\begin{equation}\{X^+,\xi\} = \frac{\d^3 X^+(\xi)}{\d X^+(\xi)}-\frac{3}{2}\left(\frac{\d^2 X^+(\xi)}{\d X^+(\xi)}\right)^2\end{equation}
is the Schwarzian derivative of $X^+(\xi)$. Note that the combination 
\begin{equation}T^\perp(\xi)-2\{X^+,\xi\}\end{equation}
is a primary operator of weight $2$. Furthermore,  the second term in \eq{LL0par} is the zero mode  of a weight $1$ primary operator and is therefore conformally invariant, with the understanding that $1/\d X^+$ counts as a primary of weight $-1$. The operator $\L$ is BRST exact 
\begin{equation}
\L = \Big[Q,\b\Big],
\end{equation}
where 
\begin{equation}\b = S b_0^\parallel S^{-1}.\end{equation}
Transforming \eq{b0parallel} we find 
\begin{equation}
\b = b_0 + ip_-\oint \frac{d\xi}{2\pi i}\frac{b(\xi)}{\d X^+(\xi)}.\phantom{\Bigg)}\label{eq:bcov}
\end{equation}
We also have the relations
\begin{equation}(\b)^2 = 0,\ \ \ \ [\L,\b] = 0.\end{equation}
The operator $\L$ was introduced by Brink and Olive in the days of dual resonance models \cite{BrinkOlive} for the purpose of defining a projection operator onto physical states. See also \cite{Ramond_proj,Corrigan}. The operator $\b$ plays a central role in Freeman and Olive's proof of the no-ghost theorem \cite{Freeman}. Let us explain the meaning of $1/\d X^+$ which appears in these expressions. It can be defined as a geometric series
\begin{equation}\frac{1}{\d X^+(\xi)} = \frac{1}{-\frac{ip_-}{\xi}+\d \widetilde{X}^+(\xi)}= \frac{i\xi}{p_-}\sum_{n=0}^\infty \left(-\frac{i\xi}{p_-}\d \widetilde{X}^+(\xi)\right)^n,
\end{equation}
which makes sense since $p_-\neq 0$ (by assumption) and $X^+$ has no self-contractions. The geometric series terminates after a finite number of terms when $\L$ and $\b$ act on a Fock space state, since $\L$ and $\b$ preserve mode number and a state at fixed level can only accommodate a finite number of lightcone oscillator excitations. With these results we obtain the desired decomposition of the covariant vector space
\begin{equation}\Hcov = \HDDF\oplus\Hlong,\end{equation}
with
\begin{equation}\HDDF = \delta(\L)\Hcov,\ \ \ \ \Hlong = (\mathbb{I}-\delta(\L))\Hcov,\end{equation} 
where $\delta(\L)$ is the projector onto the kernel of $\L$.

\subsection{Integrating out longitudinal states}
\label{subsec:integrate}

We consider a generic covariant open bosonic string field theory with string products $Q,m_2,m_3,...$ defining a cyclic $A_\infty$ algebra. The string field $\Psicov\in\Hcov$ lives in the covariant vector space, and the equations of motion are
\begin{equation}Q\Psicov = -m_2(\Psicov,\Psicov)-m_3(\Psicov,\Psicov,\Psicov)+\text{higher orders}.\end{equation}
We separate the string field into transverse and longitudinal parts:
\begin{equation}\Psicov = S\Psilc+\Psilong,\label{eq:covdecomp}\end{equation}
where
\begin{equation}\Psilc\in\Hperplc,\ \ \ \ S\Psilc\in\HDDF,\ \ \ \ \Psilong\in\Hlong.\end{equation}
The lightcone effective field theory is defined by integrating out $\Psilong$. The remaining degrees of freedom are encoded in the string field $\Psilc$ which lives in the restricted lightcone vector space. In this way, the dynamical variable of the lightcone effective field theory is the same kind of object which appears in Kaku and Kikkawa's action. The equations of motion for $\Psilong$ are given by multiplying the full equations of motion by the projector onto longitudinal states:
\begin{equation}
Q\Psilong = -(\mathbb{I}-\delta(\L))\Big[m_2(\Psicov,\Psicov)+m_3(\Psicov,\Psicov,\Psicov)+\text{higher orders}\Big].\label{eq:longEOM}
\end{equation}
These equations can be solved perturbatively in powers of the string field to determine $\Psilong$ as a function of $\Psilc$. To find a solution it is helpful to fix a gauge. The natural gauge choice is
\begin{equation}\b \Psilong = 0.\label{eq:longgauge}\end{equation}
This gauge is always reachable perturbatively. To see this, note the identity
\begin{equation}\Psilong = \left[Q,\frac{\b}{\L}\right]\Psilong.\end{equation}
Since $\L$ is nonvanishing in $\Hlong$ (by definition) its inverse is well-defined. We can rearrange this equation to read
\begin{equation}\frac{\b}{\L} Q\Psilong = \Psilong - Q\left(\frac{\b}{\L}\Psilong\right).
\end{equation}
The state on the left hand side lives in $\Hlong$ and satisfies the gauge condition \eq{longgauge}. The right hand side shows that this can be achieved by a linearized gauge transformation. An alternative and more traditional gauge choice would be Siegel gauge $b_0\Psilong = 0$. However, technically Siegel gauge is not always achievable, since $\Hlong$ contains states in the kernel of $L_0$, for example 
\begin{equation}  c_0 e^{ik\cdot X(0,0)}|0\rangle,\ \ \ \ k^2=0.\end{equation}
The solution for $\Psilong$ in Siegel gauge is expected to exhibit singularities related to the breakdown of the gauge condition on the kernel of $L_0$. Since the kernel of $L_0$ is part of a continuous spectrum, it may be possible to work around this for some purposes. However, the first choice of gauge \eq{longgauge} is more natural. 

The gauge fixed equations of motion for $\Psilong$ are obtained by multiplying \eq{longEOM} by $\b$ and dividing by $\L$:
\begin{eqnarray}
\Psilong = -\frac{\b}{\L}\bigg[m_2(\Psicov,\Psicov) +m_3(\Psicov,\Psicov,\Psicov) +\text{higher orders}\bigg].\  \ \ \ \ \ \ \ \ \ \label{eq:gflongEOM}
\end{eqnarray}
We do not need to include the projection onto longitudinal states explicitly here, since any state proportional to $\b$ is automatically longitudinal. The operator which multiplies the whole expression is analogous to a propagator, but it contains only longitudinal degrees of freedom as intermediate states. Therefore we call it the {\it longitudinal propagator}. To solve for $\Psilong$, we substitute $\Psicov = S\Psilc+\Psilong$ on the right hand side and recursively substitute the equation into itself until $\Psilong$ is eliminated out to any desired order in the string field. This determines $\Psilong$ perturbatively as a function of $\Psilc$:
\begin{eqnarray}
\Psilong\lineup  = -\frac{\b}{\L}\Bigg[m_2(S \Psilc,S\Psilc)+m_3(S \Psilc,S \Psilc,S \Psilc)\nonumber\\
\lineup\ \ \ \ \ \ \ \ \ \ \ \ \ \ \ \ \ \ \ \ \ \ \ \ \ \ \ \ \ \ \  - m_2\left(S\Psilc, \frac{\b}{\L}m_2(S\Psilc,S\Psilc)\right)\nonumber\\
\lineup\ \ \ \ \ \ \ \ \ \ \ \ \ \ \ \ \ \ \ \ \ \ \ \ \ \ \ \ \ \ \   - m_2\left(\frac{\b}{\L}m_2(S\Psilc,S\Psilc), S\Psilc\right)+\text{higher orders}\Bigg].\ \ \ \ \ \ \label{eq:I3}
\end{eqnarray}
Plugging this solution into the action,
\begin{eqnarray}
S_\text{cov} \lineup = -\frac{1}{2}\omega(\Psicov,Q\Psicov) -\frac{1}{3}\omega(\Psicov,m_2(\Psicov,\Psicov))\nonumber\\
\lineup\ \ \ \ \ \ \ \ \ \ \ \ \ \ \ \ \ \ \ \ \ \ \ \ \ \,-\frac{1}{4}\omega(\Psicov,m_3(\Psicov,\Psicov,\Psicov))+\text{higher orders},
\end{eqnarray}
gives the lightcone effective action:
\begin{eqnarray}
S_\text{lc}\lineup  = -\frac{1}{2}\omega(\Psilc,c_0L_0\Psilc)-\frac{1}{3}\omega(\Psilc,S^{-1}m_2(S\Psilc,S\Psilc))\nonumber\\
\lineup\ \ \ -\frac{1}{4}\Bigg[\omega(\Psilc,S^{-1}m_3(S\Psilc,S\Psilc,S\Psilc))\nonumber\\
\lineup\ \ \ \ \ \ \ \ \ \ -\omega\left(\Psilc,S^{-1}m_2\left(S\Psilc, \frac{\b}{\L}m_2(S\Psilc,S\Psilc)\right)\right)\nonumber\\
\lineup\ \ \ \ \ \ \ \ \ \ \left.-\omega\left(\Psilc,S^{-1}m_2\left(\frac{\b}{\L}m_2(S\Psilc,S\Psilc),S\Psilc\right)\right)\right]\nonumber\\
\lineup\ \ \ +\text{higher  orders}.
\end{eqnarray}
The terms with the longitudinal propagator are ``effective" vertices generated by integrating out the longitudinal degrees of freedom. There is only one effective vertex at quartic order which we have split into two terms by writing
\begin{eqnarray}
\lineup \omega\left(m_2(S\Psilc,S\Psilc),\frac{\b}{\L}m_2(S\Psilc,S\Psilc)\right) \nonumber\\
\lineup \ \ \ \ \ \ \ \ = -\omega\left(\Psilc,S^{-1}m_2\left(S\Psilc, \frac{\b}{\L}m_2(S\Psilc,S\Psilc)\right)\right) \nonumber\\
\lineup \ \ \ \ \ \ \ \ = -\omega\left(\Psilc,S^{-1}m_2\left(\frac{\b}{\L}m_2(S\Psilc,S\Psilc),S\Psilc\right)\right).
\end{eqnarray}
The point of writing it in this way is to display the cyclic string products of the lightcone effective action:
\begin{eqnarray}
m_1^\text{eff}\lineup  = c_0 L_0,\phantom{\bigg)}\\
m_2^\text{eff}\lineup = \delta(L_0^\parallel)S^{-1}m_2 \, S\otimes S,\phantom{\bigg)}\\
m_3^\text{eff}\lineup = \delta(L_0^\parallel)S^{-1} m_3\, S\otimes S\otimes S  - \delta(L_0^\parallel)S^{-1} m_2 \left( \mathbb{I}\otimes \frac{\b}{\L}m_2 
 + \frac{\b}{\L}m_2  \otimes\mathbb{I} \right)S\otimes S\otimes S,\phantom{\bigg)}\ \ \ \ \ \ \ \ \ \ \\
\lineup\ \vdots\ .\nonumber
\end{eqnarray} 
The lightcone effective action is then written
\begin{equation}
S_\text{lc} = -\frac{1}{2}\omega(\Psilc,c_0L_0\Psilc)-\frac{1}{3}\omega(\Psilc,m_2^\text{eff}(\Psilc,\Psilc))-\frac{1}{4}\omega(\Psilc,m_3^\text{eff}(\Psilc,\Psilc,\Psilc))+\text{higher orders}.\ \ \ \ 
\end{equation}
The string products of the lightcone effective action define a cyclic $A_\infty$ algebra, but as in the Kaku-Kikkawa theory, this follows trivially from the fact that the restricted lightcone vector space only allows states at ghost numbers $1$ and $2$. The vertices of the lightcone effective action can be organized into a Feynman graph expansion. For example, the quartic vertex may be understood as a sum of an elementary quartic interaction together with $s$ and $t$ channel diagrams containing the longitudinal propagator. This structure reflects the fact that effective vertices must ``add back in" the intermediate states which are missing from the lightcone gauge propagator when computing string amplitudes through Feynman diagrams. This is illustrated in figure \ref{fig:lightcone9}. All of this is exactly parallel to the derivation of low energy effective actions from string field theory described in \cite{SenEffective}. The only difference is we are integrating out a different subset of states.

We have written the lightcone effective action explicitly up to quartic order. The structure at higher orders can be described very efficiently using the idea of homotopy transfer~\cite{Markl}. The relation between homotopy transfer and effective field theory has been the subject of several recent works in string field theory \cite{ToruMatsunaga,Okawa_eff,Vosmera}, and in more general BV field theories \cite{Hohm}. See also \cite{Kajiura,Konopka,Jurco}. We will only state the results, and refer to these references for explanations and derivations. To start, we note that the covariant and restricted lightcone vector spaces are chain complexes related by chain maps
\begin{eqnarray}
\lineup \iota:\Hperplc \to \Hcov;\ \ \ \ \ \iota = S,\\
\lineup \tau:\Hcov \to \Hperplc;\ \ \ \ \tau = \delta(L_0^\parallel) S^{-1},
\end{eqnarray}
which satisfy
\begin{equation}Q\iota  = \iota \Qlc,\ \ \ \ \tau Q = \Qlc \tau.\end{equation}
The map $\iota$ is called the {\it inclusion} and $\tau$ is sometimes called the projection. In the present situation $\tau$ is not a projection operator, so instead we refer to it as the {\it transfer} map. The isomorphism of cohomologies follows from the existence of a chain homotopy $Q^\dag: \Hcov\to \Hcov;$ which satisfies
\begin{equation}[Q,Q^\dag] = \mathbb{I}-\iota\tau.\end{equation}
The chain homotopy in our setup is the longitudinal propagator:
\begin{equation}Q^\dag = \frac{\b}{\L}.\end{equation}
The chain homotopy and chain maps satisfy so-called {\it side conditions}
\begin{equation}(Q^\dag)^2= 0,\ \ \ \tau Q^\dag = Q^\dag\iota = 0,\end{equation}
which imply that the chain maps compose in both directions to give projectors:
\begin{equation}(\iota\tau)^2 = \iota \tau,\ \ \ \ \ (\tau\iota)^2 = \tau\iota.\end{equation}
The first is the projector onto DDF states, 
\begin{equation}\iota\tau = \delta(\L),\end{equation}
and the second is simply the identity operator on $\Hperplc$,
\begin{equation}\tau\iota = \mathbb{I}.\end{equation}
The objects $\iota,\tau,Q,\Qlc,Q^\dag$ characterize the relation between free covariant and free lightcone string field theories. The idea behind homotopy transfer is to deform these objects in such a way as to describe the relation between the {\it interacting} theories. Doing this requires working on the tensor algebras $T\Hcov$ and $T\Hperplc$, and introducing corresponding operators 
\begin{eqnarray}
\bm{\iota}: \lineup T\Hperplc\to T\Hcov,\\
\bm{\tau}:\lineup  T\Hcov\to T\Hperplc,\\
\QQ:\lineup  T\Hcov \to T\Hcov,\\
\QQlc:\lineup  T\Hlc \to T\Hlc,\\
\QQ^\dag:\lineup  T\Hcov\to T\Hcov.
\end{eqnarray} 
The operators $\bm{\iota}$ and $\bm{\tau}$ are cohomomorphisms defined by
\begin{eqnarray}
\bm{\iota}\pi_n \lineup = (\underbrace{\iota\otimes ... \otimes \iota}_{n \text{ times}}) \pi_n = (\underbrace{S \otimes ... \otimes S}_{n \text{ times}})\pi_n,  \\
\bm{\tau}\pi_n \lineup = (\underbrace{\tau\otimes ... \otimes \tau}_{n \text{ times}}) \pi_n = \Big(\underbrace{\delta(L_0^\parallel)\otimes ... \otimes \delta(L_0^\parallel)}_{n \text{ times}}\Big)\Big(\underbrace{S^{-1}\otimes ... \otimes S^{-1}}_{n \text{ times}}\Big)\pi_n .
\end{eqnarray}
The operators $\QQ$ and $\QQlc$ are coderivations defined by
\begin{eqnarray}
\QQ\pi_n \lineup = \left(\sum_{k=0}^{n-1}\mathbb{I}^{\otimes k}\otimes Q \otimes\mathbb{I}^{\otimes n-k-1}\right)\pi_n,\\
\QQlc\pi_n \lineup = \left(\sum_{k=0}^{n-1}\mathbb{I}^{\otimes k}\otimes \Qlc \otimes\mathbb{I}^{\otimes n-k-1}\right)\pi_n,
\end{eqnarray}
and finally
\begin{equation}
\QQ^\dag\pi_n = \left(\sum_{k=0}^{n-1}\mathbb{I}^{\otimes k}\otimes Q^\dag \otimes(\iota\tau)^{\otimes n-k-1}\right)\pi_n.
\end{equation}
We may confirm that all of the important relations given above hold analogously on the tensor algebra:
\begin{eqnarray}
\lineup\QQ\bm{\iota}  = \bm{\iota}\QQlc,\ \ \ \ \ \bm{\tau}\QQ = \QQlc\bm{\tau},\\
\lineup \ \ \ \  [\QQ,\QQ^\dag]  = \mathbb{I}_{T\Hcov} - \bm{\iota}\bm{\tau},\\
\lineup \  (\QQ^\dag)^2 = \bm{\tau} \QQ^\dag = \QQ^\dag\bm{\iota} = 0,\\
\lineup\ \ \ (\bm{\iota}\bm{\tau})^2 = \bm{\iota}\bm{\tau},\ \ \ \ \ \bm{\tau} \bm{\iota} = \mathbb{I}_{T\Hperplc}.
\end{eqnarray}
At the interacting level, $Q$ and $\Qlc$ are replaced by $A_\infty$ structures represented by coderivations $\mm$ and $\mmeff$. The transfer and inclusion maps $\tau$ and $\iota$ are replaced by interacting versions represented by cohomomorphisms $\TT$ and $\II$. The chain homotopy $Q^\dag$ is replaced by an interacting version $\mm^\dag$ with analogous coalgebra properties to $\QQ^\dag$. This is done so that all of the above relations continue to hold on the tensor algebra at the interacting level:
\begin{eqnarray}
\lineup\mm\II   = \II\mmeff,\ \ \ \ \ \TT\mm = \mmeff\TT, \\
\lineup \ \ \ \  [\mm,\mm^\dag]  = \mathbb{I}_{T\Hcov} - \II\TT,\\
\lineup \  (\mm^\dag)^2 = \TT\mm^\dag = \mm^\dag\II  = 0,\\
\lineup\ \ \ (\II\TT)^2=\II\TT,\ \ \ \ \ \TT\II = \mathbb{I}_{T\Hperplc}.
\end{eqnarray}
Explicit formulas can be given in terms of the interacting part of the $A_\infty$ structure of the covariant theory, 
\begin{equation}\delta\mm = \mm-\QQ,\end{equation}
as follows:
\begin{eqnarray}
\mmeff\lineup = \QQlc + \bm{\tau} \delta \mm\frac{1}{\mathbb{I}_{T\Hcov}+\QQ^\dag\delta\mm}\bm{\iota},\\
\TT\lineup = \bm{\tau} \frac{1}{\mathbb{I}_{T\Hcov} +\delta\mm\QQ^\dag},\\
\II\lineup = \frac{1}{\mathbb{I}_{T\Hcov}+\QQ^\dag\delta\mm}\bm{\iota},\label{eq:II}\\
\mm^\dag \lineup = \QQ^\dag\frac{1}{\mathbb{I}_{T\Hcov}+\delta\mm \QQ^\dag}.
\end{eqnarray}
The first equation gives a closed form expression for the string products of the lightcone effective action to all orders. The Feynman diagrams appear when expanding the second piece as a geometric series. If 
\begin{equation}\frac{1}{1-\Psilc} \,=\, 1_{T\Hlc} \,+\,  \Psilc\, +\, \Psilc\!\otimes\!\Psilc \,+\, \Psilc\!\otimes\!\Psilc\!\otimes\!\Psilc +\text{higher orders}\end{equation}
is the group-like element of the tensor algebra generated by $\Psilc$, the relation 
\begin{equation}\Psicov = \pi_1 \II\frac{1}{1-\Psilc} \label{eq:lccov}\end{equation}
tells us how the string field in the covariant theory is written as a function of the lightcone string field. In particular, this encodes the solution to the longitudinal equations of motion, which we worked out to cubic order in \eq{I3}. 

\subsection{Lightcone gauge}
\label{subsec:lightcone}

The discussion so far portrays lightcone string field theory as a kind of effective field theory. But this characterization seems peculiar. It would appear more natural if lightcone string field theory had emerged by gauge fixing a covariant string field theory, analogous to lightcone gauge fixing of Yang-Mills theory. In fact, the theory can be understood in this way. We have not approached it from this point of view up to now, since it is not initially obvious what the lightcone gauge condition should be. We claim it is 
\begin{equation}\b \Psicov = 0,\ \ \ \ \ \ \text{(lightcone gauge)}.\label{eq:lcgauge}\end{equation}
This condition leads directly to lightcone effective field theory, and can be viewed as a string theory extension of the lightcone gauge condition in Yang-Mills theory.

\begin{figure}
\begin{center}
\resizebox{5in}{3in}{\includegraphics{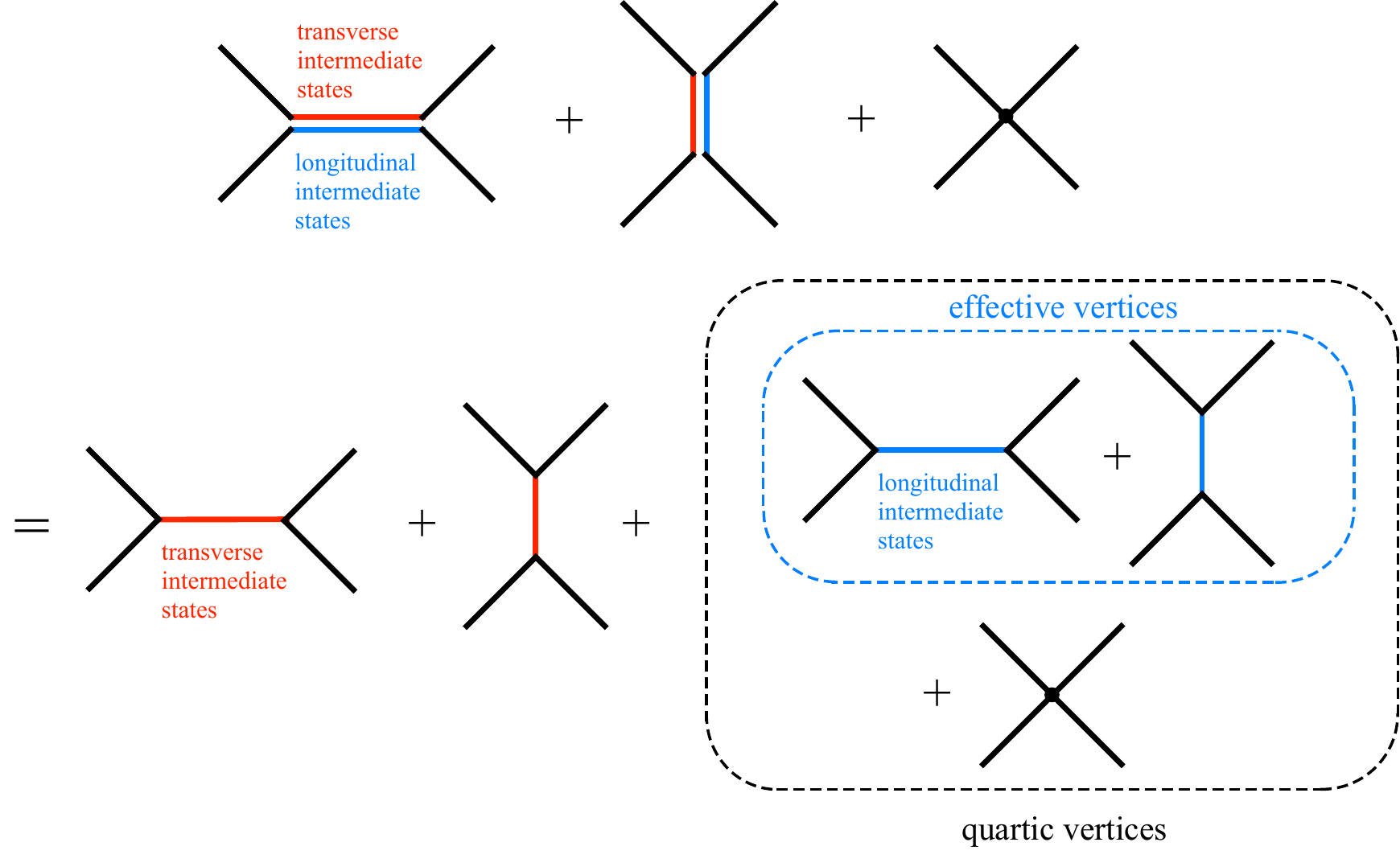}}
\end{center}
\caption{\label{fig:lightcone9} The Feynman graphs of covariant string field theory fixed to lightcone gauge can be reorganized into graphs only containing transverse states in the propagators, with the contribution from longitudinal intermediate states absorbed in the vertices.}
\end{figure}

We consider the gauge fixing at the classical level, without ghost fields and the BV formalism (which in any case should not be needed in lightcone gauge). Let us first recall the story in Siegel gauge. It is helpful to introduce a projector onto the gauge slice
\begin{equation}P_\text{Siegel} = b_0c_0,\end{equation}
so that the Siegel gauge condition can be expressed
\begin{equation}P_\text{Siegel}\Psicov=\Psicov.\end{equation}
If the string field is in Siegel gauge, we can simplify the kinetic term in the action as follows: 
\begin{eqnarray}\omega(\Psicov,Q\Psicov) \lineup = \omega(P_\text{Siege}\Psicov, Q P_\text{Siegel}\Psicov)\nonumber\\
\lineup =\omega\Big(\Psicov,(P_\text{Siegel}^\star QP_\text{Siegel})\Psicov\Big),
\end{eqnarray}
where $\star$ denotes BPZ conjugation. Noting
\begin{equation}P^\star_\text{Siegel}QP_\text{Siegel} = c_0L_0,\end{equation}
we obtain the familiar kinetic term for the string field in Siegel gauge
\begin{equation}\frac{1}{2}\omega(\Psicov,c_0L_0\Psicov).\end{equation}
The Siegel gauge propagator inverts the gauge fixed kinetic operator on the subspace of Siegel gauge states:
\begin{equation}\frac{b_0}{L_0}c_0L_0 = P_\text{Siegel}.\label{eq:Siegelprop}\end{equation}
Now consider the analogous story in lightcone gauge. The projector onto the gauge slice is slightly more complicated in this case because $\b$ supports cohomology. This is most readily seen in the lightcone vector space. One can check that all states in $\Hperplc$ are annihilated by $b_0^\parallel$. Moreover, such states cannot be $b_0^\parallel$ of something else, since acting $b_0^\parallel$ on a state always produces plus oscillators and/or $b$-ghost oscillator excitations. On the other hand, any $b_0^\parallel$-closed state in $\Hparallellc$ is automatically $b_0^\parallel$-exact since we can write
\begin{equation}\Psi = \frac{b_0^\parallel}{L_0^\parallel}\Qlc\Psi,\ \ \ \Psi\in\Hparallellc.\label{eq:bllco}\end{equation}
Therefore the cohomology of $b_0^\parallel$ is $\Hperplc$, and after applying the similarity transformation, we learn that the cohomology of $\b$ is the subspace of off-shell DDF states. This implies that the projector onto the lightcone gauge slice must be the identity operator when acting on the DDF subspace, and it cannot be proportional to $\b$. When acting on the longitudinal subspace, the projector onto the gauge slice can be given by transformation of \eq{bllco}. Adding these pieces together gives 
\begin{equation}P_\text{lc} = \delta(\L) +  \frac{\b}{\L}Q .\label{eq:proj1}\end{equation}
The projector onto the gauge slice is not unique, and this expression is not optimal since it does not directly display any simplification of the gauge fixed kinetic operator. It is somewhat analogous to choosing $(b_0/L_0) Q$ as the projector onto the gauge slice in Siegel gauge. What we are missing is the analogue of $c_0$ in lightcone gauge. To get it, note that all ghost number 1 states in the DDF subspace satisfy the Siegel gauge condition. Therefore we can multiply the first term in the projector by $b_0c_0$ without changing anything. For the second term, consider the operator 
\begin{eqnarray}\Delta^- \lineup = S\delta^-S^{-1},\nonumber\\
\lineup = \delta^-+\frac{\widehat{Q}+b_0 M}{p_-}-\frac{L_0^\text{lc}-1}{2p_-^2}\delta^+,
\end{eqnarray}
where the operators on the right hand side are given in \eq{deltam}-\eq{hatQ}. This satisfies
\begin{equation}[\b,p_-\Delta^-]=\L.\end{equation}
Therefore we can use it in place of the BRST operator in the second term of \eq{proj1}. In this way we arrive at a projector onto the lightcone gauge slice of the form
\begin{equation}P_\text{lc} = b_0 c_0 \delta(\L) +  \frac{\b}{\L}p_-\Delta^-, \label{eq:proj2}\end{equation}
and the lightcone gauge condition may be expressed 
\begin{equation}P_\text{lc}\Psicov = \Psicov.\end{equation}
This leads to a gauge fixed kinetic operator 
\begin{equation}
P_\text{lc}^\star Q P_\text{lc} = \delta(\L)c_0 L_0 +p_-\Delta^-,
\end{equation}
and paralleling \eq{Siegelprop} the propagator is found to be
\begin{equation}
\left({\text{lightcone gauge}\atop\text{propagator}}\right)=\ \frac{b_0}{L_0}\delta(\L) +\frac{\b}{\L}.\label{eq:lcprop}
\end{equation}
This is the Siegel gauge propagator restricted to DDF states plus the longitudinal propagator. It is clear that the Feynman graph expansion derived from this propagator can be reorganized into diagrams where only DDF states appear inside the propagators, and the contributions from longitudinal propagators are subsumed into the vertices. Again this is illustrated in figure \ref{fig:lightcone9}. This is precisely the Feynman graph expansion of the lightcone effective field theory. Another way to understand this is from the gauge fixed kinetic term 
\begin{equation}
\frac{1}{2}\omega(\Psicov,Q\Psicov) = \frac{1}{2}\omega(\Psi_\text{DDF},c_0L_0\Psi_\text{DDF}) + \frac{1}{2}\omega(\Psilong,p_-\Delta^-\Psilong),
\label{eq:gfkinetic}
\end{equation}
where we separated the string field into DDF and longitudinal parts. The essential observation is that the operator $p_-\Delta^-$ contains no derivatives with respect to lightcone time $x^+$, so the string field $\Psilong$ is not dynamical. Therefore we can integrate it out, and this leads precisely to the lightcone effective field theory as described in the last subsection.

Now let's make the connection to lightcone gauge in Yang-Mills theory. For present purposes it is sufficient to consider a free Maxwell field $A_\mu$. The lightcone gauge condition is 
\begin{equation}
A_- = 0.
\end{equation}
The Maxwell action in this gauge takes the form
\begin{eqnarray}
S \lineup = -\frac{1}{4}\int d^D x F_{\mu\nu}F^{\mu\nu}\nonumber\\
\lineup = \int d^D x\left[
\frac{1}{2}A_i \Box A_i -\frac{1}{2}(\d_i A_i)^2 -\d_- A_+ \d_i A_i -\frac{1}{2}(\d_- A_+)^2\right].
\end{eqnarray}
We note that $A_+$ does not appear with derivatives with respect to $x^+$, and so is nondynamical. We may integrate it out by solving its equation of motion:
\begin{equation}\d_-^2A_+ = -\d_-(\d_i A_i).\end{equation}
Since we must assume $p_-\neq 0$, this is equivalent to
\begin{equation}\d_-A_+ = -\d_i A_i, \end{equation}
and plugging back into the action gives 
\begin{equation}
S = \frac{1}{2}\int d^D x\, A_i \Box A_i.\label{eq:gfYM}
\end{equation}
To compare this to string field theory in lightcone gauge, we consider the massless part of the string field
\begin{equation}
\Psicov = \int \frac{d^{26} k}{(2\pi)^{26}}\left[A_\mu(k)\alpha_{-1}^\mu c_1 + \frac{i}{\sqrt{2}}\beta(k) c_0\right]e^{ik\cdot X(0,0)}|0\rangle,
\end{equation}
where $A_\mu$ is the gauge field and $\beta$ is an auxiliary field. Evaluating the kinetic term gives
\begin{equation}
-\frac{1}{2}\omega(\Psicov,Q\Psicov) = \int d^{26} x\left[\frac{1}{2} A_\mu \Box A^\mu + \beta \d_\mu A^\mu -\frac{1}{2}\beta^2\right].
\end{equation}
The operator $\b$ takes the form
\begin{eqnarray}
\b = \frac{1}{\sqrt{2}p_-}\alpha_1^+b_{-1} + \text{other terms} ,
\end{eqnarray}
where the other terms annihilate the massless part of the string field. Imposing $\b\Psicov = 0$ therefore leads to 
\begin{equation}A_-=0,\end{equation}
which is precisely the lightcone gauge condition. Note that the auxiliary field $\beta$ is unaffected by the gauge condition, unlike Siegel gauge which requires $\beta=0$. The massless kinetic term in this gauge is
\begin{equation}
\frac{1}{2}\omega(\Psicov,Q\Psicov) = \int d^{26} x\left[\frac{1}{2}A_i \Box A_i +\beta(\d_i A_i +\d_- A_+)-\frac{1}{2}\beta^2\right].\label{eq:mixing}
\end{equation}
Now there are two nondynamical fields $A_+$ and $\beta$. Assuming $p_-\neq 0$ their respective equations of motion are
\begin{equation}\beta = 0,\ \ \ \beta = \d_i A_i +\d_- A_+.\end{equation}
Plugging back into the action gives \eq{gfYM}.

One might notice that the action \eq{mixing} contains mixing between dynamical and nondynamical fields, while in \eq{gfkinetic} there is no mixing. This is because the conventional oscillator basis does not cleanly separate the transverse and longitudinal parts of the string field. It is useful to see the transverse and longitudinal parts of the string field explicitly up to the massless level. We may define a basis for the transverse and longitudinal parts by transforming the corresponding basis in the lightcone vector space:
\begin{eqnarray}
S c_1 e^{ik\cdot X(0,0)}|0\rangle \lineup = c_1e^{ik\cdot X(0,0)}|0\rangle,\\
S \alpha_{-1}^i c_1 e^{ik\cdot X(0,0)}|0\rangle \lineup = \left(\alpha^i_{-1}-\frac{k^i}{k_-}\alpha_{-1}^+\right)c_1e^{ik\cdot X(0,0)}|0\rangle,\\
S \alpha_{-1}^+ c_1e^{ik\cdot X(0,0)}|0\rangle\lineup = \alpha_{-1}^+ c_1 e^{ik\cdot X(0,0)}|0\rangle,\\
S\alpha_{-1}^- c_1 e^{ik\cdot X(0,0)}|0\rangle \lineup = \left(\alpha_{-1}^- +\frac{k_i}{k_-}\alpha^i_{-1} -\frac{k_i k_i}{2k_-^2}\alpha_{-1}^+\right)c_1 e^{ik\cdot X(0,0)}|0\rangle,\\
S c_0 e^{ik\cdot X(0,0)}|0\rangle \lineup = \left(c_0-\frac{1}{\sqrt{2}p_-}\alpha_{-1}^+ c_1\right)e^{ik\cdot X(0,0)}|0\rangle.
\end{eqnarray}
If we label the fields of this basis respectively as $\Tlc,\Alc_\mu,\betalc$ we find that
\begin{eqnarray}
T \lineup = \Tlc,\\
A_i \lineup = \Alc_i+\frac{k_i}{k_-}\Alc_-,\label{eq:AiAilc}\\
A_+ \lineup = \Alc_+ -\frac{k_i\Alc_i}{k_-}-\frac{k_ik_i}{2k_-^2} \Alc_--\frac{i}{2 k_-}\betalc,\\
A_-\lineup = \Alc_-\\
\beta \lineup = \betalc.\label{eq:betalong}
\end{eqnarray}
Here we see that the tachyon is purely transverse, $A_i$ and $A_+$ are mixtures of transverse and longitudinal degrees of freedom, while $A_-$ and $\beta$ are purely longitudinal. In lightcone gauge $A_-= 0$, both the tachyon and $A_i$ are purely transverse, $\beta$ is purely longitudinal, but $A_+$ is still a mixture of transverse and longitudinal components. 

\section{From Kugo-Zwiebach to lightcone string field theory}
\label{sec:KZ}

In this section we show that Kaku and Kikkawa's lightcone string field theory appears as the lightcone effective action derived from the Kugo-Zwiebach string field theory. This requires demonstrating that lightcone vertices are unchanged by integrating out longitudinal degrees of freedom. Specifically, the 2-string product must satisfy 
\begin{equation}
\delta(L_0^\parallel)\mlc_2  =  \delta(L_0^\parallel) S^{-1}   \mlc_2 \,S\otimes S,\label{eq:3transfer}
\end{equation}
the 3-string product must satisfy
\begin{equation}
\delta(L_0^\parallel)\mlc_3  =  \delta(L_0^\parallel)S^{-1} \mlc_3 \, S\otimes S\otimes S-\delta(L_0^\parallel) S^{-1}\mlc_2\left(\mathbb{I}\otimes \frac{\b}{\L}\mlc_2 + \frac{\b}{\L}\mlc_2\otimes\mathbb{I}\right)\, S\otimes S\otimes S,
\end{equation}
and all higher products of the lightcone effective action, generated by joining lightcone vertices with longitudinal propagators, must vanish identically. We refer to this property as {\it (homotopy) transfer invariance}, where presently the notion of homotopy transfer is given by integrating out the longitudinal states. It turns out that  lightcone vertices are the only surface state vertices (up to attaching stubs) which are transfer invariant in this sense. This gives another way to understand the special status of Mandelstam's interacting string picture.  

Before going further, let us rule out what seems like the simplest explanation for transfer invariance.  This is the idea that longitudinal states carry some kind of parity which forces them to appear in pairs in lightcone vertices. If this were true, we could solve their equations of motion by setting them equal to zero, and there would be no ``backreaction" to alter the couplings between transverse states. However, one can check that the cubic coupling between two tachyons and the auxiliary field $\beta$ does not vanish. (As shown in \eq{betalong}, the auxiliary field $\beta$ is purely longitudinal). Therefore longitudinal states are not forced to interact in pairs, and the mechanism behind transfer invariance is different.

\subsection{Transfer invariance of lightcone vertices}
\label{subsec:transfer}

We start by demonstrating transfer invariance of the cubic lightcone vertex. In terms of the unshifted vertex (see appendix \ref{app:vertex}), this can be stated as
\begin{equation}\Vlc_3(SA,SB,SC) = \Vlc_3(A,B,C),\ \ \ \ A,B,C\in \Hperplc.\label{eq:V3transfer}\end{equation}
The basic mechanism is this: If the cubic lightcone vertex couples states without minus oscillator excitations, all plus oscillator excitations in those states can be set to zero. Since the only thing which distinguishes a state in $\HDDF$ from a state in $\Hperplc$ are additive contributions containing plus oscillators (and no minus oscillators), this implies transfer invariance of the cubic vertex. 

Applying a plus creation operator to the vertex brings down a term linear in momenta and a term linear in plus annihilation operators:
\begin{equation}
\langle\Vlc_3|\alpha^{+,s}_{-n} = n\langle \Vlc_3|\Big(\alpha^{+,r}_m\bar{N}_{mn}^{rs}+\sqrt{2}p_-^r \bar{N}_{0n}^{rs}\Big),\label{eq:apV}
\end{equation}
where $m,n=1,...,\infty$ are mode labels and $r,s=1,2,3$ are vector space labels, $\bar{N}_{mn}^{rs},\bar{N}_{0n}^{rs}$ are Neumann coefficients, and repeated labels are summed. If we evaluate \eq{apV} against states which do not contain minus oscillators, the first term will vanish. If the second term also vanishes,
\begin{equation}p_-^r \bar{N}_{0n}^{rs}=0,\label{eq:oscid}\end{equation}
the cubic lightcone vertex will be transfer invariant, as explained above. With the convention 
\begin{equation}\bar{N}_{0m}^{1s}+\bar{N}_{0m}^{2s}+\bar{N}_{0m}^{3s} = 0,\end{equation}
the relevant Neumann coefficients take the form \cite{Mandelstam1,Cremmer}
\begin{equation}\bar{N}_{0m}^{rs} = M^{rt}p_-^t\bar{N}_m^s,\label{eq:Neumann}\end{equation}
where $\bar{N}_m^s$ is independent of $r$ and 
\begin{equation}M^{rt}=\frac{1}{3}\left(\begin{matrix}0 & 1 & -1 \\ -1  &  0 &  1 \\ 1  & -1 &0\end{matrix}\right)^{rt}.\end{equation}
Now it is clear that \eq{oscid} holds because the matrix $M^{rt}$ is antisymmetric. 

Let us understand this result from a different angle which generalizes more readily to the higher vertices. The statement that plus oscillators vanish when minus oscillators are absent is equivalent to the statement that an insertion of $\d X^+$ in the vertex can be replaced with an insertion of its zero mode,
\begin{equation}\Vlc_3\Big(\d X^+(\xi) A,B,C\Big) = -\frac{i}{\xi}\Vlc_3(p_-A,B,C),\label{eq:claim3}\end{equation}
when minus oscillators are absent. The left hand side of \eq{claim3} may be represented as a correlation function on the upper half plane,
\begin{equation}
\Vlc_3\Big(\d X^+(\xi) A,B,C\Big) = \left(\frac{d (\flc{1})^{-1}(u)}{d u}\right)^{-1} \Big\langle \d X^+(u) \,\flc{1}\circ A(0) \flc{2}\circ B(0) \flc{3}\circ C(0)  \Big\rangle_\text{UHP},
\end{equation}
where the conformal transformations $\flc{1},\flc{2},\flc{3}$ are derived from the Mandelstam mapping and are given in appendix \ref{app:lc3_vertex}, and $u = \flc{1}(\xi)$. The conformally transformed vertex operators take the form
\begin{equation}
\flc{1}\circ A(0) =(...) e^{i k^A \cdot X(1,1)},\ \ \ \flc{2}\circ B(0) = (...) e^{ik^B\cdot X(0,0)},\ \ \ \  \flc{3}\circ C(0)=(...) e^{ik^C\cdot X(\infty,\infty)},
\end{equation}
where  $(...)$ represent operator insertions which characterize the states $A,B$ and $C$. The only important thing about these insertions is that they do not generate contractions with $\d X^+$, since $A,B$ and $C$ do not contain minus oscillator excitations. The operator $\d X^+$ only produces contractions with the plane wave vertex operators at the punctures. This allows us to eliminate $\d X^+$ using the identity \eq{dXdelete} given in appendix \ref{app:free}. The result is 
\begin{equation}
\Vlc_3\Big(\d X^+(\xi) A,B,C\Big) = - i\left(\frac{d (\flc{1})^{-1}(u)}{du}\right)^{-1}\left(\frac{k_-^A}{u-1}+\frac{k_-^B}{u}\right) \Vlc_3(A,B,C).
\end{equation}
Evaluating the derivative using \eq{flc31} we find that the factor generated by contractions with $\d X^+$ cancels out, leaving
\begin{equation}
\Vlc_3\Big(\d X^+(\xi) A,B,C\Big) = -\frac{ik_-^A}{(\flc{1})^{-1}(u)}\Vlc_3(A,B,C) =  -\frac{i k_-^A}{\xi}\Vlc_3(A,B,C),
\end{equation}
which establishes transfer invariance. This result relies on a seemingly accidental relation between the geometry of the cubic lightcone vertex and correlation functions of $\d X^+$ in the presence of plane wave vertex operators.

\begin{figure}
\begin{center}
\resizebox{4in}{3.2in}{\includegraphics{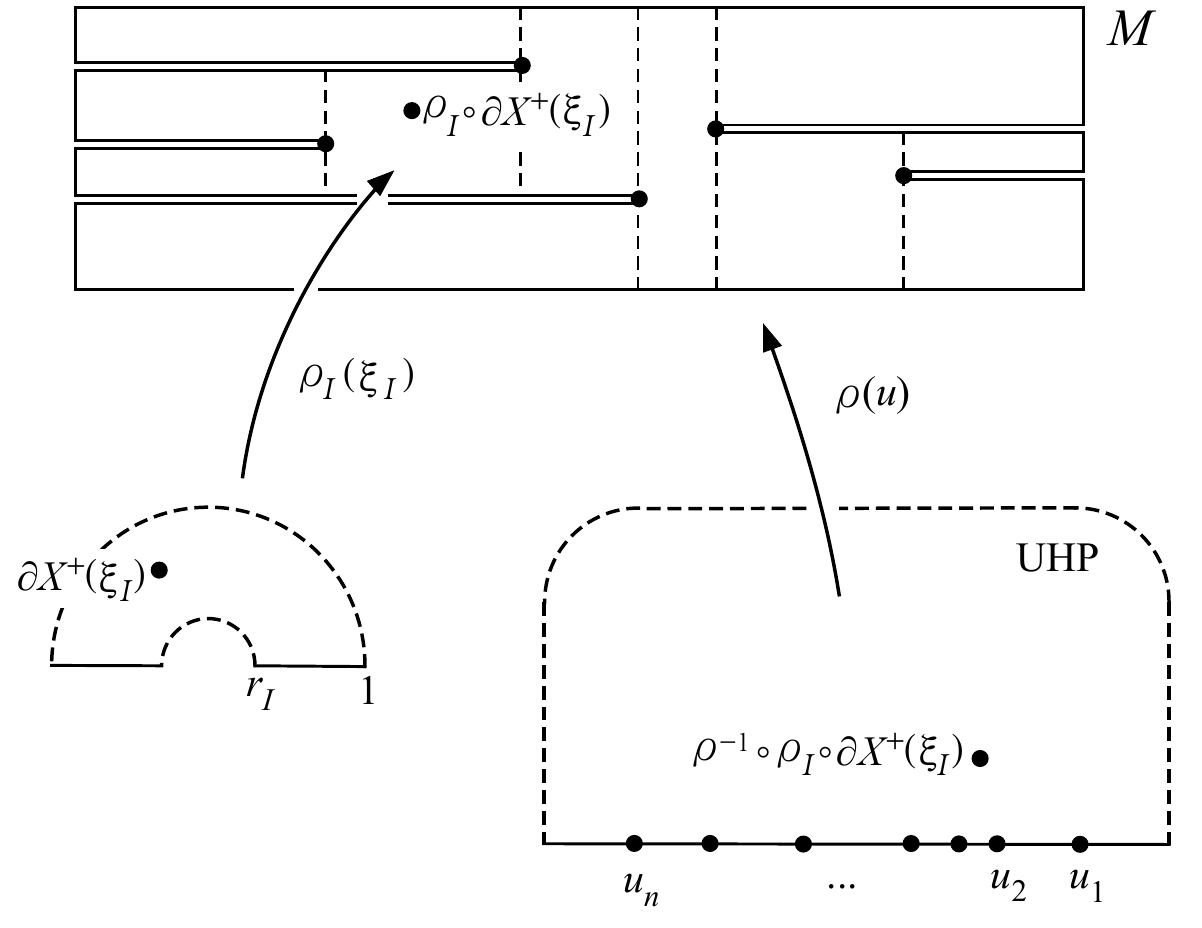}}
\end{center}
\caption{\label{fig:lightcone10} The relation between the surface $M$ of a generic Mandelstam diagram, the local coordinates $\xi_I$ on the strip domains, and the upper half plane.}
\end{figure}

This generalizes as follows. Consider a correlation function on a surface $M$ corresponding to a generic tree-level Mandelstam diagram. This is illustrated in figure \ref{fig:lightcone10}. The interaction coordinate $\rho$ on $M$ is related to the upper half plane coordinate $u$ through the Mandelstam mapping,
\begin{equation} \rho(u) = \sum_{I=1}^n \frac{k_-^I}{\pi}\ln(u-u_I),\label{eq:Mandelstam}\end{equation}
where $k_-^I$ are the minus momenta of the $n$ external states of the diagram and $u_I$ are the locations of the corresponding punctures in the upper half plane. The surface $M$ is given by gluing together some number of strip domains $\rho_I$. For $I=1,...,n$ we take these to be the strips which extend to infinity  and represent the external states. The remainder of the strip domains for $I\geq n+1$ represent propagators. On each strip domain we place a local coordinate $\xi_I$ which belongs to the unit half-disk minus a smaller concentric half-disk
\begin{equation}r_I\leq |\xi_I|\leq 1, \ \ \ \mathrm{Im}(\xi_I)>0.\end{equation}
For the external states we take $r_I = 0$ for $I=1,...,n$, since the full unit half-disk is needed to cover strips which extend to infinity. In this way the strip domain is represented as
\begin{equation}\rho_I = \rho_I(\xi_I) = \tau_I + \frac{k_-^I}{\pi}\ln\xi_I,\end{equation}
where $k_-^I$ is the width of the strip, which corresponds to the minus momentum flowing through that part of the diagram, and $\tau_I$ is a shift which appropriately places the strip domain within the larger surface. We now consider a correlation function on $M$ of the following form:
\begin{itemize}
\item Plane wave vertex operators $e^{i k^I\cdot X(0,0)}$ are inserted on the external strips with the maps $\rho_I(\xi_I)$ for $I=1,...,n$.
\item The operator $\d X^+(\xi_I)$ is inserted on one strip domain with the map $\rho_I(\xi_I)$ for some $I$.
\item Various other operators appear which do not generate contractions with $\d X^+(\xi_I)$.
\end{itemize}
We claim that this correlation function is unchanged if we replace $\d X^+(\xi_I)$ with its zero mode $-ik_-^I/\xi_I$. The correlation function may be represented in the upper half plane as
\begin{equation}
\Big\langle\big(\rho^{-1}\circ \rho_I\circ\d X^+(\xi_I)\big) e^{ik^1\cdot X(u_1,u_1)}\,...\,e^{ik^n\cdot X(u_n,u_n)}\big(\, ...\,\big)\Big\rangle_\text{UHP},
\end{equation}
where $(...)$ represent operators which do not have contractions with $\d X^+$. Defining $u = \rho^{-1}(\rho_I(\xi_I))$ this can be rewritten as
\begin{equation}
\frac{d\rho_I(\xi_I)}{d\xi_I} \left(\frac{d \rho(u)}{du}\right)^{-1}\Big\langle\d X^+(u)\, e^{ik^1\cdot X(u_1,u_1)}\,...\,e^{ik^n\cdot X(u_n,u_n)}\big(\, ...\,\big)\Big\rangle_\text{UHP}.
\end{equation}
Using \eq{dXdelete} we can remove the $\d X^+$ insertion from the correlator:
\begin{equation}
-i\frac{d\rho_I(\xi_I)}{d\xi_I} \left(\frac{d \rho(u)}{du}\right)^{-1}\left(\sum_{I=1}^n\frac{k_-^I}{u-u_I}\right)\Big\langle e^{ik^1\cdot X(u_1,u_1)}\,...\,e^{ik^n\cdot X(u_n,u_n)}\big(\, ...\,\big)\Big\rangle_\text{UHP}.
\end{equation}
It is immediately clear that the sum in front is precisely the derivative of the Mandelstam mapping \eq{Mandelstam} times $\pi$. Thus the correlator is expressed as
\begin{equation}
-i\pi\frac{d\rho_I(\xi_I)}{d\xi_I}\Big\langle e^{ik^1\cdot X(u_1,u_1)}\,...\,e^{ik^n\cdot X(u_n,u_n)}\big(\, ...\,\big)\Big\rangle_\text{UHP}=-\frac{i k_-^I}{\xi_I}\Big\langle e^{ik^1\cdot X(u_1,u_1)}\,...\,e^{ik^n\cdot X(u_n,u_n)}\big(\, ...\,\big)\Big\rangle_\text{UHP},
\end{equation}
which establishes the result. Note that if $\d X^+$ is inserted in the global coordinate $\rho$ on the Mandelstam diagram, instead of the local coordinate $\xi_I$, the above discussion implies that it can be replaced by a constant equal to $i\pi$.

From this it follows that the quartic lightcone vertex is preserved by the similarity transformation,
\begin{equation}\Vlc_4(S A,SB,SC,SD) = \Vlc_4(A,B,C,D),\end{equation}
when the states live in $\Hperplc$. To see what happens to contributions from the longitudinal propagator, we consider the Schwinger representation
\begin{equation}
\frac{\b}{\L} = \b \int_0^\infty dt\,e^{-t\L}.
\end{equation}
Using \eq{LL0par}, we can write $\L$ in terms of the total energy-momentum tensor as
\begin{equation}
\L = L_0 -2p_+p_- +i p_-\oint\frac{d\xi}{2\pi i}\frac{T^\perp(\xi)-2\{X^+,\xi\}}{\d X^+(\xi)}.
\end{equation}
Then the longitudinal propagator is expressed
\begin{equation}
\frac{\b}{\L} =\int_0^\infty dt\, \left(\b e^{-t\left(-2p_+p_- +i p_-\oint\frac{d\xi}{2\pi i}\frac{T^\perp(\xi)-2\{X^+,\xi\}}{\d X^+(\xi)}\right)}\right)e^{-t L_0}.
\end{equation}
The integrand can be interpreted as a conventional propagator strip containing a (rather complicated) operator insertion. This implies that the vertices of the lightcone effective action can be represented as correlation functions on Mandelstam diagrams, where the internal propagator strips contain this complicated operator. Since the operator does not generate contractions with $\d X^+$, we are in the situation discussed in the previous paragraph, and we can replace $\d X^+$ with its zero mode. When we do this we find that 
\begin{equation}
\b = b_0 +ip_-\oint \frac{d\xi}{2\pi i}\frac{b(\xi)}{\d X^+(\xi)} = b_0 -\oint\frac{d\xi}{2\pi i}\xi b(\xi) = 0.
\end{equation}
Therefore the longitudinal propagators vanish. This completes the proof of transfer invariance of the lightcone vertices.

\subsection{Lightcone gauge and the soft string problem}
\label{subsec:genericcubic}

It is interesting to ask what happens when we evaluate the lightcone effective action in a generic covariant string field theory. A full answer to this question requires understanding how to evaluate correlators containing the longitudinal propagator, a question which we will not address in this paper. However, as a first step we can evaluate the cubic vertex of the lightcone effective action. If $V_3$ is the vertex of the parent theory, the transferred cubic vertex is given by 
\begin{equation}
\Veff_3(A,B,C) = V_3(SA,SB,SC),
\end{equation}
where $A,B,C\in\Hperplc$. We will evaluate this in the oscillator basis. We write the states as 
\begin{eqnarray}
A \lineup = \alpha_{-n}^{i}\alpha_A  c_1 e^{ik^A\cdot X(0,0)}|0\rangle,\nonumber\\
B \lineup = \alpha_B c_1 e^{ik^B\cdot X(0,0)}|0\rangle,\nonumber\\
C \lineup = \alpha_C  c_1 e^{ik^C\cdot X(0,0)}|0\rangle,
\end{eqnarray}
where $\alpha_A,\alpha_B$ and $\alpha_C$ are collections of transverse oscillators which we do not write out explicitly. Presently, our focus is understanding what happens to the first oscillator $\alpha_{-n}^i$ in the state $A$. Applying the similarity transformation gives
\begin{eqnarray}
\langle \Veff_3|A\otimes B\otimes C\lineup = \langle V_3| SA\otimes SB\otimes SC\nonumber\\
 \lineup = \langle V_3| \left(A_{-n}^{i} S \alpha_A S^{-1} c_1 e^{i\left(k^A+\frac{n}{2k_-^A}\delta^+\right)\cdot X(0,0)}|0\rangle\right)\nonumber\\
\lineup\ \ \ \ \ \ \ \ \ \ \otimes\Big( S \alpha_B S^{-1}c_1 e^{ik^B\cdot X(0,0)}|0\rangle\Big)\otimes\Big( S\alpha_C S^{-1}c_1 e^{ik^C\cdot X(0,0)}|0\rangle \Big).\label{eq:CT1}
\end{eqnarray}
The similarity transformation turns the oscillator $\alpha_{-n}^i$ into the DDF operator $A_{-n}^{i}$. The shift in the momentum of the first state comes from the position zero mode prefactor in \eq{alphaDDF}. $\delta_\mu^+$ is the 1-form which extracts the plus component of a vector, $\delta^+\cdot v = v^+$ for a vector $v^\mu$. 

\begin{figure}
\begin{center}
\resizebox{4in}{2.3in}{\includegraphics{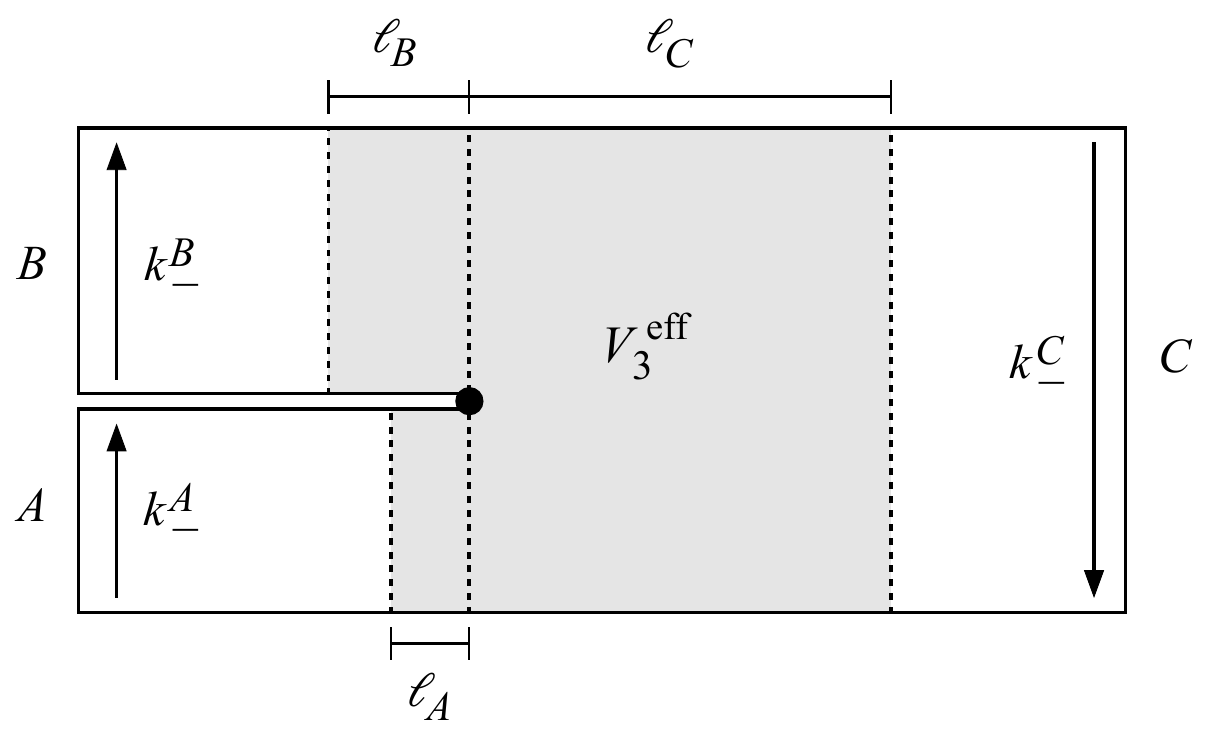}}
\end{center}
\caption{\label{fig:lightcone24} The cubic vertex of a lightcone effective field theory is the traditional lightcone vertex attached to stubs whose length depends on the minus momenta of the states interacting and the local dilatation around each puncture provided by the cubic vertex of the parent covariant string field theory.}
\end{figure}

We assume that $V_3$ is a surface state defined by local coordinate maps $\f{1},\f{2},\f{3}$ which place punctures at $1,0$ and $\infty$ in the upper half plane. We may then evaluate \eq{CT1} as a correlation function on the upper half plane:
\begin{eqnarray}
\langle \Veff_3|A\otimes B\otimes C \lineup = \Bigg\langle  \f{1}\circ\bigg(A_{-n}^{i}  S^{-1}\alpha_A S c e^{i\left(k^A+\frac{n}{2k_-^A}\delta^+\right)\cdot X(0,0)}\bigg)\nonumber\\
\lineup\ \ \ \times \f{2}\circ\Big(S^{-1}\alpha_B S c e^{i k^B\cdot X(0,0)}\Big)  
\f{3}\circ\Big(S^{-1}\alpha_C S c e^{i k^C\cdot X(0,0)}\Big)\Bigg\rangle_\mathrm{UHP}.
\end{eqnarray}
Evaluating the conformal transformations of the DDF operator and the plane wave vertex operators at the punctures leads to
\begin{eqnarray}
\langle \Veff_3|A\otimes B\otimes C \lineup =
\big( \fp{1}(0)\big)^{(k^A)^2-1}\big(\fp{2}(0)\big)^{(k^B)^2-1}\big((I\circ \f{3})'(0)\big)^{(k^C)^2-1}\big( \fp{1}(0)\big)^{n}\nonumber\\
\lineup\ \ \  \times \left\langle \left(A_{-n}^{i}c e^{i\left(k^A+\frac{n}{2k_-^A}\delta^+\right)\cdot X(1,1)} \right)\Big(ce^{i k^B\cdot X(0,0)}\Big)\Big( I\circ c e^{i k^C\cdot X(0,0)}\Big)\right.\nonumber\\
\lineup \ \ \ \ \ \ \ \ \ \ \times \f{1}\circ\big(S^{-1}\alpha_AS\big)\f{2}\circ\big(S^{-1}\alpha_B S\big)\f{3}\circ\big(S^{-1}\alpha_C S\big)\Bigg\rangle_\text{UHP},\label{eq:gc1}\ \ \ \ \ \ \ \ \ \ \ \ 
\end{eqnarray}
where the contour of the DDF operator surrounds the puncture at $u=1$. Since the DDF operator is conformally invariant, it does not otherwise retain information about the transformation with $\f{1}$. The DDF operator contains the chiral plane wave vertex operator $e^{-i (n/k_-^A) X^+(u)}$ in its integrand. The only objects in the correlator which generate contractions with this operator are the boundary plane wave vertex operators at the punctures. Using the recursive formula \eq{eXdelete}, we may eliminate the chiral plane wave vertex operator from the integrand of the DDF operator. This leads to
\begin{eqnarray}
A_{-n}^{i}e^{i\left(k^A+\frac{n}{2k_-^A}\delta^+\right)\cdot X(1,1)} \lineup  = e^{i\left(k^A+\frac{n}{2k_-^A}\delta^+\right)\cdot X(1,1)}  \oint_1\frac{du}{2\pi i} i\sqrt{2}\d X^{i} e^{-i \frac{n}{k_-^A} X^+(u)}\nonumber\\
\lineup =  e^{i k^A \cdot X(1,1)}\oint_1\frac{du}{2\pi i} (u-1)^{-n} u^{-n\frac{k_-^B}{k_-^A}} i\sqrt{2}\d X^{i}(u).
\end{eqnarray}
Following our description of the cubic lightcone vertex in appendix \ref{app:lc3_vertex}, we introduce {\it kinematic moduli} 
\begin{equation}\lambda_1=-\frac{k_-^A}{k_-^C},\ \ \ \ \lambda_2=-\frac{k_-^B}{k_-^C},\end{equation}
and rewrite this as
\begin{equation}A_{-n}^{i}e^{i\left(k^A+\frac{n}{2k_-^A}\delta^+\right)\cdot X(1,1)} =  e^{i k^A \cdot X(1,1)} \oint_1\frac{du}{2\pi i} \frac{1}{\left((u-1) u^{\frac{\lambda_2}{\lambda_1}}\right)^{n}}i\sqrt{2} \d X^{i_1}(u) .
\end{equation}
The factor in the integrand is closely related to the conformal transformation defining the cubic lightcone vertex, \eq{flc31}. In fact, we can write
\begin{eqnarray}
A_{-n}^{i}e^{i\left(k^A+\frac{n}{2k_-^A}\delta^+\right)\cdot X(1,1)} \lineup =  e^{i k^A \cdot X(1,1)} \frac{1}{\left(|\lambda_1||\lambda_2|^{\frac{\lambda_2}{\lambda1}}\right)^{n}}\oint_1\frac{du}{2\pi i} \frac{1}{(\flc{1})^{-1}(u)^{n}}i\sqrt{2} \d X^{i}(u) \nonumber\\
\lineup =   e^{i k^A \cdot X(1,1)} \frac{1}{\big(\flcp{1}(0)\big)^{n}}\flc{1}\circ\alpha_{-n}^{i}.
\end{eqnarray}
Interestingly, regardless of the initial choice of vertex, the oscillator behaves as though it is being transformed by the local coordinate maps of the lightcone vertex. Plugging into \eq{gc1} we obtain 
\begin{eqnarray}
\langle \Veff_3|A\otimes B\otimes C \lineup = \big(\fp{1}(0)\big)^{(k^A)^2-1}\big(\fp{2}(0)\big)^{(k^B)^2-1}((I\circ \f{3})'(0))^{(k^C)^2-1}\left(\frac{\fp{1}(0)}{\flcp{1}(0)}\right)^{n}\nonumber\\
\lineup\ \ \  \times \Big\langle \big(\flc{1}\circ \alpha_{-n}^{i}\big)\big(c e^{i k^A\cdot X(1,1)}\big)\big(ce^{i k^B\cdot X(0,0)}\big)\big( I\circ c e^{i k^C\cdot X(0,0)}\big)\nonumber\\
\lineup \ \ \ \ \ \ \  \f{1}\circ\big(S^{-1}\alpha_A S\big)\f{2}\circ\big(S^{-1}\alpha_B S\big)\f{3}\circ\big(S^{-1}\alpha_C S\big)\Big\rangle_\text{UHP}.\nonumber\\
\end{eqnarray}
Now this works in a similar way for the other oscillators. Writing them out explicitly as 
\begin{eqnarray}
A \lineup = \alpha_{-l_1}^{i_1}...\alpha_{-l_L}^{i_L} c_1 e^{ik^A\cdot X(0,0)}|0\rangle,\nonumber\\
B \lineup = \alpha_{-m_1}^{j_1}...\alpha_{-m_M}^{j_M} c_1 e^{ik^B\cdot X(0,0)}|0\rangle,\nonumber\\
C \lineup =\alpha_{-n_1}^{k_1}...\alpha_{-n_N}^{k_N}  c_1 e^{ik^C\cdot X(0,0)}|0\rangle,
\end{eqnarray}
we find
\begin{eqnarray}
\lineup\!\!\!\!\!\!\!\!\! \langle \Veff_3|A\otimes B\otimes C \nonumber\\
\lineup  \!\!\!\!\!\!\!\!\!=\big(\fp{1}(0)\big)^{(k^A)^2-1}\big(\fp{2}(0)\big)^{(k^B)^2-1}((I\circ \f{3})'(0))^{(k^C)^2-1}\nonumber\\
\lineup  \!\!\!\!\!\!\!\!\!\times \left(\frac{\fp{1}(0)}{\flcp{1}(0)}\right)^{l_1+...+l_L}\!\!\left(\frac{\fp{2}(0)}{\flcp{2}(0)}\right)^{m_1+...+m_M}\!\!\left(\frac{(I\circ \f{3})'(0)}{(I\circ\flc{3})'(0)}\right)^{n_1+...+n_N}\nonumber\\
\lineup\!\!\!\!\!\!\!\! \!\times \Big\langle\! c e^{i k^A\cdot X(0,0)}c e^{i k^B\cdot X(0,0)}c e^{i k^C\cdot X(0,0)}\flc{1}\!\circ\!\big(\alpha_{-l_1}^{i_1}...\alpha_{-l_L}^{i_L}\big)\!\flc{2}\!\circ\!\big(\alpha_{-m_1}^{j_1}...\alpha_{-m_M}^{j_M}\big)\!\flc{3}\!\circ\!\big(\alpha_{-n_1}^{k_1}...\alpha_{-n_N}^{k_N}\big)\!\Big\rangle_\text{\!UHP}.\nonumber\\
\end{eqnarray}
Now we would like to simplify the conformal factors in front of the correlator. It is helpful to pull the plane wave vertex operators inside the lightcone local coordinate maps: 
\begin{eqnarray}
\lineup\!\!\!\!\!\!\!\!\! \langle \Veff_3|A\otimes B\otimes C \nonumber\\
\lineup  \!\!\!\!\!\!\!\!\!=\left(\frac{\fp{1}(0)}{\flcp{1}(0)}\right)^{(k^A)^2-1+l_1+...+l_L}\!\!\left(\frac{\fp{2}(0)}{\flcp{2}(0)}\right)^{(k^B)^2-1+m_1+...+m_M}\!\!\left(\frac{(I\circ \f{3})'(0)}{(I\circ\flc{3})'(0)}\right)^{(k^C)^2-1+n_1+...+n_N}\nonumber\\
\lineup\!\!\!\!\!\!\!\! \!\times \Big\langle \!\flc{1}\!\circ\!\big(\alpha_{-l_1}^{i_1}...\alpha_{-l_L}^{i_L}c e^{i k^A\cdot X(0,0)}\big)\!\flc{2}\!\circ\!\big(\alpha_{-m_1}^{j_1}...\alpha_{-m_M}^{j_M}c e^{i k^B\cdot X(0,0)}\big)\!\flc{3}\!\circ\!\big(\alpha_{-n_1}^{k_1}...\alpha_{-n_N}^{k_N}c e^{i k^C\cdot X(0,0)}\big)\!\Big\rangle_\text{\!UHP}.\nonumber\\
\end{eqnarray}
We further introduce three scale transformations,
\begin{eqnarray}
s_{(3,1)}(\xi) \lineup = \frac{\fp{1}(0)}{\flcp{1}(0)}\,\xi,\\
s_{(3,2)}(\xi)\lineup = \frac{\fp{2}(0)}{\flcp{2}(0)}\,\xi,\\
s_{(3,3)}(\xi)\lineup = \frac{(I\circ \f{3})'(0)}{(I\circ\flc{3})'(0)}\,\xi,
\end{eqnarray}
which allow us to write
\begin{eqnarray}
\langle \Veff_3|A\otimes B\otimes C \lineup = \Big\langle \flc{1}\!\circ\! s_{(3,1)}\!\circ\!\big(\alpha_{-l_1}^{i_1}...\alpha_{-l_L}^{i_L}c e^{i k^A\cdot X(0,0)}\big)\nonumber\\
\lineup \ \ \ \ \ \ \flc{2}\!\circ\!s_{(3,2)}\!\circ\!\big(\alpha_{-m_1}^{j_1}...\alpha_{-m_M}^{j_M}c e^{i k^B\cdot X(0,0)}\big)\nonumber\\
\lineup\ \ \ \ \ \ \flc{3}\!\circ\!s_{(3,3)}\!\circ\!\big(\alpha_{-n_1}^{k_1}...\alpha_{-n_N}^{k_N}c e^{i k^C\cdot X(0,0)}\big) \Big\rangle_\text{\!UHP}.
\end{eqnarray}
At the end the effective cubic vertex is a surface state, though  the local coordinate maps are quite different  from the  parent  cubic vertex. We can express this result equivalently as 
\begin{equation}
\langle \Veff_3| = \langle \Vlc_3|\left(\frac{\fp{1}(0)}{\flcp{1}(0)}\right)^{L_0}\otimes\left(\frac{\fp{2}(0)}{\flcp{2}(0)}\right)^{L_0}\otimes \left(\frac{(I\circ \f{3})'(0)}{(I\circ\flc{3})'(0)}\right)^{L_0}.
\end{equation}
We make the following observations: 
\begin{itemize}
\item Fixing lightcone gauge in a generic covariant SFT produces the traditional lightcone cubic vertex composed with certain scale transformations of the external states. The scale transformations ensure that the local dilatations around the punctures of the effective cubic vertex match those of the cubic vertex of the parent string field theory. This is the only information about the parent vertex which is retained after integrating out the longitudinal degrees of freedom. The scale transformations can be interpreted geometrically as strips of worldsheet (``stubs") attached to the lightcone vertex whose length (in general) depends on the minus momenta of the states in the vertex. In the interaction coordinate \eq{Mandelstam3} the stubs have length
\begin{eqnarray}
\ell_A \lineup = \frac{|k_-^A|}{\pi}\ln\left(\frac{|\lambda_1||\lambda_2|^{\frac{\lambda_2}{\lambda1}}}{\fp{1}(0)}\right),\\
\ell_B \lineup = \frac{|k_-^B|}{\pi}\ln\left(\frac{|\lambda_2||\lambda_1|^{\frac{\lambda_1}{\lambda2}}}{\fp{2}(0)}\right),\\
\ell_C\lineup= \frac{|k_-^C|}{\pi}\ln\left( \frac{1}{(I\circ \f{3})'(0)|\lambda_1|^{\lambda_1}|\lambda_2|^{\lambda_2}}\right),
\end{eqnarray}
for the states $A,B$ and $C$ respectively. This is illustrated in figure \ref{fig:lightcone24}. If the parent covariant vertex is already the lightcone vertex, as for the Kugo-Zwiebach theory, it follows from \eq{dflc31}-\eq{dflc33} that the stub lengths vanish, consistent with the result of the last section. 
\item If the initial cubic vertex is Lorentz invariant,  the derivatives of the local coordinate maps at the origin will be constant numbers. For small enough $\lambda_1$ or $\lambda_2$, this implies that one or both of the stub lengths $\ell_A$ or $\ell_B$ will be {\it negative}, and the effective vertex fails to be normalizable. The reason this happens is that when a state approaches zero momentum its local coordinate patch in the lightcone vertex becomes arbitrarily small, and an arbitrarily large scale transformation is needed to match the local dilatation provided by the cubic vertex of the parent theory. This means that lightcone effective actions are generically singular. 
\end{itemize}

\noindent Let us illustrate this last point concretely. Consider the cubic coupling between two tachyons and a state at mass level $n$ created by a single oscillator:
\begin{eqnarray}
\lineup\!\!\!\!\!\!\! \langle \Veff_3|\Big(\alpha_{-n}^i c_1 e^{ik^1\cdot X(0,0)}|0\rangle\Big)\otimes \Big(c_1 e^{ik^2\cdot X(0,0)}|0\rangle\Big)\otimes \Big(c_1 e^{ik^3\cdot X(0,0)}|0\rangle\Big)\nonumber\\
\lineup = \big( \sqrt{2}  k_i^r M^{rs} k_-^s \big)n \bar{N}_n^1 \left(\frac{\fp{1}(0)}{\lambda_1\lambda_2^{\lambda_2/\lambda_1}}\right)^n 
\langle V_3|\Big(c_1 e^{ik^1\cdot X(0,0)}|0\rangle\Big)\!\otimes\! \Big(c_1 e^{ik^2\cdot X(0,0)}|0\rangle\Big)\!\otimes\! \Big(c_1 e^{ik^3\cdot X(0,0)}|0\rangle\Big).\nonumber\\
\end{eqnarray}
The prefactors come from the Neumann coefficient \eq{Neumann} of the lightcone vertex and the dilatation of the oscillator. We also noted that the cubic tachyon coupling of the lightcone effective field theory is equal to that of the parent theory. The Neumann coefficient takes  the form \cite{Mandelstam1,Cremmer}
\begin{equation}
\bar{N}_{n}^{1}=\frac{1}{k_-^1}(\lambda_1\lambda_2^{\lambda_2/\lambda_1})^n  \frac{\Gamma(-n\lambda_2/\lambda_1)}{n!\Gamma(-n/\lambda_1+1)}.
\end{equation}
Using the Sterling approximation we can extract the large $n$ behavior,
\begin{equation}
n \bar{N}_n^1 \left(\frac{\fp{1}(0)}{\lambda_1\lambda_2^{\lambda_2/\lambda_1}}\right)^n \sim \frac{1}{k_-^3}\frac{(-1)^{n}}{\sqrt{2\pi \lambda_2}}\frac{1}{n^{1/2}} \left(\frac{\fp{1}(0)}{\lambda_1\lambda_2^{\lambda_2/\lambda_1}}\right)^n .
\end{equation}
If $\fp{1}(0)$ is independent of the momenta, the 3-point amplitude grows exponentially with the mass level for small enough $\lambda_1$. Ordinarily, we expect that couplings to highly excited string states will be suppressed in finite energy processes. This is necessary to ensure that sums over intermediate states converge in higher point amplitudes. Therefore it appears that lightcone gauge is singular for generic momenta in typical covariant string field theories. Since the difficulty appears when one string has low minus momentum relative to others, we refer to this as the {\it soft string problem} of lightcone gauge.

\section{From Kugo-Zwiebach to Witten's string field theory}

\label{sec:Kaku}

The final step of our analysis is finding a field redefinition which relates the Kugo-Zwiebach string field theory and the Witten string field theory. There is a known strategy for dealing with this kind of problem \cite{HataZwiebach}. We look for a continuous 1-parameter family of covariant string field theories which connect the initial and final theories of interest, and integrate the infinitesimal field redefinition which moves along the family. There is little doubt that such a field redefinition exists \cite{Costello1}. However, it will not be unique, since it will depend on the path we follow to connect the initial and final theories. This ambiguity corresponds to the freedom to modify the field redefinition by symmetries of the respective actions. The interesting question is if there is a reasonable way to fix this ambiguity by chosing a preferred path connecting the Kugo-Zwiebach and Witten theories. A natural proposal was made by Kaku \cite{Kaku}, and we discuss this next.

\begin{figure}
\begin{center}
\resizebox{2.3in}{1.5in}{\includegraphics{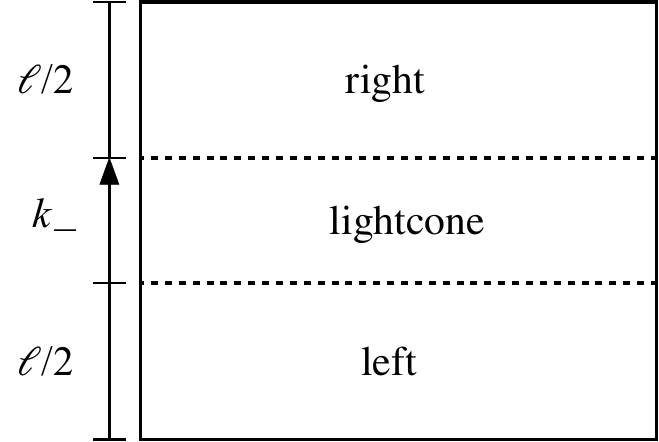}}
\end{center}
\caption{\label{fig:lightcone11} The propagator strip of a lightcone string attached to strips representing Chan-Paton indices.}
\end{figure}

\subsection{Kaku's deformation}
\label{subsec:Kaku}

Open string interactions in Witten's theory are loosely described as those of a scalar particle in the adjoint representation of a gigantic Lie algebra. The ``Lie algebra index" in this case corresponds to one half of an open string. It is possible to attach this kind of ``index" to the endpoints of a lightcone open string, where  it can be viewed as a Chan-Paton factor. In this case, the open string propagator strip is formed by stacking three strips on top of each other.  Through the middle we have a strip representing the lightcone string with width $k_-$, where $k_-$ is the minus momentum of the state passing through the propagator. On the top and bottom we have strips representing ``Chan-Paton indices" of width $\ell/2$.  This is shown in figure \ref{fig:lightcone11}. 
The lightcone portion of the propagator strip splits and joins in the usual way through the cubic and quartic lightcone vertices, and the Chan-Paton strips are glued together in the process as would be implied by tracing over an internal Lie algebra index. The resulting cubic and quartic vertices will be called {\it Kaku vertices}, and the associated string field theory will be called a {\it Kaku string field theory}. It is intuitively clear that Kaku vertices define a consistent string field theory, since they simply utilize the freedom to add Chan-Paton labels to the endpoints of a lightcone string. In the limit $\ell\to 0$ the Chan-Paton factors disappear, and we recover the lightcone-style interactions of the Kugo-Zwiebach theory. In the limit $\ell\to\infty$, we can scale the coordinates by $1/\ell$ and the lightcone interactions are squeezed to the midpoint. Then we recover the interactions of Witten's string field theory.  

\begin{figure}
\begin{center}
\resizebox{3.7in}{1.2in}{\includegraphics{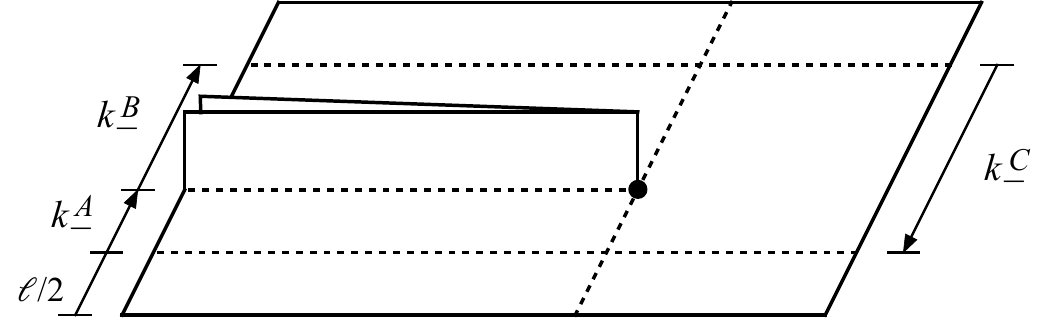}} 
\end{center}
\caption{\label{fig:lightcone12} Geometry of the cubic Kaku vertex.}
\end{figure}

Our present task is to derive explicit expressions for the Kaku vertices. We write the unshifted cubic vertex as 
\begin{eqnarray} V^\ell_3(A,B,C) \lineup = \Big\langle  \fl{1}\circ A(0) \fl{2}\circ B(0)  \fl{3}\circ C(0)\Big\rangle_\text{UHP}\nonumber\\
\lineup = (-1)^{|A|+|B|}\omega(A, m^\ell_2(B,C)).
\end{eqnarray}
The superscript indicates the dependence on $\ell$ so that
\begin{equation}\lim_{\ell\to 0}V_3^\ell = \Vlc_3,\ \ \ \ \  \lim_{\ell\to\infty}V_3^\ell = \VW_3.\end{equation}
We characterize the vertex assuming the {\it standard confirguation} (see appendix \ref{app:lc3_vertex}), that is, we assume that all states have definite minus momenta and the third state $C$ has minus momentum of largest magnitude. The surface of the interaction is shown in figure \ref{fig:lightcone12}. To derive the local coordinate maps we need to find a conformal transformation which relates this surface to the upper half plane. This requires a version of the Schwarz-Christoffel map which applies to polygons containing conical singularities \cite{Gilbarg}: 
\begin{SC} Consider an $m$-sided polygon bounding a surface containing $n$ conical singularities, but  is otherwise flat. Following the boundary of the polygon counterclockwise, at each corner we turn consecutively by angles $\phi_1,...,\phi_m$. The conical singularities have deficit angles $\theta_1,...,\theta_n$. The geometry of the surface constrains the angles to satisfy
\begin{equation}\phi_1+...+\phi_m+\theta_1+...+\theta_n =2\pi.\end{equation}
The surface is related to the upper half plane through the conformal transformation
\begin{equation}
f(u) = A+ C\int du \frac{1}{(u-a_1)^{\frac{\phi_1}{\pi}}}...\frac{1}{(u-a_m)^{\frac{\phi_m}{\pi}}}\left(\frac{1}{(u-b_1)(u-\overline{b}_1)}\right)^{\frac{\theta_1}{2\pi}}...\left(\frac{1}{(u-b_n)(u-\overline{b}_n)}\right)^{\frac{\theta_n}{2\pi}},
\end{equation}
where $a_1,..,a_m\in\mathbb{R}$ are the preimages of the corners on the real axis, $b_1,..,b_n$ are the preimages of the conical singularities in the interior of the upper half plane, and $A$ and $C$  are constants which fix the origin and the overall size of the polygon.
\end{SC}
\begin{figure}
\begin{center}
\resizebox{2.6in}{2in}{\includegraphics{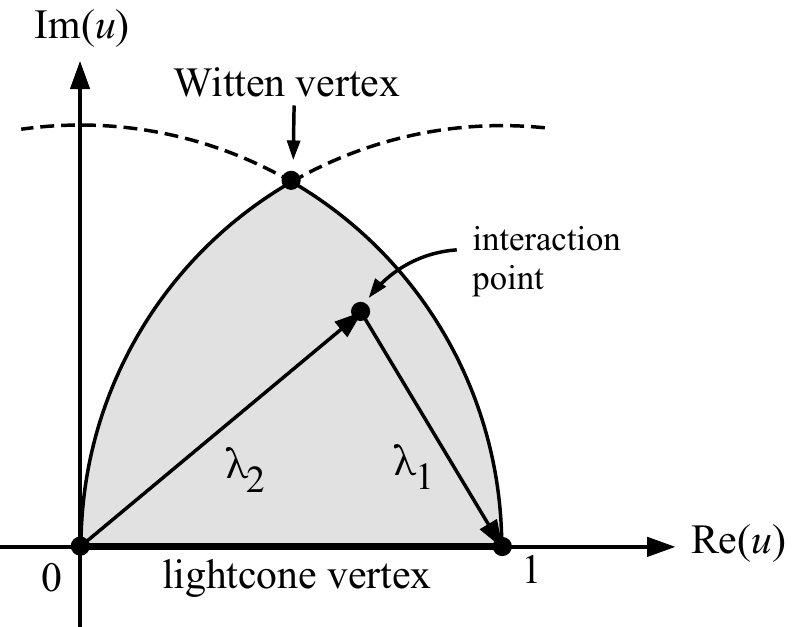}}
\end{center}
\caption{\label{fig:lightcone14} The kinematic moduli of the cubic Kaku vertex.}
\end{figure}

\noindent In the current situation, we have a three sided polygon with turning angle $\pi$ at each corner and one conical singularity with deficit angle $-\pi$. The punctures of $A,B$ and $C$ are the preimages of the corners, and are placed respectively at $1,0$ and $\infty$. Using the above Schwarz-Christoffel mapping, we find that the interaction coordinate $\rho$ on the surface is related to the upper half plane coordinate $u$ through
\begin{eqnarray}
\rho(u)\lineup = \frac{|\lambda_1|}{\pi}\ln\left(\frac{1-u}{|\lambda_1|}\frac{i\mathrm{Im}(\lambda_1)}{U(1-u,\lambda_1)+(1-u)\frac{\mathrm{Re}(\lambda_1)}{|\lambda_1|}-|\lambda_1|}\right)\nonumber\\
\lineup\ \ \ +\frac{|\lambda_2|}{\pi}\ln\left(\frac{u}{|\lambda_2|}\,\frac{i\mathrm{Im}(\lambda_2)}{U(u,\lambda_2)+u\frac{\mathrm{Re}(\lambda_2)}{|\lambda_2|}-|\lambda_2|}\right)\nonumber\\
\lineup\ \ \ -\frac{1}{\pi}\ln\left(\frac{i \mathrm{Im}(\lambda_2)}{U(u,\lambda_2)+u-\mathrm{Re}(\lambda_2)}\right),\label{eq:Kaku}
\end{eqnarray}
where
\begin{equation}
U(u,u') = (u-u')\sqrt{\frac{u-\overline{u}'}{u-u'}}.
\end{equation}
We have normalized the coordinate $\rho$ so that the outgoing strip has unit width. To obtain the normalization of figure \ref{fig:lightcone12}, we should scale by $k_-^C+\mathrm{sgn}(k_-^C)\ell$. We have also incorporated a shift which places the conical singularity at the origin of the coordinate $\rho$.
The map depends on two kinematic moduli $\lambda_1$ and $\lambda_2$. They are determined as the unique complex numbers with positive imaginary part whose magnitudes are given by the ratios of the widths of the incoming strips to the outgoing strip, 
\begin{equation}|\lambda_1| = \frac{|k_-^A|+\ell}{|k_-^C|+\ell},\ \ \ \ |\lambda_2| = \frac{|k_-^B|+\ell}{|k_-^C|+\ell},\end{equation}
and which satisfy the constraint
\begin{equation}\lambda_1+\lambda_2=1.\end{equation}
\begin{figure}
\begin{center}
\resizebox{1.3in}{1.3in}{\includegraphics{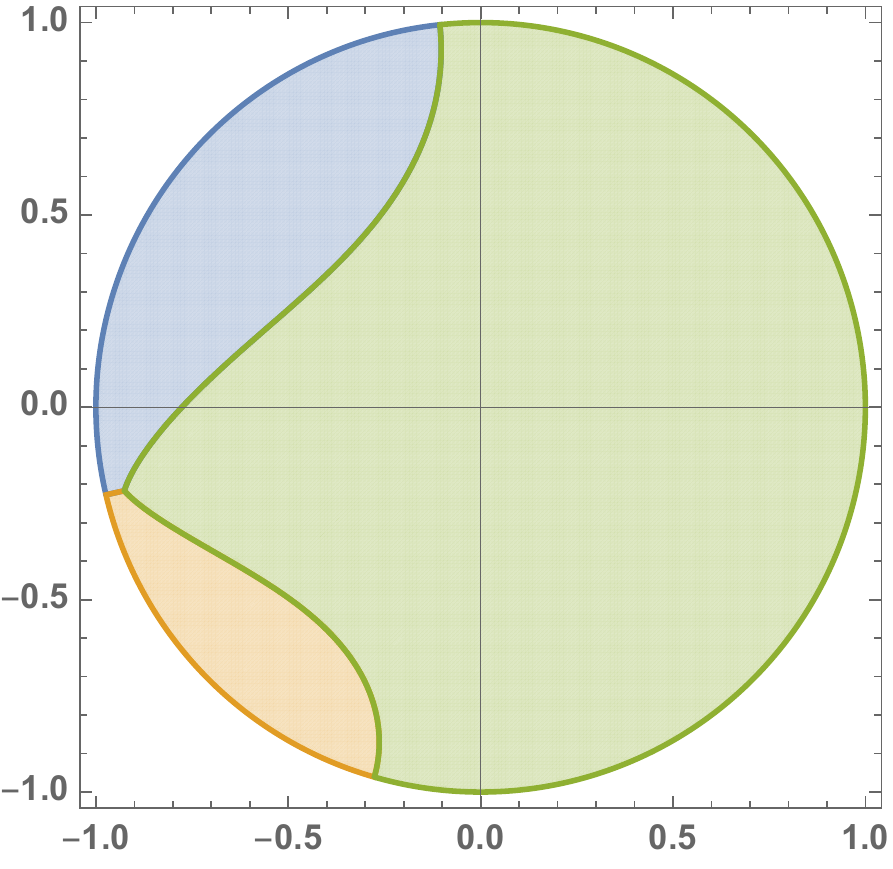}}\ \ \
\resizebox{1.3in}{1.3in}{\includegraphics{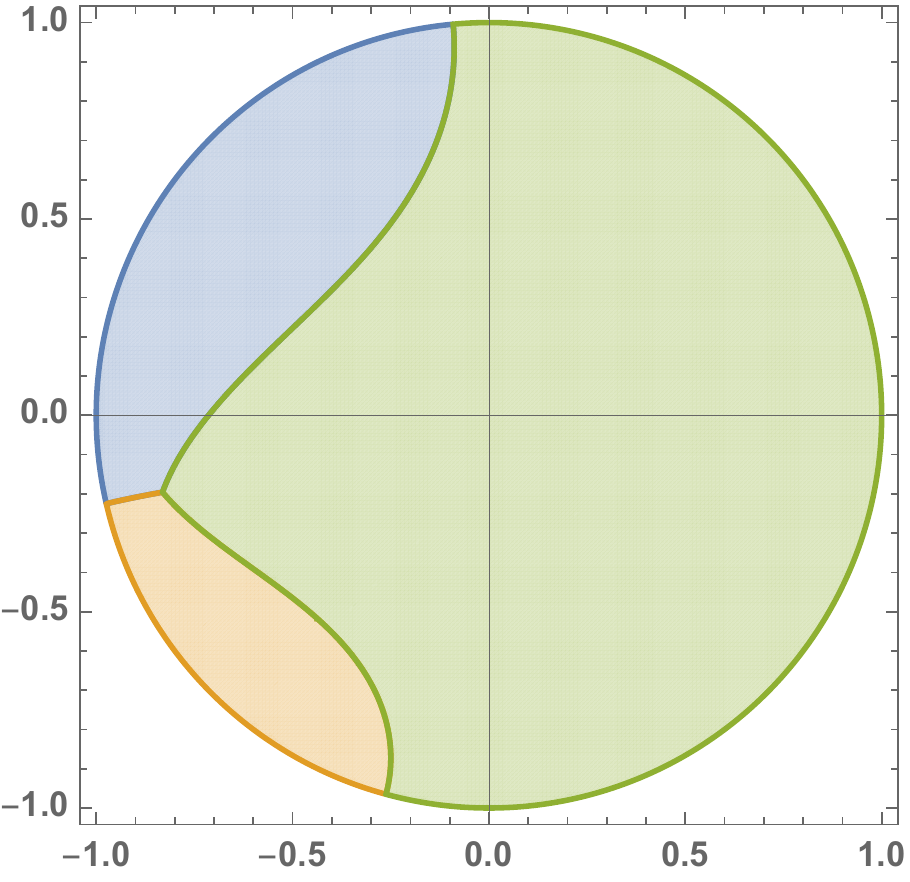}}\ \ \
\resizebox{1.3in}{1.3in}{\includegraphics{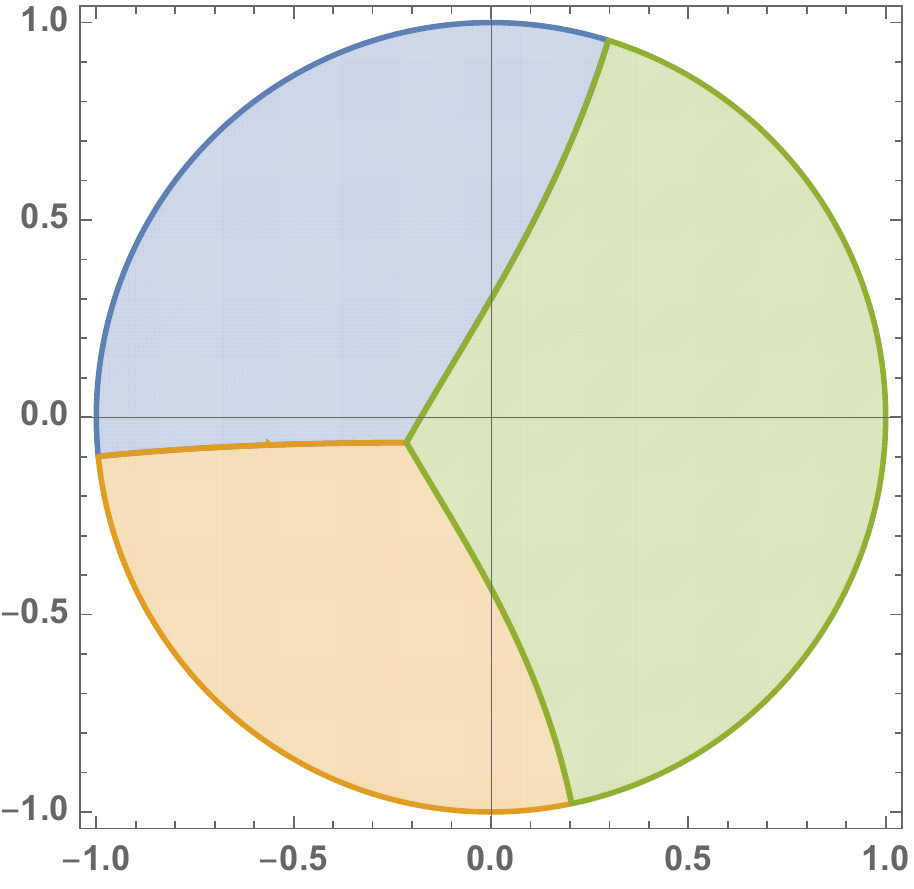}}\ \ \
\resizebox{1.3in}{1.3in}{\includegraphics{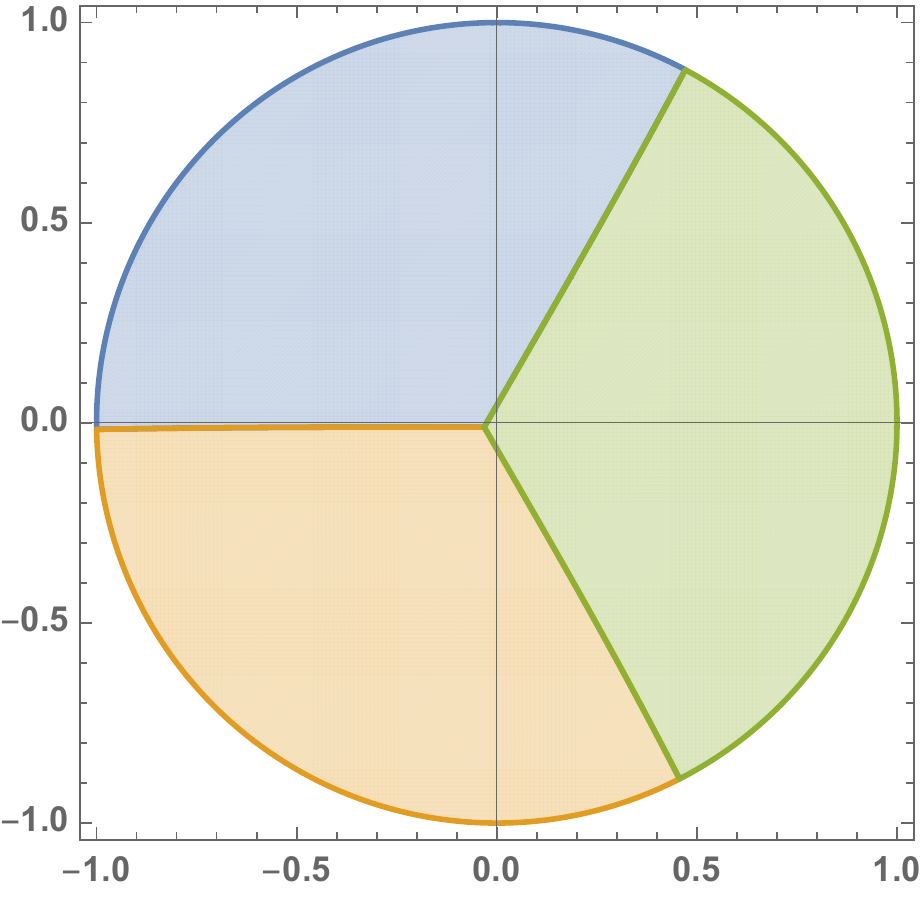}}
\end{center}
\caption{\label{fig:lightcone15} Local coordinate patches of the cubic Kaku vertex on the unit disk in the representation of figure \ref{fig:lightcone1} and \ref{fig:lightcone3}. We take the normalized ratios $K=.6$ and $L = .001,.1,1$ and $10$.}
\end{figure}

\noindent Solving the trigonometry problem gives 
\begin{eqnarray}
\lambda_2 = \frac{L^2 + 2 K (1 + 2 L)}{2 (1 + L)^2}+i \frac{\sqrt{8 (1 - K) K L + 4 (1 + 3 K - 3 K^2) L^2 + 8 L^3 + 
 3 L^4}}{2 (1 + L)^2},
\end{eqnarray}
where we introduce dimensionless ratios
\begin{equation}K = \frac{|k_-^B|}{|k_-^C|},\ \ \ \ L = \frac{\ell}{|k_-^C|}.\label{eq:KL}\end{equation}
In the upper half plane, the point
\begin{equation}u=\lambda_2\end{equation}
is the preimage of the conical singularity of the interaction surface, and is also the ``interaction point" where all three local coordinate patches touch. Since $k_-^C$ has largest magnitude, $|\lambda_2|$ and $|\lambda_1|$ must be less than 1, which means that $u=\lambda_2$ is bounded in the region of overlap between a unit disk centered at $u=0$ and a unit disk centered at $u=1$. For fixed finite and nonzero $\ell$, the interaction point can sit anywhere in this region depending on the minus momenta of the states. In the limit where the minus momenta become zero, $\lambda_2$ approaches the point of intersection of the bounding circles at $u=e^{i\pi/3}$. This point represents the Witten vertex. In the limit where the minus momenta become infinite,  $\lambda_2$ approaches the real axis between $0$ and $1$. This segment of the real axis represents the kinematic moduli \eq{l1l2} of the cubic lightcone vertex. All of this is illustrated in figure \ref{fig:lightcone14}.

We cover each strip on the interaction surface with a local coordinate $\xi$ through the appropriate scaling and translation of $\ln(\xi)$. In this way we find the (inverse) local coordinate maps:
\begin{eqnarray}
 (\fl{1})^{-1}(u)\lineup = \frac{u-1}{|\lambda_1|} \frac{\mathrm{Im} (\lambda_1)}{U_-(1-u,\lambda_1) + (1 - u) \frac{\mathrm{Re}(\lambda_1)}{|\lambda_1|}- |\lambda_1|}  \nonumber\\
\lineup \ \ \ \  \times \left(\frac{u}{|\lambda_2|} \frac{\mathrm{Im}(\lambda_2)}{U_+(u,\lambda_2) + u \frac{\mathrm{Re}(\lambda_2)}{|\lambda_2|} -|\lambda_2|}\right)^{\frac{|\lambda_2|}{|\lambda_1|}}\!\!\left(\frac{u + U_+ (u,\lambda_2) - \mathrm{Re}(\lambda_2)}{\mathrm{Im}(\lambda_2)}\right)^{\frac{1}{|\lambda_1|}},\ \ \ \ \ \ \ \\
 (\fl{2})^{-1}(u)\lineup = -\frac{u}{|\lambda_2|}\frac{\mathrm{Im}(\lambda_2)}{U_-(u,\lambda_2)+ u\frac{\mathrm{Re}(\lambda_2)}{|\lambda_2|}- |\lambda_2| }\left(-\frac{u + U_-(u,\lambda_2) -\mathrm{Re}(\lambda_2)}{\mathrm{Im}(\lambda_2)}\right)^{\frac{1}{|\lambda_2|}}\nonumber\\
\lineup\ \ \ \ \times \left(\frac{u-1}{|\lambda_1|}\frac{\mathrm{Im}(\lambda_1)}{U_+(1-u, \lambda_1) + (1 - u) \frac{\mathrm{Re}(\lambda_1)}{|\lambda_1|} -|\lambda_1|}\right)^{\frac{|\lambda_1|}{|\lambda_2|}},\\
 (\fl{3})^{-1}(u)\lineup = -\frac{\mathrm{Im}(\lambda_2)}{u + U(u, \lambda_2) - \mathrm{Re}(\lambda_2)} \left(-\frac{|\lambda_1|}{1 - u}\frac{U(1- u,  \lambda_1)+ (1 - u) \frac{\mathrm{Re}(\lambda_1)}{|\lambda_1|} -|\lambda_1|}{\mathrm{Im}(\lambda_1)}\right)^{|\lambda_1|}\nonumber\\
\lineup\ \ \ \ \times \left(\frac{|\lambda_2|}{u} \frac{U(u, \lambda_2) + u\frac{\mathrm{Re}(\lambda_2)}{|\lambda_2|} -|\lambda_2|}{\mathrm{Im}(\lambda_2)}\right)^{|\lambda_2|}.
\end{eqnarray}
Here we define
\begin{equation}U_+(u,u') = \sqrt{(u-u')(u-\overline{u}')}  = -U_-(u,u'),\end{equation}
with the understanding  that the branch cut of the square root is taken along the negative real axis. The functions $U,U_\pm$ are different branches of the same multi-valued  function. We plot the local coordinate patches on the unit disk for fixed minus momenta and varying $\ell$ in figure \ref{fig:lightcone15}. As expected, the local coordinate patches continuously deform between those of the cubic lightcone vertex and those of the Witten vertex, as shown in figures \ref{fig:lightcone1} and \ref{fig:lightcone3}.

\begin{figure}
\begin{center}
\resizebox{3in}{2in}{\includegraphics{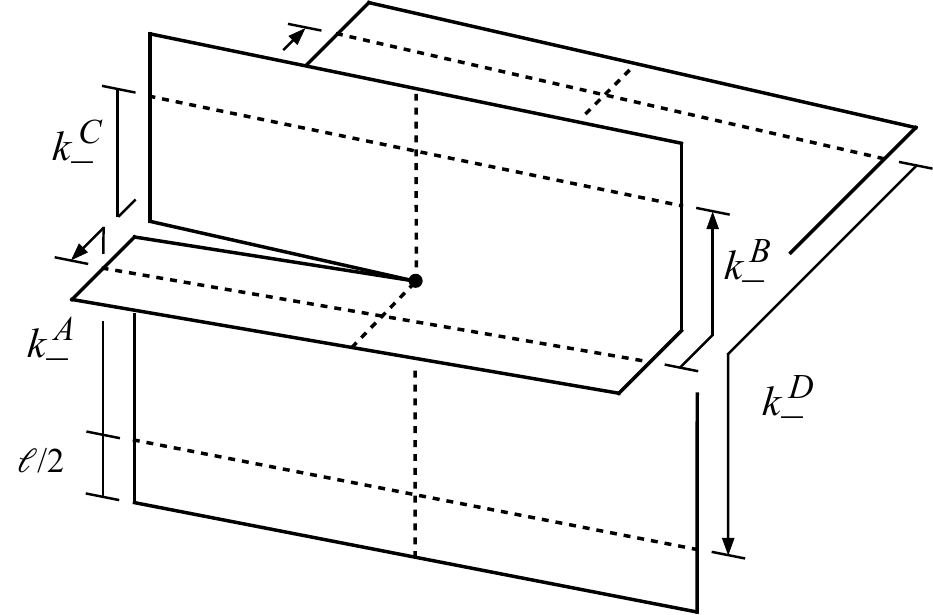}}
\end{center}
\caption{\label{fig:lightcone21} Geometry of the quartic Kaku vertex.}
\end{figure}

Next we consider the quartic vertex. The story here is somewhat simpler, since attaching Chan-Paton strips does not alter the geometry of the quartic interaction surface. See figure \ref{fig:lightcone21}. The only thing that changes is the relation between the width of the strips and the minus momenta of the states. The unshifted quartic Kaku vertex may be written
\begin{equation} V^\ell_4(A,B,C,D)=(-1)^{1+|C|}\omega(A, m^\ell_3(B,C,D)).\end{equation}
\begin{figure}
\begin{center}
\resizebox{3in}{2in}{\includegraphics{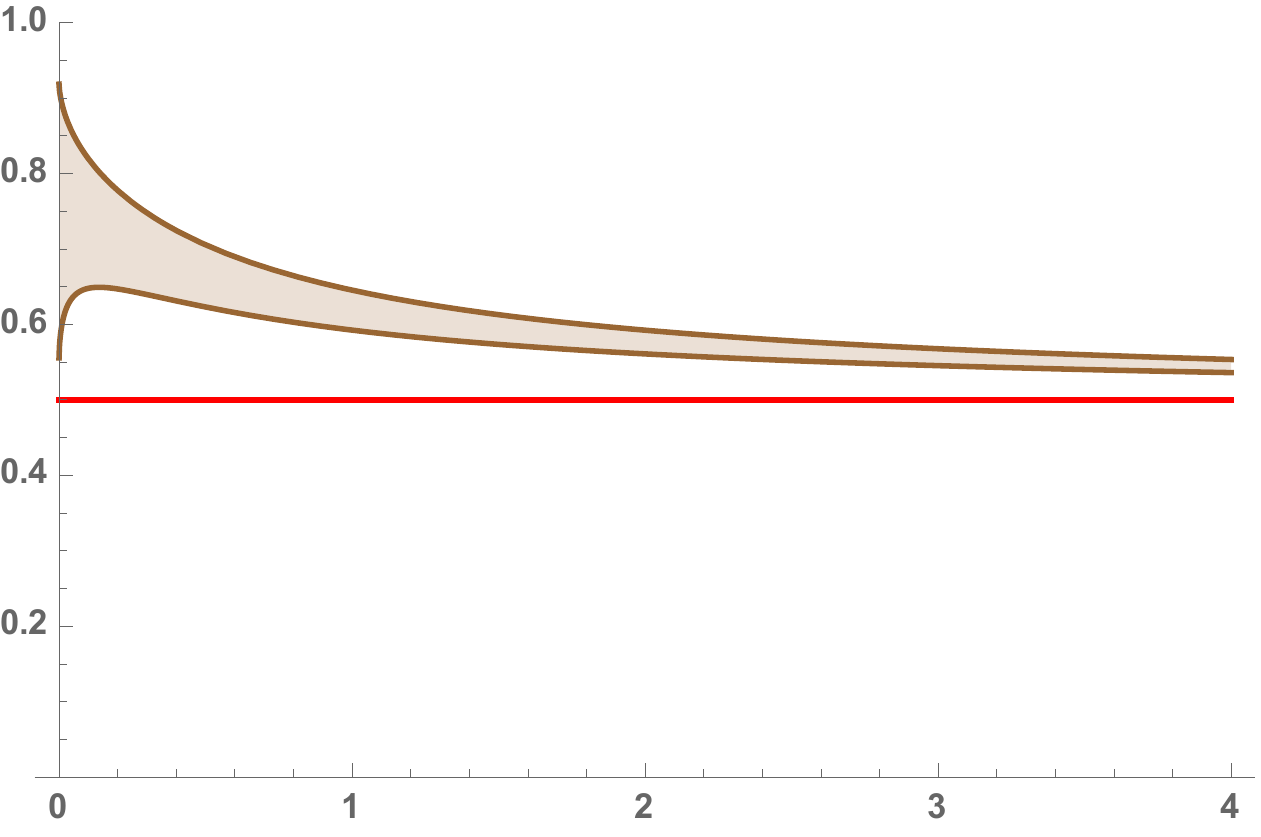}}\ \ \ \ \ \ \ \ \ \ \ \ 
\resizebox{3in}{2in}{\includegraphics{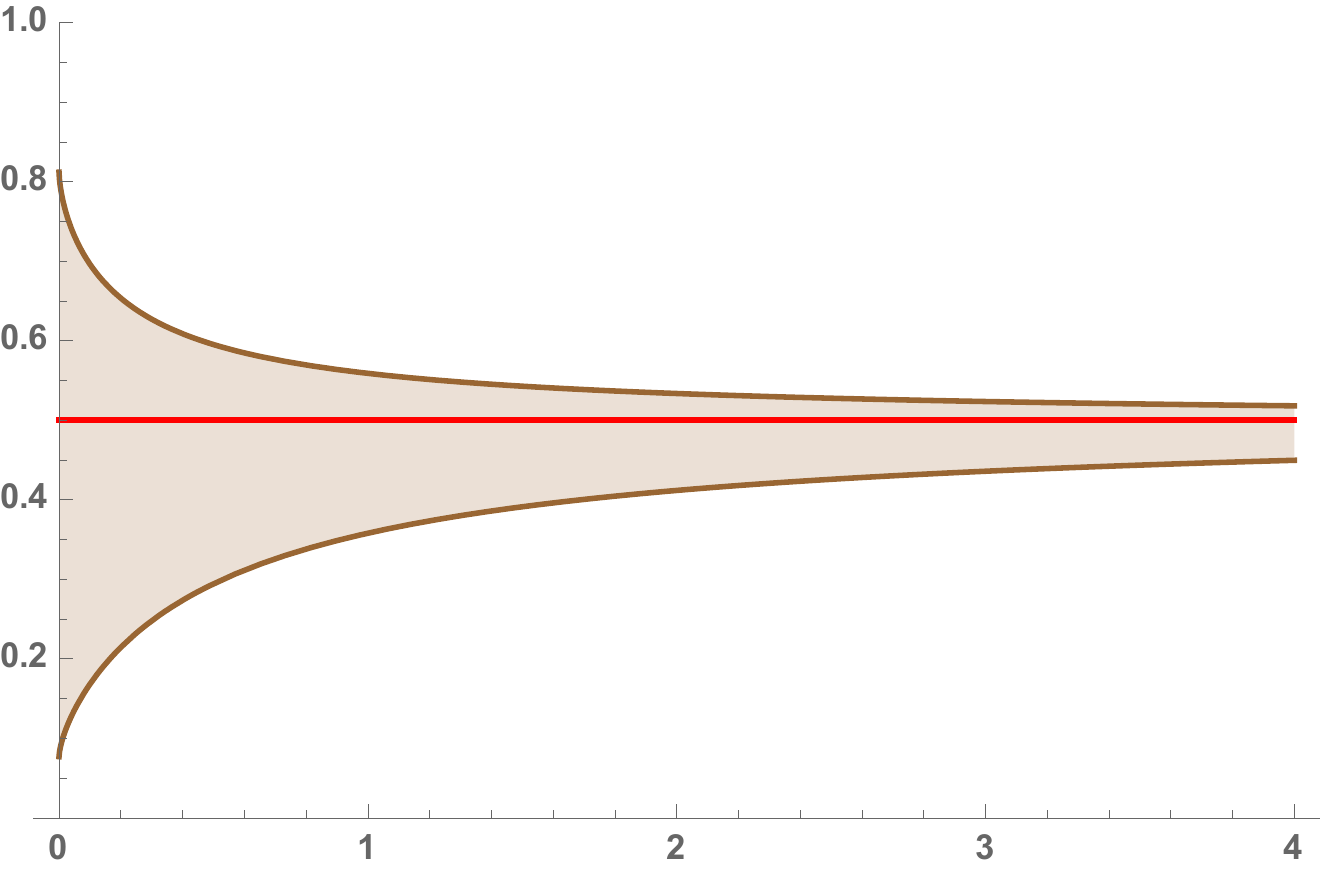}}
\end{center}
\caption{\label{fig:lightcone19} The region of moduli space $m\in[m_-^\ell,m_+^\ell]$ provided by the quartic Kaku vertex as a function of $\ell/|k_-^D|$. Left shows the region when $\frac{|k_-^A|}{|k_-^D|} = .7$ and $\frac{|k_-^C|}{|k_-^D|} = .5$, and right when $\frac{|k_-^A|}{|k_-^D|} = .25$ and $\frac{|k_-^C|}{|k_-^D|} = .8$.}
\end{figure}

\noindent We place the punctures of $A,B,C,D$ respectively at $1,m\in[0,1],0$ and $\infty$. In this case the local coordinate maps are the same as those of the quartic lightcone vertex \eq{LCquartic1}-\eq{LCquartic4} with the understanding that the kinematic moduli are now given by 
\begin{eqnarray}
\lambda_1  \lineup = \frac{|k_-^A|+\ell}{|k_-^D|+\ell},\label{eq:4l1}\\
\lambda_2  \lineup = \frac{|k_-^B|+\ell}{|k_-^D|+\ell},\\
\lambda_3  \lineup = \frac{|k_-^C|+\ell}{|k_-^D|+\ell}.\label{eq:4l3}
\end{eqnarray}
What does change, however, is the region of moduli space supplied by the quartic vertex. This region may be characterized by the range of possible positions of the second puncture in the upper half plane:
\begin{equation}m_-^\ell<m<m_+^\ell.\end{equation}
For the quartic lightcone vertex, the upper and lower bounds are given in \eq{mpm}. For the quartic Kaku vertex, the upper and lower bounds must be determined implicitly by requiring that the conical singularity on the interaction surface ranges from the bottom to the top of the lightcone portion of the strip of the state $B$. This leads to the inequality
\begin{equation}\frac{\ell/2}{|k_-^D|+\ell}< -\frac{\lambda_1}{\pi}\mathrm{Arg}(1-u_*)+\frac{\lambda_2}{\pi}\mathrm{Arg}(u_*-m)-\frac{\lambda_3}{\pi}\mathrm{Arg}(u_*)<\frac{|k_-^B|+\ell/2}{|k_-^D|+\ell},\label{eq:impmpm}\end{equation}
where $u_*$ is given in \eq{ustar}. The quartic vertex region of the moduli space is shown in figure \ref{fig:lightcone19} in a couple of examples as a function of the width of the Chan-Paton strips. It is clear that as the width becomes large the range of $m$ narrows down to $m=1/2$, and the quartic vertex region of moduli space disappears. This is expected since the Feynman diagrams generated by the Witten vertex already provide a complete single cover of the moduli space, and there is no need for a quartic vertex to fill in gaps.

\subsection{Field redefinition}
\label{subsec:infinitesimal}

The dynamical string field of the Kaku theory with half-string length $\ell$ will be denoted $\Psi_\ell$. An infinitesimal change of $\ell$ should give another string field which is related by infinitesimal field redefinition:
\begin{equation}
\Psi_{\ell+\eps} = \Psi_\ell -\eps\Big[\mu_2^\ell(\Psi_\ell,\Psi_\ell)+\mu_3^\ell(\Psi_\ell,\Psi_\ell,\Psi_\ell)+\mu_4^\ell(\Psi_\ell,\Psi_\ell,\Psi_\ell,\Psi_\ell)+...\Big],\label{eq:inf_red}
\end{equation}
where $\mu_2^\ell,\mu_3^\ell,...$ are multi-string products whose form we describe in a moment. We can assume that the string fields are equal at linearized order (i.e. $\mu_1^\ell=0$) since the kinetic term is independent of $\ell$. In coalgebra language we may equivalently express the infinitesimal field redefinition as 
\begin{equation}
\Psi_{\ell+\eps}=\Psi_\ell -\eps \pi_1\mmu^\ell\frac{1}{1-\Psi_\ell},
\end{equation}
where $\mmu^\ell$ is the coderivation formed by the string products of the infinitesimal field redefinition
\begin{equation}\mmu^\ell=\mmu_2^\ell+\mmu_3^\ell+\mmu_4^\ell +...\end{equation}
By a continuous iteration of the infinitesimal field redefinition, we can effect a large shift in the half string length. In this way we find a field redefinition between Witten and Kugo-Zwiebach string field theories of the form 
\begin{equation}
\PsiW = \pi_1\GG\frac{1}{1-\PsiKZ},
\end{equation}
where $\GG$ is a cohomomorphism given by a path-ordered exponential
\begin{equation}
\GG = \overleftarrow{\mathcal{P}}\exp\left[-\int_0^\infty d\ell \, \mmu^\ell\right],\label{eq:GG}
\end{equation}
where the arrow over $\mathcal{P}$ indicates that compositions of $\mmu^\ell$ are ordered from right to left in sequence of increasing $\ell$.
The reverse field redefinition is given by
\begin{equation}
\PsiKZ = \pi_1\GG^{-1}\frac{1}{1-\PsiW},
\end{equation}
where
\begin{equation}
\GG^{-1} = \overrightarrow{\mathcal{P}}\exp\left[\int_0^\infty d\ell \, \mmu^\ell\right],
\end{equation}
and the arrow indicates that $\mmu^\ell$ are ordered from left to right in sequence of increasing $\ell$. Unpackaging the coalgebra notation, the field redefinition can be written explicitly up to cubic order as
\begin{eqnarray}
\PsiW\lineup = \PsiKZ - \int_0^\infty d\ell\, \mu_2^\ell(\PsiKZ,\PsiKZ) -\int_0^\infty d\ell\, \mu_3^\ell(\PsiKZ,\PsiKZ.\PsiKZ)\nonumber\\
\lineup\ \ \ \ \ \ \ \ \ \ +\int_0^\infty d\ell'\int_0^{\ell'} d\ell\left[\mu_2^{\ell'}\Big(\mu_2^\ell(\PsiKZ,\PsiKZ),\PsiKZ\Big)+\mu_2^{\ell'}\Big(\PsiKZ,\mu_2^\ell(\PsiKZ,\PsiKZ)\Big)\right]\nonumber\\
\lineup\ \ \ \ \ \ \ \ \ \ + \text{higher orders}.\phantom{\bigg)}
\end{eqnarray}
What remains is to find the string products $\mu_n^\ell$. 

This is a particular occurrence of the general problem solved in \cite{HataZwiebach}. We will describe the results without derivation. The $\mu_n^\ell$s will generate the infinitesimal field redefinition \eq{inf_red} if two conditions are satisfied. First, they must be cyclic,
\begin{equation}\langle \omega|\pi_2\mmu^\ell = 0,\end{equation}
and second they must satisfy the differential equation
\begin{equation}
\frac{d}{d\ell} \mm^\ell = [\mm^\ell ,\mmu^\ell],
\end{equation}
where 
\begin{equation}\mm^\ell = \QQ+\mm_2^\ell +\mm_3^\ell\end{equation}
is the coderivation representing the cyclic $A_\infty$ structure of the Kaku theory with half string length~$\ell$. Expanded into component products the differential equation implies
\begin{eqnarray}
\frac{d}{d\ell}\mm^\ell_2 \lineup = [\QQ,\mmu_2^\ell],\label{eq:dmK2}\\
\frac{d}{d\ell}\mm^\ell_3 \lineup = [\QQ,\mmu_3^\ell]+[\mm_2^\ell,\mmu_2^\ell],\label{eq:dmK3}\\
0\lineup = [\QQ,\mmu_4^\ell] +[\mm_2^\ell,\mmu_3^\ell]+[\mm_3^\ell,\mmu_2^\ell],\phantom{\bigg)}\label{eq:dmK4}\\
\lineup \ \vdots\  .\nonumber
\end{eqnarray}
Since $m_3^\ell$ is the highest nonvanishing product of the Kaku theory, the left hand side of \eq{dmK4} is zero. We will give the solution in terms of unshifted vertices
\begin{eqnarray}
e_3^\ell(A,B,C) \lineup  = (-1)^{1+|B|}\omega(A,\mu_2^\ell(B,C)),\\
e_4^\ell(A,B,C,D) \lineup = (-1)^{|A|+|C|}\omega(A,\mu_3^\ell(B,C,D)),\\
e_5^\ell(A,B,C,D,E)\lineup = (-1)^{1+|B|+|D|}\omega(A,\mu_4^\ell(B,C,D,E)),\\
\lineup\ \vdots\ ,\ \ \ \nonumber
\end{eqnarray}
where the signs are determined by \eq{unshift}. 

The solution is characterized by Schiffer vector fields which alter the half-string length of the cubic and quartic Kaku vertices. Specifically, the Schiffer vector fields determine certain energy-momentum contour integrals satisfying the relations
\begin{eqnarray}
\frac{d}{d\ell}\langle V_3^\ell|\lineup = -\langle V_3^\ell| T_{(3),\ell}^\ell,\\
\frac{\d}{\d\ell}\langle \Sigma_4^{\ell,m}|\lineup =-\langle\Sigma_4^{\ell,m}|T_{(4),\ell}^{\ell,m},\\
\frac{\d}{\d m}\langle \Sigma_4^{\ell,m}|\lineup = -\langle\Sigma_4^{\ell,m}|T_{(4),m}^{\ell,m},
\end{eqnarray}
where $\langle \Sigma_4^{\ell,m}|$ is the surface state of the quartic Kaku vertex with half string length $\ell$ and at a point $m\in[m_-^\ell,m_+^\ell]$ in the moduli space. On the right hand side we have a set of contour integrals of the energy-momentum tensor operating on the inputs:
\begin{eqnarray}
T_{(3),\ell}^\ell\lineup = T_{(3,1),\ell}^\ell\otimes\mathbb{I}\otimes\mathbb{I}\, +\, \mathbb{I}\otimes T_{(3,2),\ell}^\ell\otimes \mathbb{I}\, +\, \mathbb{I}\otimes\mathbb{I}\otimes T_{(3,3),\ell}^\ell,\nonumber\\ \\
T_{(4),\ell}^{\ell,m}\lineup = T_{(4,1),\ell}^{\ell,m}\otimes\mathbb{I}\otimes\mathbb{I}\otimes\mathbb{I}\, +\, \mathbb{I}\otimes T_{(4,2),\ell}^{\ell,m}\otimes \mathbb{I}\otimes\mathbb{I}\,+\,\mathbb{I}\otimes\mathbb{I}\otimes T_{(4,3),\ell}^{\ell,m}\otimes\mathbb{I}\, +\, \mathbb{I}\otimes\mathbb{I}\otimes\mathbb{I}\otimes T_{(4,4),\ell}^{\ell,m},\nonumber\\ \\
T_{(4),m}^{\ell,m}\lineup = T_{(4,1),m}^{\ell,m}\otimes\mathbb{I}\otimes\mathbb{I}\otimes\mathbb{I}\, +\, \mathbb{I}\otimes T_{(4,2),m}^{\ell,m}\otimes \mathbb{I}\otimes\mathbb{I}\,+\,\mathbb{I}\otimes\mathbb{I}\otimes T_{(4,3),m}^{\ell,m}\otimes\mathbb{I}\, +\, \mathbb{I}\otimes\mathbb{I}\otimes\mathbb{I}\otimes T_{(4,4),m}^{\ell,m}.\ \ \ \ \ \ \ \label{eq:T4mlm}\nonumber\\
\end{eqnarray}
The Schiffer vector fields determine the form of these energy-momentum contour integrals via
\begin{eqnarray}
f_{(3,i)}^\ell\circ T_{(3,i),\ell}^\ell \lineup = \oint_{f_{(3,i)}^\ell(0)}\frac{du}{2\pi i}v_{(3,i),\ell}^\ell(u) T(u),\\
f_{(4,i)}^{\ell,m}\circ T_{(4,i),\ell}^{\ell,m} \lineup = \oint_{f_{(4,i)}^{\ell,m}(0)}\frac{du}{2\pi i}v_{(4,i),\ell}^{\ell,m}(u) T(u),\\
f_{(4,i)}^{\ell,m}\circ T_{(4,i),m}^{\ell,m} \lineup = \oint_{f_{(4,i)}^{\ell,m}(0)}\frac{du}{2\pi i}v_{(4,i),m}^{\ell,m}(u) T(u).\label{eq:T4imlm}
\end{eqnarray}
Here $v_{(3,i),\ell}^\ell(u),v_{(4,i),\ell}^{\ell,m}(u)$ and $v_{(4,i),m}^{\ell,m}(u)$ are the Schiffer vector fields expressed in the upper half plane coordinate of the cubic and quartic vertices. We claim that their explicit form is given by 
\begin{eqnarray}
v_{(3,i),\ell}^\ell(u)\lineup = \frac{\d  (\fl{i})^{-1}(u)}{\d\ell}\left(\frac{\d  (\fl{i})^{-1}(u)}{\d u}\right)^{-1},\\
v_{(4,i),\ell}^{\ell,m}(u)\lineup = \frac{\d   (\flm{i})^{-1}(u)}{\d \ell}\left(\frac{\d  (\flm{i})^{-1}(u)}{\d u}\right)^{-1} ,\\
v_{(4,i),m}^{\ell,m}(u)\lineup = \frac{\d   (\flm{i})^{-1}(u)}{\d m}\left(\frac{\d  (\flm{i})^{-1}(u)}{\d u}\right)^{-1},
\end{eqnarray}
where $f_{(4,i)}^{\ell,m}$ are the local coordinate maps of the quartic Kaku vertex. This result can be derived, for example, paralleling the computation in section 3 of \cite{ClosedSFT_Erler}.

The Schiffer vector fields $v_{(4,i),m}^{\ell,m}$ provide the measure for integration over the quartic vertex region of moduli space. In particular, the quartic Kaku vertex may be expressed
\begin{equation}
\langle V_4^\ell| = \int_{m_-^\ell}^{m_+^\ell} dm\,\langle \Sigma_4^{\ell,m}|b_{(4),m}^{\ell,m},
\end{equation}
where $b_{(4),m}^{\ell,m}$ is given as in $T_{(4),m}^{\ell,m}$ in \eq{T4mlm} and \eq{T4imlm} after replacing the energy-momentum tensor with the $b$ ghost. We define $b_{(3),\ell}^{\ell}$ and $b_{(4),\ell}^{\ell,m}$ similarly. The measure can be simplified by recognizing that all four Schiffer vector fields $v_{(4,i),m}^{\ell,m}$ are equal to the same globally defined meramorphic vector field on the upper half plane:
\begin{equation}
v_{(4,i),m}^{\ell,m}(u) = \frac{u(1-u)}{m(1-m)}\frac{-|u_*|^2+m\mathrm{Re}(u_*)+(\mathrm{Re}(u_*)-m)u}{(u-u_*)(u-\overline{u}_*)},
\end{equation}
where $u_*$ is the interaction point of the quartic vertex, given in \eq{ustar} (with the kinematic moduli defined as for the quartic Kaku vertex). This means that the $b$-ghost contours around each puncture can be joined into a single contour surrounding the pole of the vector field at the interaction point, and, using the doubling trick, its conjugate in the lower half plane. Picking out the residue of the pole gives  \eq{V4complete}. We did not find any noteworthy simplification of the vector fields $v_{(3,i),\ell}^\ell$ and~$v_{(4,i),\ell}^{\ell,m}$.

With these preparations, the unshifted vertices of the field redefinition can be taken as
\begin{eqnarray}
\langle e_3^\ell | \lineup = -\langle V_3^\ell |b_{(3),\ell}^\ell,\\
\langle e_4^\ell| \lineup = \int_{m_-^\ell}^{m_+^\ell}dm \, \langle \Sigma^{\ell,m}_4|b_{(4),m}^{\ell,m}b_{(4),\ell}^{\ell,m},\\
\langle e_n^\ell| \lineup = 0,\ \ \ n\geq 5.
\end{eqnarray}
The last equation implies that the infinitesimal field redefinition terminates at cubic order, i.e. $\mu_n^\ell =0$ for $n\geq 4$. It is not difficult to verify that $\langle e_3^\ell|$ satisfies \eq{dmK2}. The higher relations are trickier since the $b$-ghost contour integrals in the various terms do not take the same form, and one must derive certain conservation laws to demonstrate the requisite cancellations. For more on this, see \cite{HataZwiebach}.

\section{The full mapping}
\label{sec:transformation}

We propose to transform the lightcone string field into the Kugo-Zwiebach string field following sections \ref{sec:effective} and \ref{sec:KZ}, and then further map the Kugo-Zwiebach string field to the Witten string field following section \ref{sec:Kaku}. The result is a transformation between the lightcone and Witten string fields:
\begin{equation}\PsiW = \pi_1 \GG\II^\text{lc}\frac{1}{1-\Psilc}\end{equation}
where $\II^\text{lc}$ is the interacting inclusion \eq{II} based on the lightcone $A_\infty$ structure $\mmlc$, and $\GG$ is given as in \eq{GG}.  Expanding to third order in $\Psilc$ gives 
\begin{eqnarray}
\PsiW \lineup = S\Psilc - \int_0^\infty d\ell\, \mu_2^\ell (S\Psilc,S\Psilc)-\frac{\b}{\L}\mlc_2(S \Psilc,S\Psilc)\nonumber\\
\lineup \ \ \ \ \ \ \ \ \ -\int_0^\infty d\ell\, \mu_3^\ell(S\Psilc,S\Psilc.S\Psilc) -\frac{\b}{\L}\mlc_3(S \Psilc,S \Psilc,S \Psilc)\nonumber\\
\lineup\ \ \ \ \ \ \ \ \ \ +\int_0^\infty d\ell'\int_0^{\ell'} d\ell\left[\mu_2^{\ell'}\Big(\mu_2^\ell(S\Psilc,S\Psilc),S\Psilc\Big)+\mu_2^{\ell'}\Big(S\Psilc,\mu_2^\ell(S\Psilc,S\Psilc)\Big)\right]\nonumber\\
\lineup\ \ \ \ \ \ \ \ \ \ +\frac{\b}{\L}\mlc_2\left(S\Psilc, \frac{\b}{\L}\mlc_2(S\Psilc,S\Psilc)\right) +\frac{\b}{\L}\mlc_2\left(\frac{\b}{\L}\mlc_2(S\Psilc,S\Psilc), S\Psilc\right)\nonumber\\
\lineup\ \ \ \ \ \ \ \ \ \ + \int_0^\infty d\ell\, \left[\mu_2^\ell\left(S\Psilc,\frac{\b}{\L}\mlc_2(S\Psilc,S\Psilc)\right)+\mu_2^\ell\left(\frac{\b}{\L}\mlc_2(S\Psilc,S\Psilc),S\Psilc\right)\right]\nonumber\\
\lineup\ \ \ \ \ \ \ \ \ \ + \text{higher orders}.\phantom{\bigg)}\label{eq:lcWexp} 
\end{eqnarray}
We now discuss a few features of this proposal.

First we observe that the Kugo-Zwiebach string field is already a nonpolynomial function of the lightcone string field, due to the higher string products contained in the  interacting inclusion~$\II^\text{lc}$. This is surprising since only the linear relation $\PsiKZ=S\Psilc$ is sufficient to map the Kugo-Zwiebach action into the lightcone action. The nonlinear terms contribute exclusively to the longitudinal part of the Kugo-Zwiebach string field, and are present because the equations of motion provide a source for the longitudinal states when the transverse states are nonzero. Thus the nonlinear corrections are necessary to ensure that an on-shell configuration of the  lightcone theory maps to an on-shell configuration of the Kugo-Zwiebach theory. The reason why this is not noticed when comparing the actions is that the longitudinal states do not ``push back" on the transverse states, and in this way do not alter their effective couplings.  This seeming violation of Newton's third law is the surprising mechanism behind transfer invariance. The arguments of subsection \ref{subsec:transfer} imply that the nonlinear corrections are only visible when contracted with states containing minus oscillators. This implies that nonlinear corrections only effect component fields which multiply states containing plus oscillators. The lowest level field which is effected by nonlinear corrections is the plus component of the Kugo-Zwiebach gauge field, $A_+^\text{KZ}$.

Considering the complete transformation, one might still ask why it should be nonpolynomial. Clearly, a polynomial field transformation of a polynomial string field theory will generate another polynomial string field theory. One might hope that the reverse is also true, that  any two  polynomial theories can be related by a polynomial transformation. Indeed we have seen that this is true for Kaku theories differing by an infinitesimal change in the half string length. If such a transformation exists between Witten and lightcone string field theories, it would establish a much more powerful connection between the theories than the one given here. 

However, the most significant issue with the present proposal becomes apparent when  we contract the  Witten string field with  a  transverse test state. Consider a test state $\phi\in\Hlc^\perp$ and $S\phi\in\HDDF$. The result of subsection \ref{subsec:genericcubic} shows that the  field transformation up to quadratic order takes the form
\begin{equation}
\langle S\phi,\PsiW\rangle = \langle\phi,\Psilc\rangle+\int_0^\infty d\ell\,\langle V_3^{\text{eff},\ell}|b_{(3),\ell}^\ell \big(\phi\otimes\Psilc\otimes\Psilc\big)+\text{higher orders},
\end{equation}
where $\langle V_3^{\text{eff},\ell}|$ is the cubic vertex of the lightcone effective field theory derived from the Kaku theory with half string length $\ell$. While the local dilatation at each puncture in the Kaku vertex is a function of the minus momenta, it does not shrink to zero as the minus momentum at the puncture is taken to be small. This is intuitively clear since the Chan-Paton strips do not disappear as the  string approaches zero momentum. Therefore the cubic vertex of the lightcone effective field theory suffers from the soft string problem, and highly excited string states give exponentially enhanced contribution to the field redefinition. It seems that the interacting inclusion produces a Kugo-Zwiebach string field in a form which is not fully suitable for further field redefinition. This is an important problem, but we will not try to address it in this paper. 

\begin{figure}
\begin{center}
\resizebox{3in}{2.5in}{\includegraphics{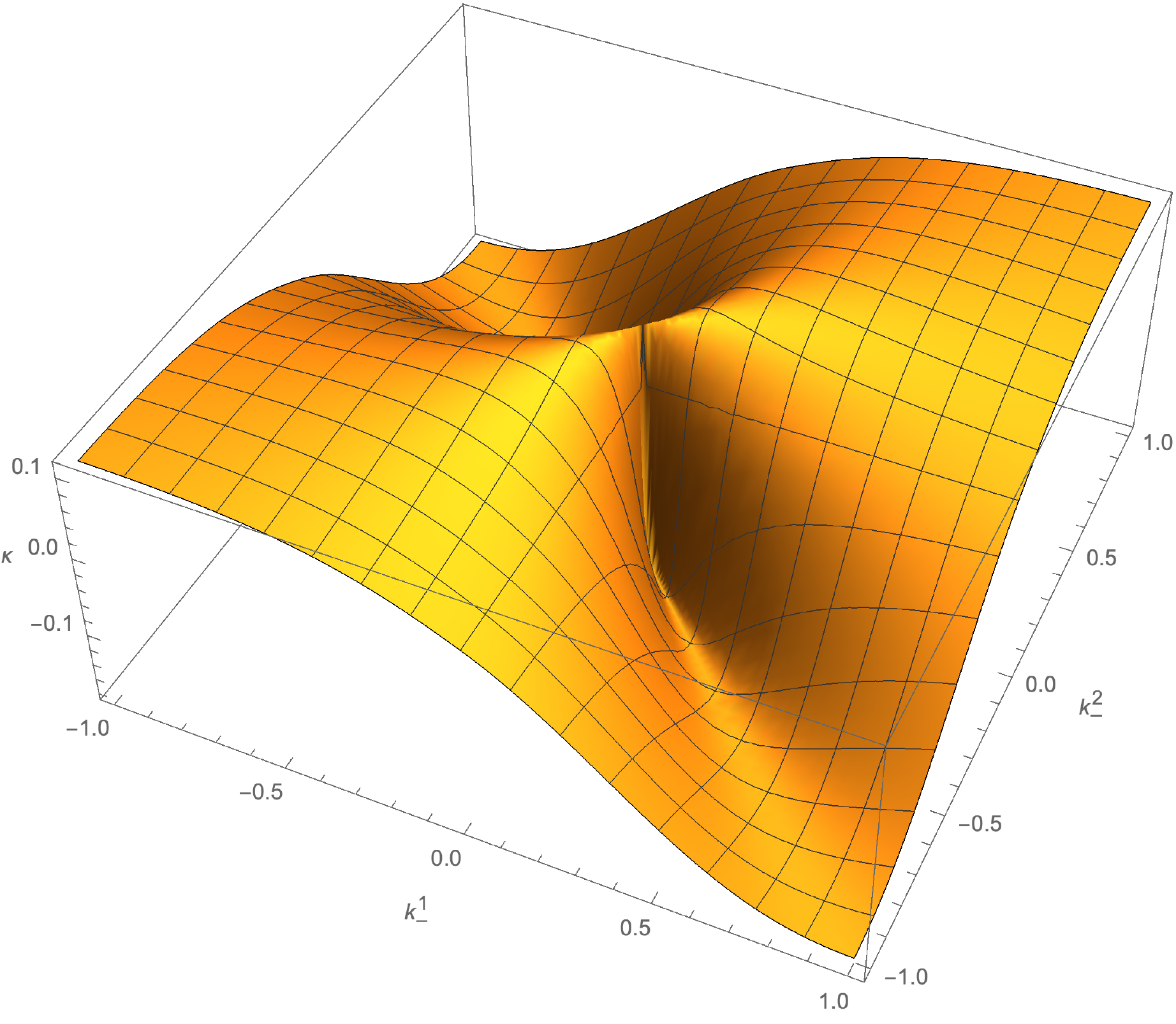}}
\resizebox{3in}{2.5in}{\includegraphics{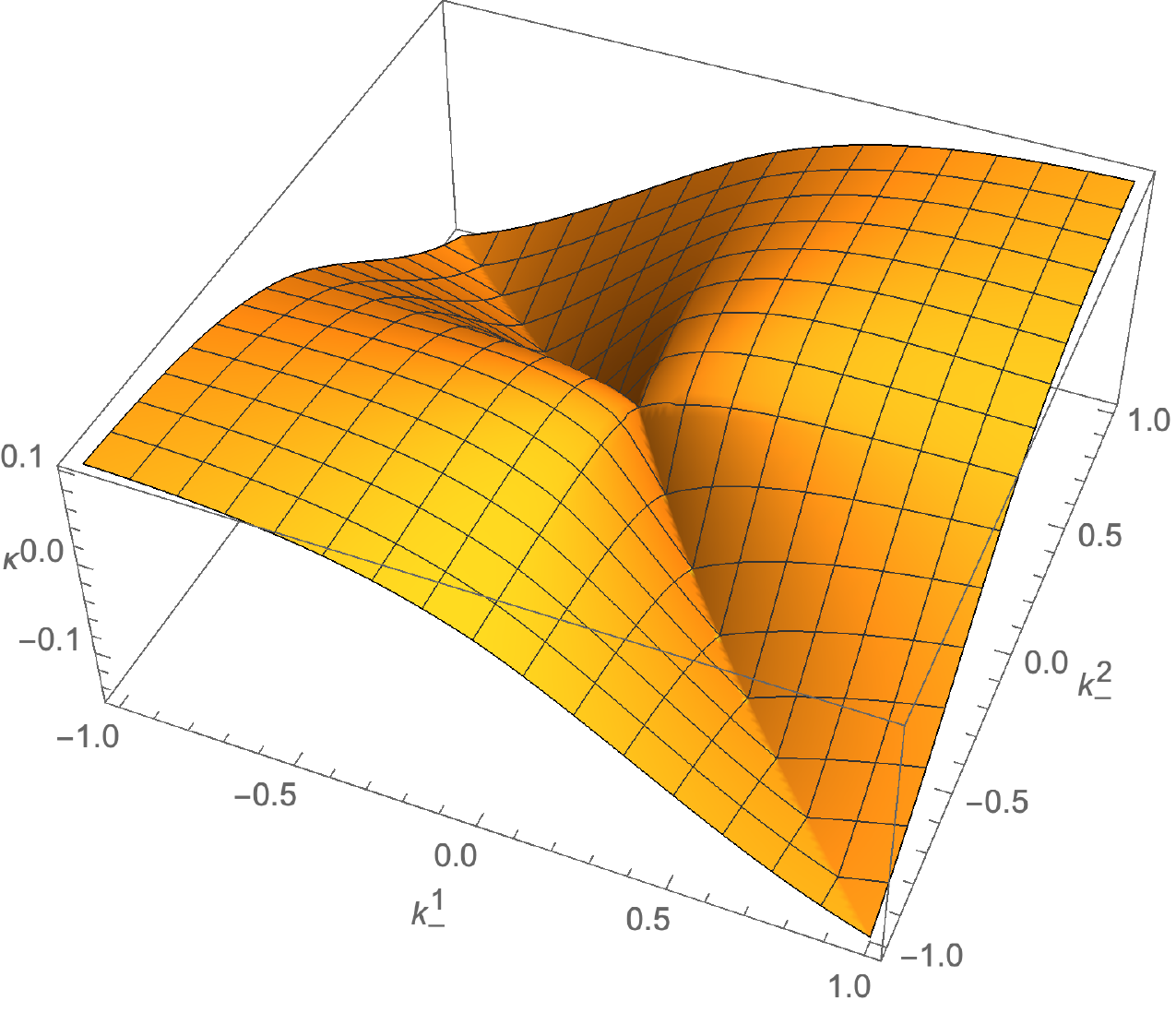}}
\end{center}
\caption{\label{fig:lightcone23} On the left is the kernel $\kappa(k^1,k^2,k^3)$ as a function of $k_-^1,k_-^2$ with fixed 
$(k^1)^2=(k^2)^2=(k^3)^2=3$. The kernel is discontinuous at the origin in the $k_-^1,k_-^2$ plane. On the right is the corresponding plot as it applies to the tachyon of the Kaku theory with half string length $\ell =.5$.  The discontinuity is removed.}
\end{figure}

Still we would like to do a concrete calculation. One thing we can calculate is the quadratic order relation between the tachyon of the  lightcone theory and the tachyon of the Witten theory. Since this does not concern highly excited string states, it is possible that the result may be unaffected by improvements to the present proposal.  The Witten tachyon field $\TW(k^3)$ is given by contracting the field transformation \eq{lcWexp} with the test state
\begin{equation}
c_0c_1e^{-ik^3\cdot X(0,0)}|0\rangle.
\end{equation}
To extract the contribution from the lightcone tachyon, we substitute
\begin{equation}S\Psilc = \int\frac{d^{26}k}{(2\pi)^{26}}\Tlc(k) c_1 e^{ik\cdot X(0,0)}|0\rangle.\end{equation}
Recall that the transformation $S$ leaves the tachyon state invariant. Up to quadratic order we find
\begin{eqnarray}
\TW(k^3) \lineup= \Tlc(k^3)-\int\frac{d^{26}k_1d^{26}k_2}{(2\pi)^{2\cdot 26}}\Tlc(k_1)\Tlc(k_2)\nonumber\\
\lineup\ \ \ \ \ \ \ \ \ \ \ \ \ \ \ \  \times \!\int_0^\infty d\ell \left.\langle e_3|\Big(c_0c_1e^{-ik^3\cdot X(0,0)}|0\rangle\Big)\!\otimes\!\Big(c_1 e^{ik^1\cdot X(0,0)}|0\rangle\Big)\!\otimes\!\Big(c_1e^{ik^2\cdot X(0,0)}|0\rangle\Big)\right|_\ell\nonumber\\
\lineup\ \ \ \ \ \ \ \ \ \ \ \ +\ \text{higher orders}. \phantom{\bigg)}
\end{eqnarray}
The integration over momenta takes the vertex out of the standard configuration. To put it into the standard configuration we assume cyclicity and write
\begin{figure}
\begin{center}
\resizebox{2.1in}{2in}{\includegraphics{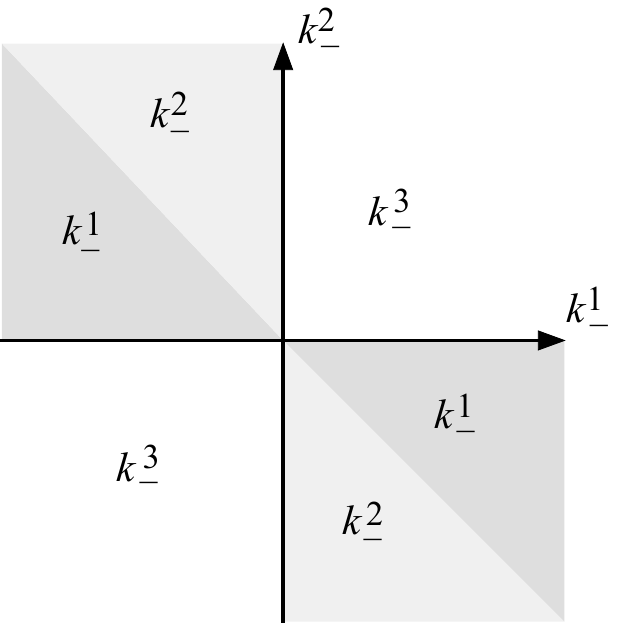}}
\end{center}
\caption{\label{fig:lightcone22} As a function of the minus momenta $k_-^1,k_-^2$, transforming the vertex $\langle e_3|$ into the standard configuration defines it piecewise on three pairs of regions in the $k_-^1,k_-^2$ plane. The regions are labeled above according to whether $k_-^1,k_-^2$, or $k_-^3=k_-^1+k_-^2$ has largest magnitude.}
\end{figure}
\begin{eqnarray}
\lineup \langle e_3|\Big(c_0c_1e^{-ik^3\cdot X(0,0)}|0\rangle\Big)\!\otimes\!\Big(c_1 e^{ik^1\cdot X(0,0)}|0\rangle\Big)\!\otimes\!\Big(c_1e^{ik^2\cdot X(0,0)}|0\rangle\Big)\nonumber\\
\lineup\ \ \ \ \ \ \ \ \ \ \ \ \ \ \  = \delta_{k_-^1/k_-^2>0} \,\langle e_3|\Big(c_1e^{ik^1\cdot X(0,0)}|0\rangle\Big)\!\otimes\!\Big(c_1 e^{ik^2\cdot X(0,0)}|0\rangle\Big)\!\otimes\!\Big(c_0c_1e^{-ik^3\cdot X(0,0)}|0\rangle\Big)\nonumber\\
\lineup\ \ \ \ \ \ \ \ \ \ \ \ \ \ \ \ \ \  -\delta_{k_-^2/k_-^3<0}\,\langle e_3|\Big(c_1e^{ik^2\cdot X(0,0)}|0\rangle\Big)\!\otimes\!\Big(c_0c_1 e^{-ik^3\cdot X(0,0)}|0\rangle\Big)\!\otimes\!\Big(c_1e^{ik^1\cdot X(0,0)}|0\rangle\Big)\nonumber\\
\lineup \ \ \ \ \ \ \ \ \ \ \ \ \ \ \ \ \ \  +\delta_{k_-^1/k_-^3<0}\,\langle e_3|\Big(c_0c_1e^{-ik^3\cdot X(0,0)}|0\rangle\Big)\!\otimes\!\Big(c_1 e^{ik^1\cdot X(0,0)}|0\rangle\Big)\!\otimes\!\Big(c_1e^{ik^2\cdot X(0,0)}|0\rangle\Big).\ \ \ \ \ \ \label{eq:M3standard}
\end{eqnarray}
The three terms correspond to three pairs of regions in the $k_-^1,k_-^2$ plane, as illustrated in figure~\ref{fig:lightcone22}. Since the Schiffer vector fields are holomorphic at the origin, the $b$-ghost operators inside $\langle e_3|$ are given as a term proportional to $b_0$ plus annihilation operators. Only the term proportional to $b_0$ contributes in \eq{M3standard}, and then only when it acts on the $c_0c_1$ vacuum. We have
\begin{eqnarray}
b_{(3,1),\ell}^\ell\lineup =-\frac{1}{f_{(3,1)}^{\ell}\!\!\! \,' (0)}\frac{d f_{(3,1)}^{\ell}\!\!\! \,' (0)}{d\ell} b_0 + \text{annihilation operators},\\
b_{(3,2),\ell}^\ell\lineup =-\frac{1}{f_{(3,2)}^{\ell}\!\!\! \,' (0)}\frac{d f_{(3,2)}^{\ell}\!\!\! \,' (0)}{d\ell}b_0 + \text{annihilation operators},\\
b_{(3,3),\ell}^\ell\lineup =\frac{1}{(I\circ f_{(3,3)}^{\ell})' (0)}\frac{d(I\circ f_{(3,3)}^{\ell})' (0)}{d\ell}b_0 + \text{annihilation operators}.
\end{eqnarray}
Bringing all this together and evaluating the correlators gives
\begin{eqnarray}
\TW(k^3)\lineup = \Tlc(k^3) - \int\frac{d^{26}k^1d^{26}k^2}{(2\pi)^{2\cdot 26}}(2\pi)^{26}\delta^{26}(k^1+k^2-k^3)\,\kappa(k^1,k^2,k^3)\,\Tlc(k^1)\Tlc(k^2)\nonumber\\
\lineup\ \ \ \ \ \ \ \ \ \ \ \ \ \ \ \ +\text{higher orders},\phantom{\Big)}
\label{eq:kappa}\end{eqnarray}
where the kernel $\kappa$ is the expression
\begin{eqnarray}
\lineup\kappa(k^1,k^2,k^3) \nonumber\\
\lineup = -\int_0^\infty dL\left[ \delta_{\frac{k_-^1}{k_-^2}> 0}\big[f_{(3,1)}^{\ell}\!\!\! \,' (0)\big]^{(k^1)^2-1}\big[f_{(3,2)}^{\ell}\!\!\! \,' (0)\big]^{(k^2)^2-1}\big[(I\circ f_{(3,3)}^{\ell})' (0)\big]^{(k^3)^2-2}\frac{ \d(I\circ f_{(3,3)}^{\ell})' (0)}{\d L}\Bigg|_{K=\frac{|k_-^2|}{|k_-^3|}}\right.\nonumber\\
\lineup\ \ \ \ \ \ \  +\delta_{\frac{k_-^2}{k_-^3}> 0}\big[f_{(3,1)}^{\ell}\!\!\! \,' (0)\big]^{(k^2)^2-1}\big[f_{(3,2)}^{\ell}\!\!\! \,' (0)\big]^{(k^3)^2-2}\big[(I\circ f_{(3,3)}^{\ell})' (0)\big]^{(k^1)^2-1}\frac{\d f_{(3,2)}^{\ell}\,\!\!' (0)}{\d L}\Bigg|_{K=\frac{|k_-^3|}{|k_-^1|}}\nonumber\\
\lineup\ \ \ \ \ \ \  +\left.\delta_{\frac{k_-^1}{k_-^3}> 0}\big[f_{(3,1)}^{\ell}\!\!\! \,' (0)\big]^{(k^3)^2-2}\big[f_{(3,2)}^{\ell}\!\!\! \,' (0)\big]^{(k^1)^2-1}\big[(I\circ f_{(3,3)}^{\ell})' (0)\big]^{(k^2)^2-1}\frac{ \d f_{(3,1)}^{\ell}\,\!\!' (0)}{\d L}\Bigg|_{K=\frac{|k_-^1|}{|k_-^2|}}\right].\label{eq:kappa2}
\end{eqnarray}
Here $L$ and $K$ are the normalized ratios \eq{KL}, evaluated in the three terms as indicated. The kernel is expressed as a function of three momenta $k^1,k^2,k^3$, but they are constrained by momentum conservation. We may evaluate the integration over $L$ numerically. In figure \ref{fig:lightcone23}, we plot the kernel as a function of $k_-^1$ and $k_-^2$ keeping the squared momenta fixed. The notable feature of this plot is a discontinuity at the origin in the $k_-^1,k_-^2$ plane. This implies that there is no well-defined mapping of the zero momentum tachyon of the lightcone theory into the zero momentum tachyon of the Witten theory. This may have been expected, since lightcone string field theory does not define the dynamics of the tachyon at zero momentum. It is interesting to compare to what happens if we had considered the tachyon of the  Kaku theory with small but nonzero half string length. The relation to the tachyon of the Witten theory is given by suitably regularizing the lower limit of the integration over $L$ in \eq{kappa2}. One finds that the discontinuity in the kernel is removed, and in fact the kernel vanishes at the origin of the $k^1_-,k^2_-$ plane. This implies that the zero momentum tachyon of the Kaku theory is equal to that of  the Witten theory (at least to this order).  This is also expected, since the Kaku theory at zero momentum is identical to Witten's string field theory and the field redefinition relating them should be  trivial.

\section{Concluding Remarks}

We conclude by listing some questions for future inquiry. 
\begin{itemize}

\item The  present form of the field transformation is problematic due to the singular behavior of highly excited DDF states in generic string vertices---what we refer to as the ``soft string problem." It seems unlikely that the soft string problem indicates a physical obstruction in relating covariant and lightcone string field theories. More likely, the field transformation can be altered as
\begin{equation}\PsiW = \pi_1\GG{\bf H}\II^\text{lc}\frac{1}{1-\Psilc}\end{equation}
where the cohomomorphism ${\bf H}$ represents a symmetry of the Kugo-Zwiebach action which prepares the transferred lightcone string field so that further field redefinition is unproblematic. A suitable construction of ${\bf H}$ will require some thought. 

\item We have discussed the open bosonic string, but we expect that the story will be parallel for the closed bosonic string, up to a point. The unresolved question is how to deform the closed string Kugo-Zwiebach theory into a canonically defined covariant closed bosonic string field theory. It seems natural to consider the theory defined by polyhedral vertices \cite{poly,Zwiebach}. The cubic Kaku vertex can be generalized without difficulty to the closed string, where it defines an intermediate between the lightcone vertex and the Witten-style trihedral vertex. Perhaps a corresponding deformation of higher vertices at genus 0 can be found by modification of length conditions on the closed string polyhedra. It may be useful to think about this in connection to recent efforts to define string vertices using hyperbolic Riemann surfaces \cite{PiusMoosavian,CostelloZwiebach,Firat}.

\item At the quantum level we encounter the puzzle that loop amplitudes computed in the closed string Kugo-Zwiebach theory appear to have too many moduli. For an $n$-point amplitude at genus $g$, we find the expected $6g-6+2n$ moduli from the Schwinger parameters and twist angles of the propagators, but it appears there are $g$ additional moduli resulting from integration over the string length parameters in loops. The overcounting of moduli however is illusory, since integrating over loop lightcone energies produces delta functions which reduce the number of independent Schwinger parameters. This is expected from the point  of view of Hamiltonian quantization, where the time coordinate of all strings is identified with the lightcone coordinate in spacetime. The mechanism is more obscure from the covariant point of view, since the required delta functions only appear with the Lorentzian definition of the propagator, and the integration over loop momenta is usually implicit as part of the worldsheet path integral. Assuming this can be  dealt with, the Kugo-Zwiebach theory may offer a useful approach to computation of loop amplitudes in covariant string field theory, since lightcone vertices lead to a very simple parameterization of the moduli space \cite{Wolpert}. 

\item For the superstring, we encounter the problem of contact divergences in the lightcone theory resulting from collisions of operators at interaction points \cite{Klinkhamer1,Klinkhamer2,Seiberg}. This is expected to be related to the issue of picture-changing operator collisions and spurious poles in covariant superstring field theories. In covariant string field theories we now have a fairly good understanding of how to deal with this problem \cite{WittenSS,SenOffShell,SenWitten,vertical}. It would be very interesting if this could be translated to address the contact divergence problem in lightcone superstring field theories. 

\end{itemize}

\noindent We hope to see progress on these questions.

\subsubsection*{Acknowledgements}

T.E. would like to thank D. Gross and I. Sachs for discussions which motivated interest in this problem, and additionally thanks I. Sachs for invitation to the LMU workshop ``Recent developments on the light front."  H.M. would like to thank H. Kunitomo for discussions. The work of H.M. is supported by the GA{\v C}R grant 19-282268X. The work of T.E. is supported by European Structural and Investment Fund and the Czech Ministry of Education, Youth and Sports (Project CoGraDS
- CZ.02.1.01/0.0/0.0/15\_ 003/0000437) and the GA{\v C}R project 18-07776S and RVO: 67985840.

\begin{appendix}

\section{Free bosons}
\label{app:free}

Here we list some formulas related to the free boson BCFT of open strings attached to a space-filling D25-brane. First we establish some notation concerning worldsheet coordinates:
\begin{itemize}
\item We use $(u,\overline{u})$ to denote holomorphic and antiholomorphic coordinates on the upper half plane $\text{Im}(u)\geq 0$. 
\item We will use $(\xi,\overline{\xi})$ to denote the holomorphic and antiholomorphic coordinates of radial quantization on the unit half-disk $\text{Im}(\xi)\geq 0$ and $|\xi|<1$. 
\item We use $(w,\overline{w})$ to denote holomorphic and antiholomorphic coordinates on the strip $\mathrm{Im}(w)\in[0,\pi]$ and $\mathrm{Re}(w)<0$. These are related to the coordinates of radial quantization through $w=\ln(\xi)$.
\item We use $(\rho,\overline{\rho})$ to denote the {\it interaction coordinate} on a Mandelstam diagram. 
\item We use $(z,\overline{z})$ to denote unspecified holomorphic and antiholomorphic coordinates on the worldsheet.
\end{itemize}
The string embedding coordinates are free bosons $X^\mu(z,\overline{z})$ with $\mu=0,...,25$ (or $\mu=+,-,1,...,24$), subject to Neumann boundary conditions.  The free bosons have OPE  
\begin{equation}X^\mu(z,\overline{z})X^\nu(z',\overline{z}') = -\frac{1}{2}\eta^{\mu\nu}\ln|z-z'|^2+\text{regular},\end{equation}
provided that $(z,\overline{z})$ does not sit on the open string boundary. We have the mode expansion 
\begin{equation}X^\mu(\xi,\overline{\xi}) = x^\mu -ip^\mu \ln|\xi|^2+\frac{i}{\sqrt{2}}\sum_{n\in \mathbb{Z}}\delta_{n\neq 0}\frac{\alpha_n^\mu}{n}\left(\frac{1}{\xi^m}+\frac{1}{\overline{\xi}^n}\right).\end{equation}
where $x^\mu$ is the position zero mode, $p_\mu$ is the momentum, and $\alpha_m^\mu$ are oscillators
\begin{equation}\alpha_m^\mu = i\sqrt{2}\oint\frac{d\xi}{2\pi i}\xi^n\d X^\mu(\xi).\end{equation}
We have commutation relations 
\begin{eqnarray}
[x^\mu,p_\nu] \lineup = i\delta^\mu_\nu,\\
\ [\alpha_m^\mu,\alpha_n^\nu] \lineup = m\eta^{\mu\nu}\delta_{m+n=0}.
\end{eqnarray}
The associated $U(1)$ current has OPE 
\begin{equation}
\d X^\mu(z)\d X^\nu(z') = -\frac{\eta^{\mu\nu}}{2}\frac{1}{(z-z')^2}+\text{regular},
\end{equation}
and mode expansion
\begin{equation}
\d X^\mu(\xi) = -\frac{i}{\sqrt{2}}\sum_{n\in\mathbb{Z}}\frac{\alpha^\mu_n}{\xi^{n+1}},
\end{equation}
where $\alpha_0^\mu = \sqrt{2}p^\mu$. It will further be useful to introduce the chiral free boson $X^\mu(z)$ with OPE
\begin{equation}
X^\mu(z)X^\nu(z') = -\frac{\eta^{\mu\nu}}{2}\ln(z-z')+\text{regular},
\end{equation} 
and mode expansion
\begin{equation}
X^\mu(\xi) = \frac{1}{2}x^\mu -ip^\mu \ln(\xi)+\frac{i}{\sqrt{2}}\sum_{n\in \mathbb{Z}}\delta_{n\neq 0}\frac{\alpha_n^\mu}{n}\frac{1}{\xi^n}.
\end{equation}
We have the relation 
\begin{equation}
X^\mu(z,\overline{z}) = X^\mu(z)+X^\mu(\overline{z}).
\end{equation}
Note that the chiral free boson is denoted with one holomorphic argument, whereas the full free boson is denoted with a holomorphic and antiholomorphic argument. 

We are interested in correlation functions of plane wave vertex operators on the upper half plane. The bulk, boundary, and chiral plane wave vertex operators have respective conformal weights
\begin{eqnarray}
e^{ik\cdot X(u,\overline{u})}:\lineup \ \ \ (h,\overline{h}) = \left(\frac{k^2}{4},\frac{k^2}{4}\right),\ \ \ \ \text{Im}(u)>0,\\
e^{ik\cdot X(y,y)}:\lineup \ \ \ h = k^2,\ \ \ \ \text{Im}(y)=0,\phantom{\bigg)}\\
e^{ik\cdot X(u)}:\lineup \ \ \ h=\frac{k^2}{4}.
\end{eqnarray}
Bulk or boundary normal ordering is kept implicit. We have OPEs
\begin{eqnarray}
e^{ik\cdot X(u,\overline{u})}e^{ik'\cdot X(u',\overline{u}')} \lineup = |u-u'|^{k\cdot k'} e^{i (k+k')\cdot X(u',\overline{u}')}+\text{regular},\phantom{\Big)}\\
e^{ik\cdot X(y,y)}e^{ik'\cdot X(y',y')} \lineup = |y-y'|^{2 k\cdot k'} e^{i (k+k')\cdot X(y',y')}+\text{regular},\phantom{\Big)}\\
e^{ik\cdot X(u)}e^{ik'\cdot X(u')} \lineup = (u-u')^{\frac{1}{2}k\cdot k'} e^{i (k+k')\cdot X(u')}+\text{regular}.\phantom{\Big)}
\end{eqnarray}
OPEs with the $U(1)$ current are given by
\begin{eqnarray}
\d X^\mu(u)e^{ik\cdot X(u',\overline{u}')} \lineup = -\frac{ik^\mu}{2}\frac{1}{u-u'}e^{ik\cdot X(u',\overline{u}')}+\text{regular},\\
\d X^\mu(u)e^{ik\cdot X(y,y)} \lineup = -\frac{ik^\mu}{u-y}e^{ik\cdot X(y,y)}+\text{regular},\\
\d X^\mu(u)e^{ik\cdot X(u')} \lineup = -\frac{ik^\mu}{2}\frac{1}{u-u'}e^{ik\cdot X(u')}+\text{regular}.
\end{eqnarray}
Correlators of bulk and boundary plane wave vertex operators can be reduced to correlators involving only chiral plane wave vertex operators. Specifically we have the identification
\begin{eqnarray}
e^{ik\cdot X(u,\overline{u})}\lineup\ \longleftrightarrow\  (-i)^{k^2/2} e^{ik\cdot X(u)}e^{ik\cdot X(\overline{u})},\ \ \ \text{Im}(u)>0,\phantom{\Big)}\\
e^{ik\cdot X(y,y)}\lineup\ \longleftrightarrow\   e^{2 ik\cdot X(y)},  \ \ \ \ \ \ \ \ \ \ \ \ \ \ \ \ \ \ \ \ \ \ \text{Im}(y)=0,\phantom{\Big)}
\end{eqnarray}
so that
\begin{eqnarray}
\lineup\!\!\!\!\!\!\!\!\!\!\Big\langle \big(e^{ik^1\cdot X(u_1,\overline{u}_1)}...e^{ik^n\cdot X(u_m,\overline{u}_m)}\big)\big(e^{ik^{m+1}\cdot X(y_{m+1},y_{n+1})}...e^{ik^{n+m}\cdot X(y_{m+n},y_{m+n})}\big)\Big\rangle_{\text{UHP}}\phantom{\bigg)}\nonumber\\
\lineup \!\!\!\!\!\!\!\!=\!(-i)^{((k^1)^2+...+(k^n)^2)/2}\!\Big\langle\!\big( e^{ik^1\cdot X(u_1)}e^{ik^1\cdot X(\overline{u}_1)}...e^{ik^n\cdot X(u_n)}e^{ik^n\cdot X,(\overline{u}_n)}\big)\!\big(e^{2ik^{n+1}\cdot X(y_{n+1})}...e^{2 ik^{n+m}\cdot X(y_{n+m})}\big)\!\Big\rangle_{\mathbb{C}}.\phantom{\bigg)} \nonumber\\ 
\end{eqnarray}
Correlators of chiral plane wave vertex operators may be characterized recursively by the formula 
\begin{eqnarray}
\lineup \Big\langle e^{i k^0 \cdot X(u_0)}\,e^{i k^1 \cdot X(u_1)}\,e^{i k^2\cdot X(u_2)}\, ...\, e^{i k^n\cdot X(u_n)} \Big\rangle_\mathbb{C} \nonumber\\
\lineup \ \ \ \ \ \
 = (u_{01})^{\frac{1}{2}k^0\cdot k^1}\prod_{j=2}^n\left(\frac{u_{0j}}{u_{1j}}\right)^{\frac{1}{2}k^0\cdot k^j}\Big\langle e^{i (k^1+k^0) \cdot X(u_1)}\,e^{i k^2\cdot X(u_2)} \,... \,e^{i k^n\cdot X(u_n)} \Big\rangle_\mathbb{C},\label{eq:eXdelete}
\end{eqnarray}
where $u_{ij}=u_i-u_j$. Note that on the right hand side the operator $e^{i k^0\cdot X(u_0)}$ has been eliminated, and its momentum has been transferred to the chiral plane wave vertex operator at $u_1$. Through repeated application of the recursion we arrive at the correlator containing only one chiral plane wave vertex operator, which is defined as
\begin{equation}
\langle e^{i k\cdot X(u)}\rangle_\mathbb{C} = (2\pi)^{26} \delta^{26}(\textstyle{\frac{1}{2}}k).
\end{equation}
We may further accommodate insertions of the $U(1)$ current to arrive at the recursive formula
\begin{eqnarray}
\lineup \Big\langle \d X^{\mu_0}(u_0)\,\d X^{\mu_1}(u_1)\,...\, \d X^{\mu_m}(u_m)\, e^{i k^{m+1}\cdot X(u_{m+1})}\, ... \, e^{i k^{m+n}\cdot X(u_{m+n})}\Big\rangle_\mathbb{C}\nonumber\\
\lineup\ \ \ \ \ \ \ \ =-\frac{1}{2}\left(\sum_{j=1}^m \frac{\eta^{\mu_0\mu_j}}{u_{0j}^2}\right)\Big\langle \d X^{\mu_1}(u_1)\,...\, \widehat{\d X^{\mu_j}(u_j)}\,...\,\d X^{\mu_m}(m_m)\, e^{i k^{m+1}\cdot X(u_{m+1})}\, ... \, e^{i k^{m+n}\cdot X(u_{m+n})}\Big\rangle_\mathbb{C}\nonumber\\
\lineup\ \ \ \ \ \ \ \ \ \ \ -\frac{i}{2}\left(\sum_{j=m+1}^{m+n}\frac{\eta^{\mu_0 \nu}k_\nu^j}{u_{0j}}\right)\Big\langle\d X^{\mu_1}(u_1)\,...\, \d X^{\mu_m}(u_m)\, e^{i k^{m+1}\cdot X(u_{m+1})}\, ... \, e^{i k^{m+n}\cdot X(u_{m+n})}\Big\rangle_\mathbb{C}.\label{eq:dXdelete}
\end{eqnarray}
The hat over $\d X$ in the first sum indicates omission. Note that on the right hand side the operator $\d X^{\mu_0}(u_0)$ has been eliminated.

\section{Witten and lightcone vertices}
\label{app:vertex}

Here we give definitions of the Witten and lightcone vertices in terms of worldsheet correlation functions in the manner of \cite{LeClair}. When dealing with correlation functions it is uncomfortable to use the shifted grading scheme. We therefore explain the signs needed to relate to the vertices defined with the Grassmann grading. Let $b_n:\H^{\otimes n}\to \H$ be an $n$-string product defined with the shifted grading. We define a corresponding {\it unshifted vertex} according to
\begin{eqnarray}
B_{n+1}(A_1,...,A_{n+1}) \lineup = (-1)^{|A_n|+|A_{n-2}|+|A_{n-4}|+...+|A_{3\,\text{or}\,2}|}(-1)^{(|b_n|+n)(|A_1|+1)+1}\nonumber\\
\lineup\ \ \ \ \times\omega(A_1,b_n(A_2,...,A_{n+1})).\label{eq:unshift}
\end{eqnarray}
The first sign appears from relating to the string product as naturally defined in the Grassmann grading (see e.g. appendix A of \cite{MoellerSachs}), and the second sign appears from ordering all Grassmann odd operators inside $b_n$ to the left in the correlation function defining the vertex. It is often convenient to describe the unshifted vertex as an $n+1$-fold bra state
\begin{equation}\langle B_{n+1}|:(s^{-1}\H)^{\otimes n+1}\to\mathbb{C},\end{equation}
so that
\begin{equation}\langle B_{n+1}|A_1\otimes...\otimes A_{n+1} = B_{n+1}(A_1,...,A_{n+1}).\end{equation}
We view the vertex as acting on the ``unshifted" vector space $s^{-1}\H$. What this means operationally is that tensor products of operators on $s^{-1}\H$ commute through tensor products of states in $s^{-1}\H$ with a sign given by Grassmann parity, rather than degree. We are specifically concerned with the 2-string product $\mW_2$ of the Witten theory and the 2- and 3-string products $\mlc_2,\mlc_3$ of the lightcone theory. By the above definitions, they are related to corresponding unshifted vertices by
\begin{eqnarray}
\VW_3(A,B,C)  \lineup = (-1)^{|A|+|B|}\omega(A,\mW_2(B,C)),\\
\Vlc_3(A,B,C)  \lineup = (-1)^{|A|+|B|}\omega(A,\mlc_2(B,C)),\\
\Vlc_4(A,B,C,D)  \lineup = (-1)^{1+|C|}\omega(A,\mlc_3(B,C,D)).
\end{eqnarray}
The unshifted vertex of the Witten theory would be more commonly written as $\langle A,B*C\rangle$. In the following we choose the cyclic ordering of punctures on the open string boundary according to the left handed convention~\cite{simple}.

\subsection{Witten Vertex}
\label{app:Witten_vertex}

\begin{figure}
\begin{center}
\resizebox{3.4in}{2in}{\includegraphics{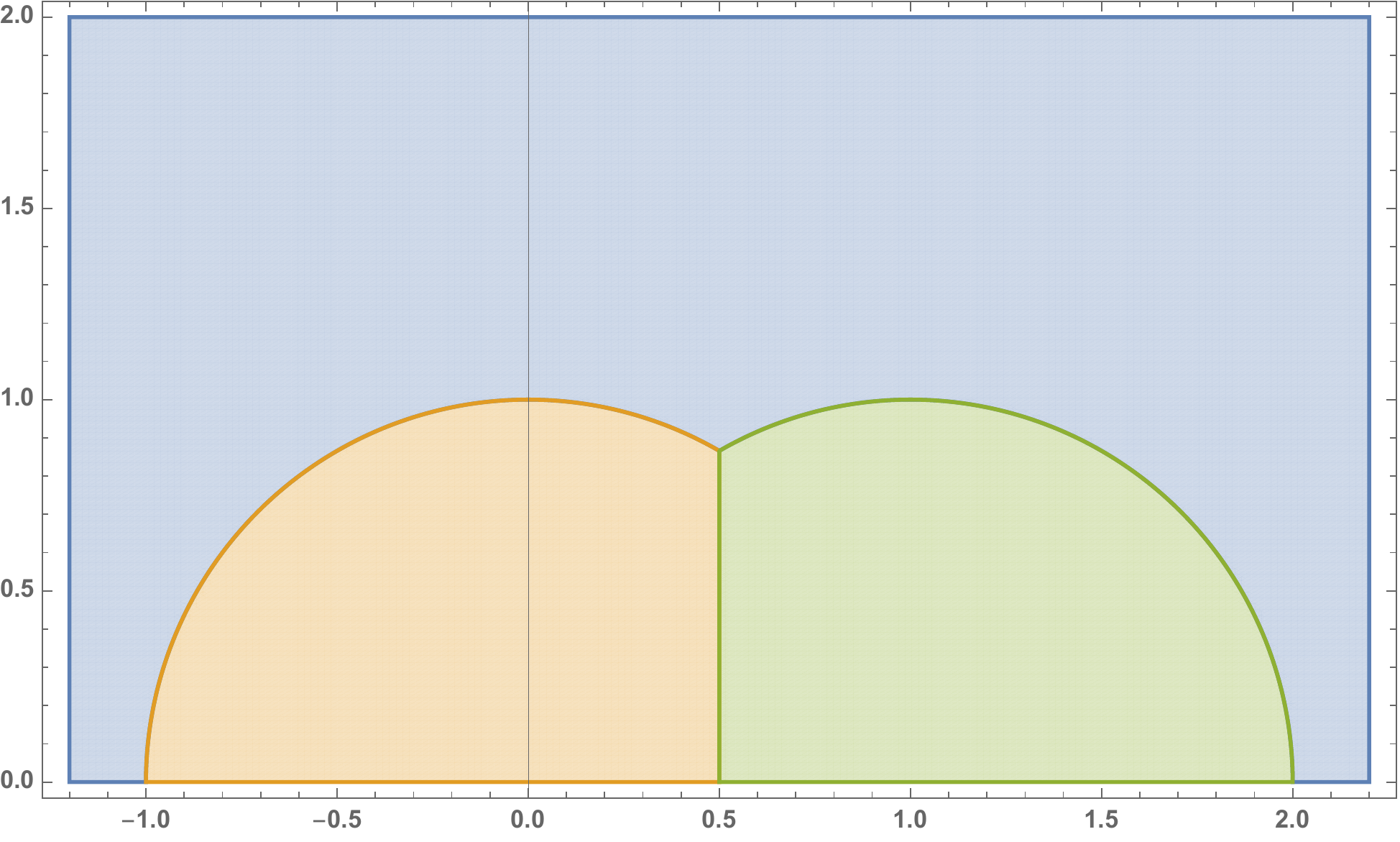}}\ \ \ \ \ 
\resizebox{2in}{2in}{\includegraphics{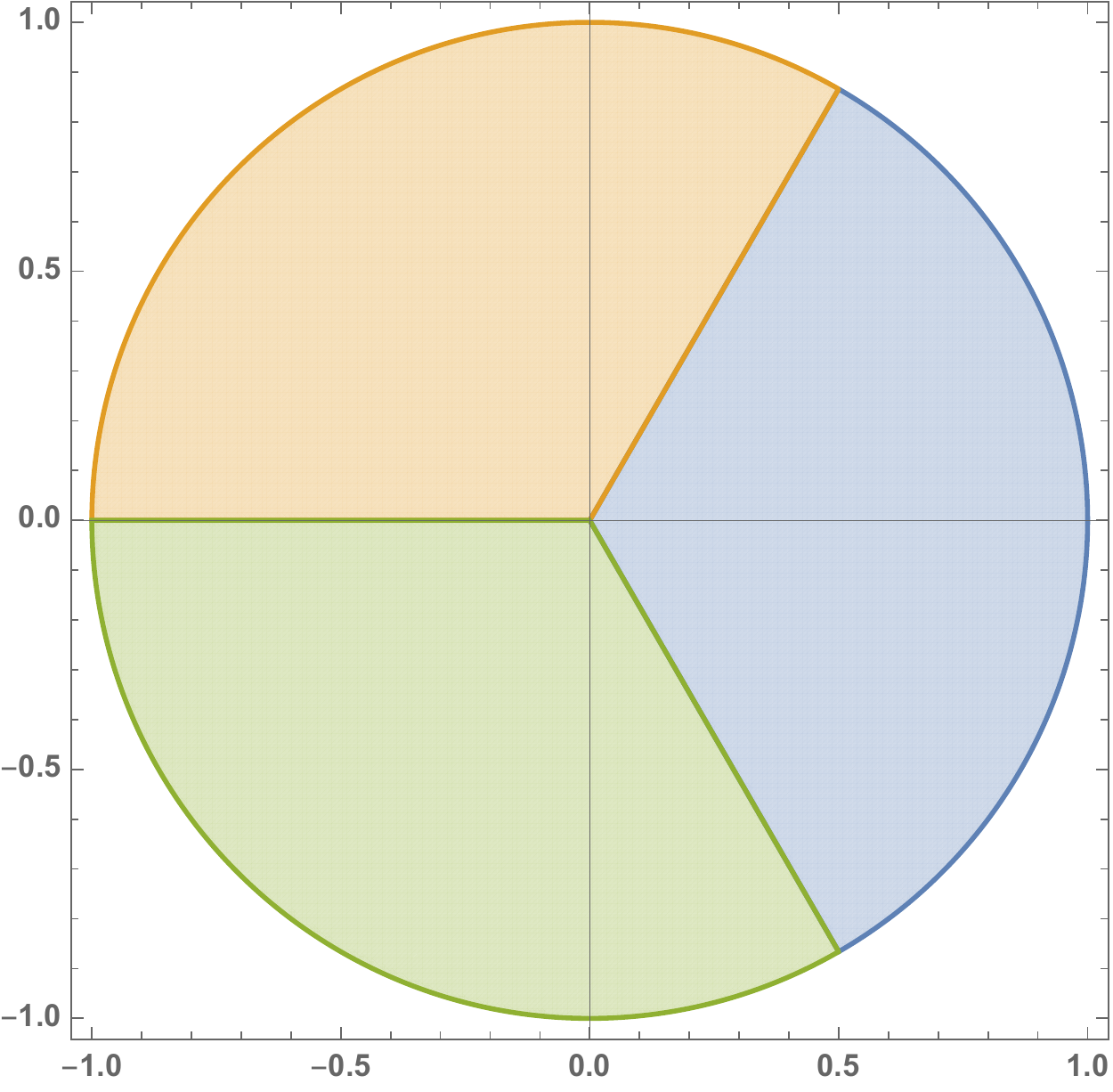}}
\end{center}
\caption{\label{fig:lightcone1} Local coordinate patches of the Witten vertex on the upper half plane and the unit disk. The unit disk in this  figure is related to the upper half plane through $d = \frac{e^{-2\pi i/3}+u}{e^{2\pi i/3}+u}$, which produces the familiar ``pie slice" picture of the Witten vertex.}
\end{figure}

The Witten vertex can be computed as a correlation function of three conformally transformed vertex operators in the upper half plane:
\begin{equation}
\VW_3(A,B,C) = \Big\langle \fW{1}\circ A(0)\fW{2}\circ B(0)\fW{3}\circ C(0)\Big\rangle_\text{UHP}.
\end{equation}
We adopt a notation for the local coordinate maps where the superscript indicates the theory, the first subscript indicates the number of  strings in the vertex, and the second subscript labels the state in the vertex. We fix the punctures of the first, second, and third states to lie at $u=1,0,\infty$. The requisite conformal maps are then
\begin{eqnarray}
\fW{1}(\xi)\lineup =\frac{1-e^{2\pi i/3}\left(\frac{1+i\xi }{1-i\xi}\right)^{2/3}}{\left(\frac{1+i\xi }{1-i\xi}\right)^{2/3}-e^{2\pi i/3}},\\
\fW{2}(\xi)\lineup =\frac{1-\left(\frac{1+i\xi }{1-i\xi}\right)^{2/3}}{e^{-2\pi i/3}\left(\frac{1+i\xi }{1-i\xi}\right)^{2/3}-e^{2\pi i/3}},\\
\fW{3}(\xi)\lineup=\frac{e^{-2\pi i/3}-e^{2\pi i/3}\left(\frac{1+i\xi }{1-i\xi}\right)^{2/3}}{\left(\frac{1+i\xi }{1-i\xi}\right)^{2/3}-1}.
\end{eqnarray}
The local coordinate patches of the Witten vertex are illustrated in figure \ref{fig:lightcone1}. 

\subsection{Cubic Lightcone Vertex}
\label{app:lc3_vertex}

Next we turn to the lightcone vertices. They are unusual from the point of view of covariant string field theory, first of all because they break covariance, but more specifically because the geometry of the vertices depends on the states which are interacting.

It will be helpful to make the following definition:
\begin{definition}
\label{def:standard3}
A vertex $V_n(A_1,A_2,...,A_n)$ is said to be evaluated in the ``standard configuration" if all states have definite minus momenta $k_-^1,k_-^2,...,k_-^n$, respectively, and the last state has minus momentum of largest magnitude.
\end{definition}
\noindent Defining the vertex in the standard configuration is enough to define it completely. We can compute the vertex for arbitrary states by taking superpositions of states with definite momenta, and if the final state in the vertex does not have minus momentum of the largest magnitude, we assume cyclicity and permute the states until the final entry does have minus momentum of largest magnitude.

\begin{figure}
\begin{center}
\resizebox{2.4in}{1.2in}{\includegraphics{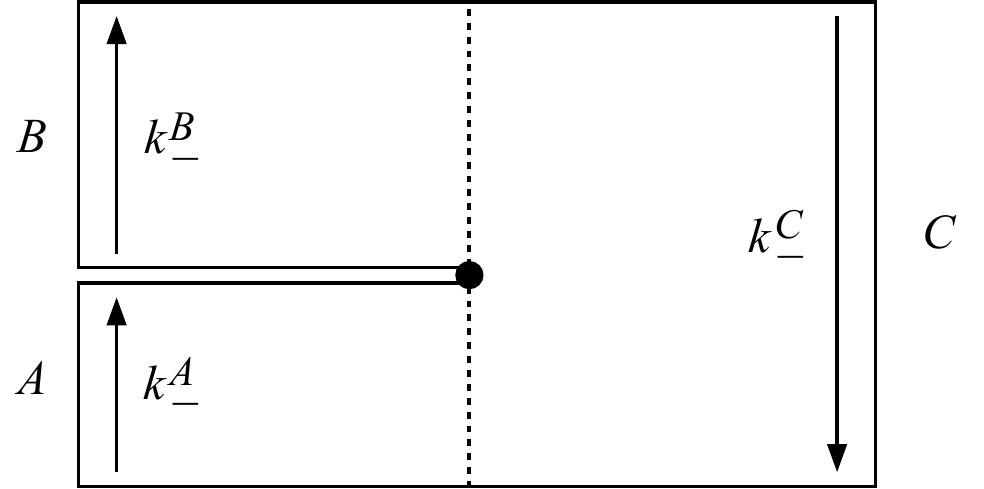}}
\end{center}
\caption{\label{fig:lightcone7} Geometry of the cubic lightcone vertex. This picture assumes $k_-^C<0$.}
\end{figure}

\begin{figure}
\begin{center}
\resizebox{3.4in}{2in}{\includegraphics{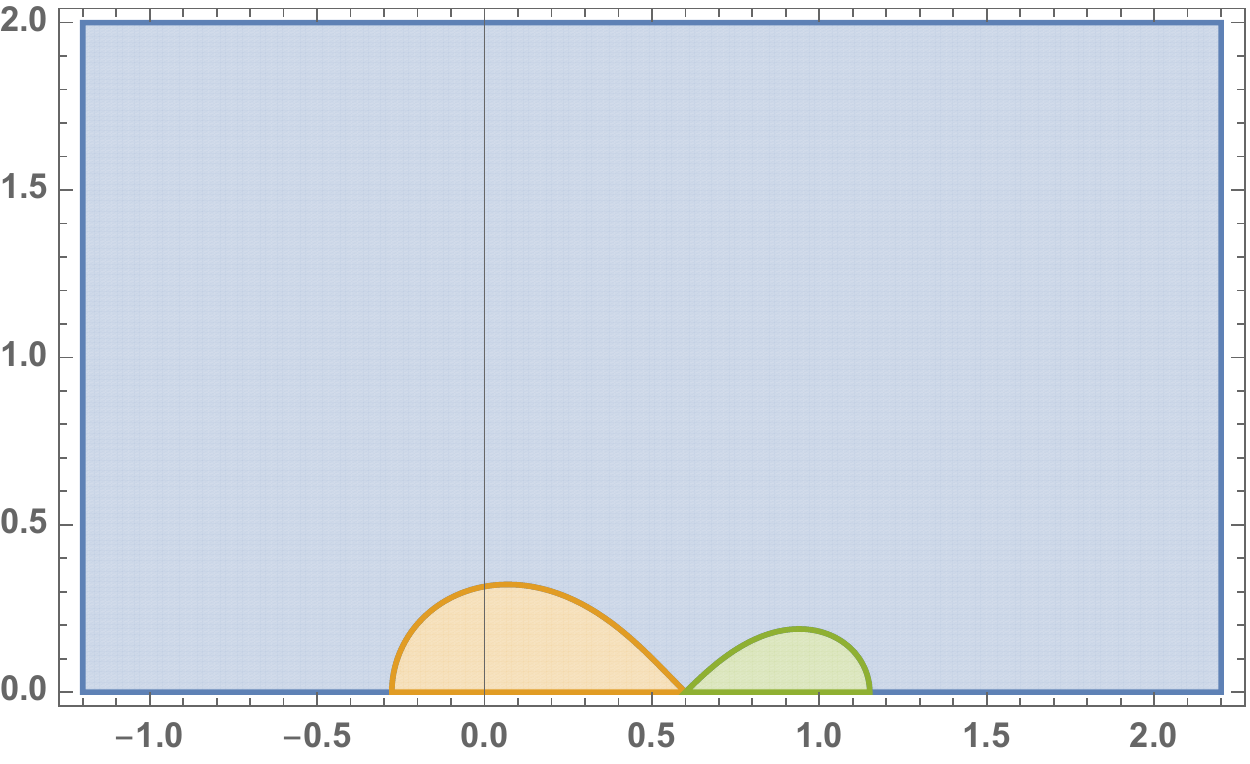}}\ \ \ \ \ 
\resizebox{2in}{2in}{\includegraphics{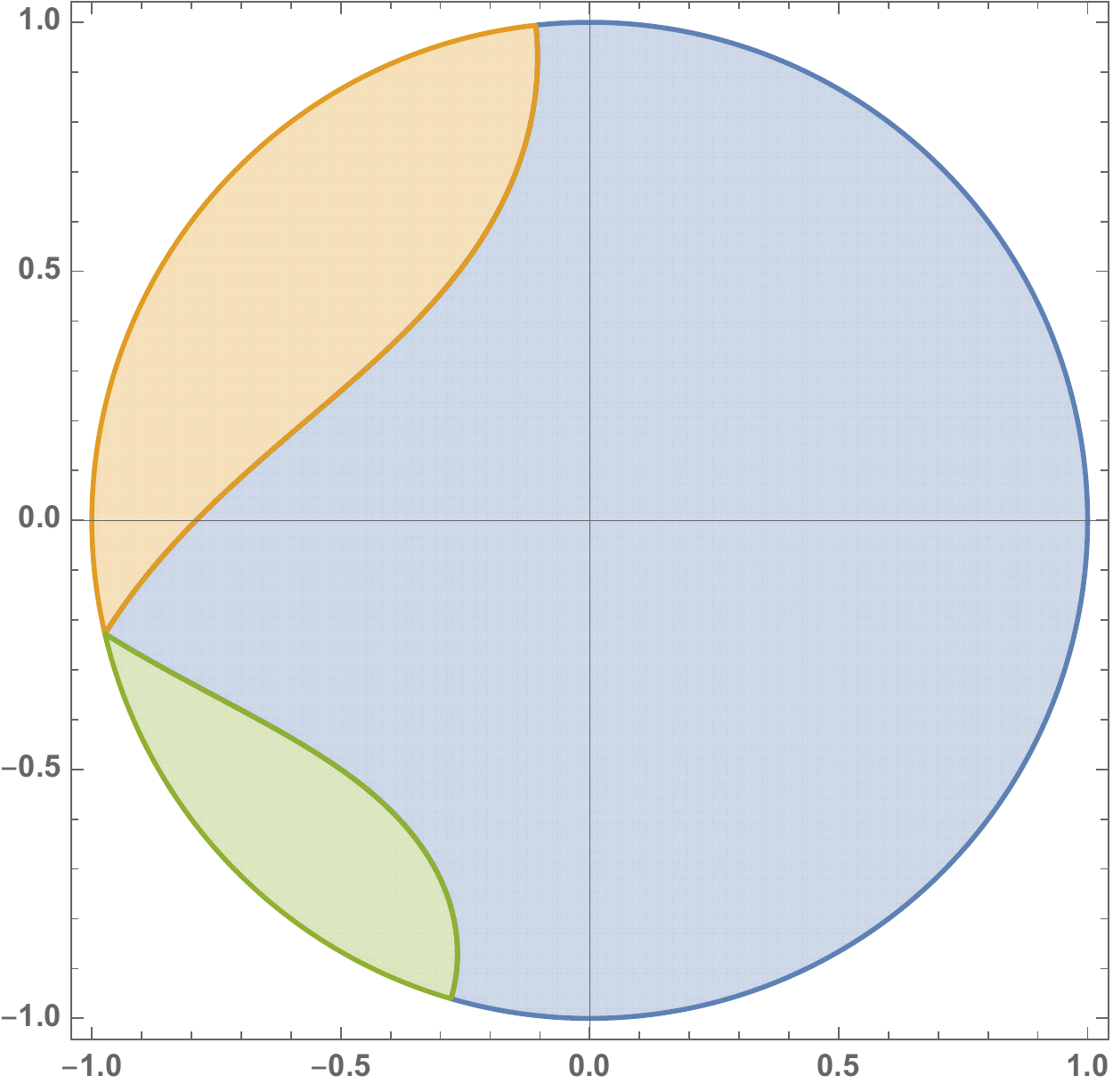}}
\end{center}
\caption{\label{fig:lightcone3} Local coordinate patches of the cubic lightcone vertex in the upper half plane and the unit disk. The unit disk is related to the upper half plane in the same way as in figure \ref{fig:lightcone1}. We have chosen the kinematic moduli $\lambda_1= .4,\lambda_2=.6$, so the minus momentum of the second state has larger magnitude than the first. }
\end{figure}

Below we characterize the cubic lightcone vertex $\Vlc_3(A,B,C)$ in the standard configuration. We fix the punctures of the states $A,B,C$ in the upper half plane to $1,0,$ and $\infty$, respectively. The vertex is most simply understood in the interaction coordinate $\rho$ representing the surface illustrated in figure \ref{fig:lightcone7}. This coordinate is related to the upper half plane through the Mandelstam mapping
\begin{equation}
\rho(u) = \frac{k_-^A}{\pi} \ln(u-1) +\frac{k_-^B}{\pi}\ln u .\label{eq:Mandelstam3}
\end{equation}
We cover each strip on this surface with the local coordinate $\xi$ of the corresponding state, where $\xi$ on the unit half-disk is related to the interaction coordinate by the appropriate scaling and translation of $\ln(\xi)$. Composing with the Mandelstam mapping then allows us to derive the local coordinate maps for the three states on the upper half plane. The local coordinate maps depend on ``kinematic moduli"
\begin{equation}
\lambda_1 = -\frac{k_-^A}{k_-^C} \in [0,1],\ \ \ \ \ \lambda_2  = -\frac{k_-^B}{k_-^C}\in [0,1].\label{eq:l1l2}
\end{equation}
Momentum conservation implies 
\begin{equation}\lambda_1+\lambda_2=1.\end{equation}
We write
\begin{equation}
\Vlc_3(A,B,C) = \big\langle\flc{1}\circ A(0)\flc{2}\circ B(0)\flc{3}\circ C(0)\big\rangle_\text{UHP},
\end{equation}
where the inverse of the local coordinate maps are given by
\begin{eqnarray}
 (\flc{1})^{-1}(u) \lineup = \frac{u-1}{|\lambda_1|}\left(\frac{u}{|\lambda_2|}\right)^{\frac{\lambda_2}{\lambda_1}},\label{eq:flc31}\\
(\flc{2})^{-1}(u) \lineup = \frac{u}{|\lambda_2|}\left(\frac{1-u}{|\lambda_1|}\right)^{\frac{\lambda_1}{\lambda_2}},\\
(\flc{3})^{-1}(u) \lineup = -\left(\frac{|\lambda_1|}{u-1}\right)^{\lambda_1}\left(\frac{|\lambda_2|}{u}\right)^{\lambda_2}.\label{eq:flc33}
\end{eqnarray}
Only the inverse local coordinate maps are expressible in closed form. The absolute values of the kinematic moduli are redundant in the standard configuration. We will explain their presence in a moment. For dealing with the third puncture at infinity it is useful to note
\begin{equation}
(I\circ\flc{3})^{-1}(u) = u|\lambda_2|^{\lambda_2}\left(\frac{|\lambda_1|}{1+u}\right)^{\lambda_1}.
\end{equation}
Also useful are the derivatives of the local coordinate maps at the punctures:
\begin{eqnarray}
\flcp{1}(0) \lineup =|\lambda_1||\lambda_2|^{\lambda_2/\lambda_1}, \label{eq:dflc31}\\
\flcp{2}(0)\lineup = |\lambda_2||\lambda_1|^{\lambda_1/\lambda_2},\\
(I\circ \flc{3})'(0)\lineup =|\lambda_1|^{-\lambda_1}|\lambda_2|^{-\lambda_2}.\label{eq:dflc33}
\end{eqnarray}
The local coordinate patches are illustrated in figure \ref{fig:lightcone3}.

If the momenta are not arranged in the standard configuration, one of the kinematic moduli  $\lambda_1,\lambda_2$ will be greater than one  and the other will be negative. However, the above formulas are still valid in this case if we replace $\lambda_1,\lambda_2$ with the respective absolute values everywhere except in the exponents. This explains the absolute values in \eq{flc31}-\eq{flc33}. To demonstrate this result we cyclically permute the cubic vertex to achieve the standard configuration,  and then perform an $SL(2,\mathbb{R})$ transformation on the upper half plane which permutes the punctures backwards until the vertex operator at infinity corresponds to the state $C$. We did not find a convenient prescription for extending the local coordinate maps of the Kaku vertices in section \ref{sec:Kaku} outside the standard configuration.

\subsection{Quartic Lightcone Vertex}
\label{app:lc4_vertex}

\begin{figure}
\begin{center}
\resizebox{5.5in}{1.5in}{\includegraphics{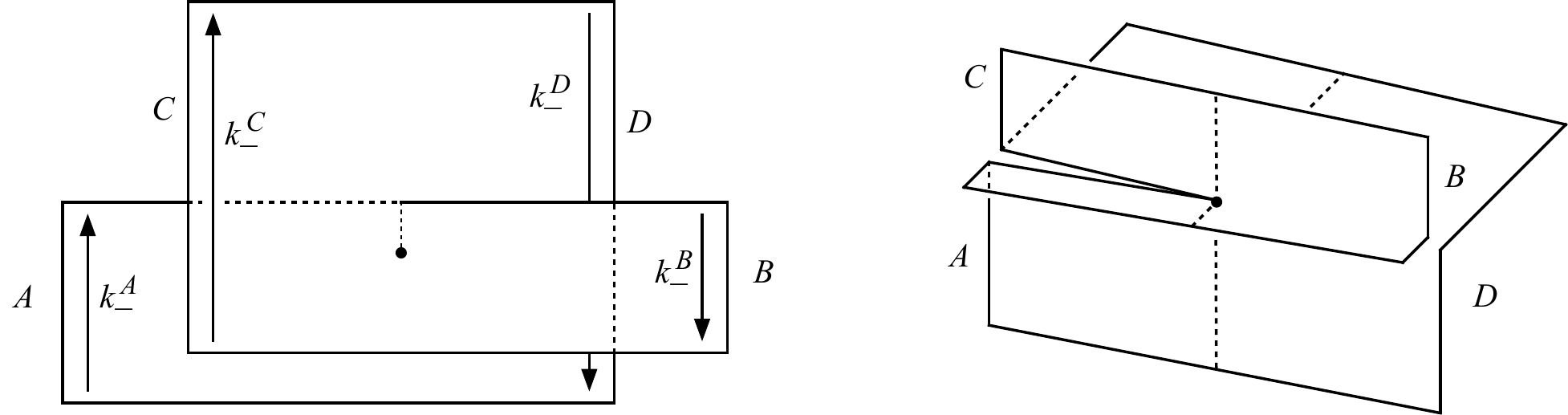}}
\end{center}
\caption{\label{fig:lightcone5} The surface of the four string interaction produced by the Mandelstam map. Represented in the complex plane, the surface has a self-intersection. The self-intersection can be removed by unfolding the surface, which gives a clearer picture of the interaction.}
\end{figure}

The quartic lightcone vertex represents a process where two open strings touch at an interior point and reconnect to a different topology, as shown in figure \ref{fig:lightcone5}. There is a continuum of such processes which differ by the point of reconnection in the interior. To account for all these processes we must integrate over the interior point. This represents integration over a portion of the moduli space of disks with boundary punctures.

Below we characterize the quartic lightcone vertex $\Vlc_4(A,B,C,D)$ in the standard configuration. Additionally, we assume that the minus momenta of the states in the vertex have alternating signs when listed in cyclic order. If this is not the case, the quartic lightcone vertex vanishes identically. Since the fourth state $D$ has minus momentum of largest magnitude, and the minus momenta have alternating signs, we know that the second state $B$ has minus momentum of smallest magnitude. We fix the punctures of the states $A,B,C,D$ in the upper half plane to $1,m\in[0,1],0$ and $\infty$, respectively. The location of the second puncture $m$ is a coordinate on the moduli space of four punctured disks. The vertex is most simply understood in the interaction coordinate $\rho$ representing the surface illustrated in figure \ref{fig:lightcone5}. This coordinate is related to the upper half plane through the  Mandelstam mapping 
\begin{equation}
\rho =\frac{k_-^A}{\pi}\ln(u-1) +\frac{k_-^B}{\pi}\ln(u-m)+\frac{k_-^C}{\pi}\ln(u).\label{eq:MandelstamV4}
\end{equation}
The incoming strings $A$ and $C$ touch and reconnect at the interaction point, which is a conical singularity on the surface with deficit angle $-2\pi$. The position of the interaction point can vary from the bottom to the top edge of the strip representing the state $B$. This assumes that $B$ has minus momentum of the smallest magnitude, since otherwise the interaction point could meet an open string boundary before reaching the top or bottom edges of the strip of $B$. We cover each strip on the surface with the local coordinate $\xi$ of the corresponding state through the appropriate scaling and translation of $\ln(\xi)$. Composing with the Mandelstam mapping then implies the form of the local coordinate maps on the upper half plane. The local coordinate maps depend on kinematic moduli
\begin{equation}\lambda_1 = -\frac{k_-^A}{k_-^D},\ \ \ \lambda_2 = \frac{k_-^B}{k_-^D},\ \ \ \lambda_3 = -\frac{k_-^C}{k_-^D}.\end{equation}

\begin{figure}
\begin{center}
\resizebox{3in}{1.5in}{\includegraphics{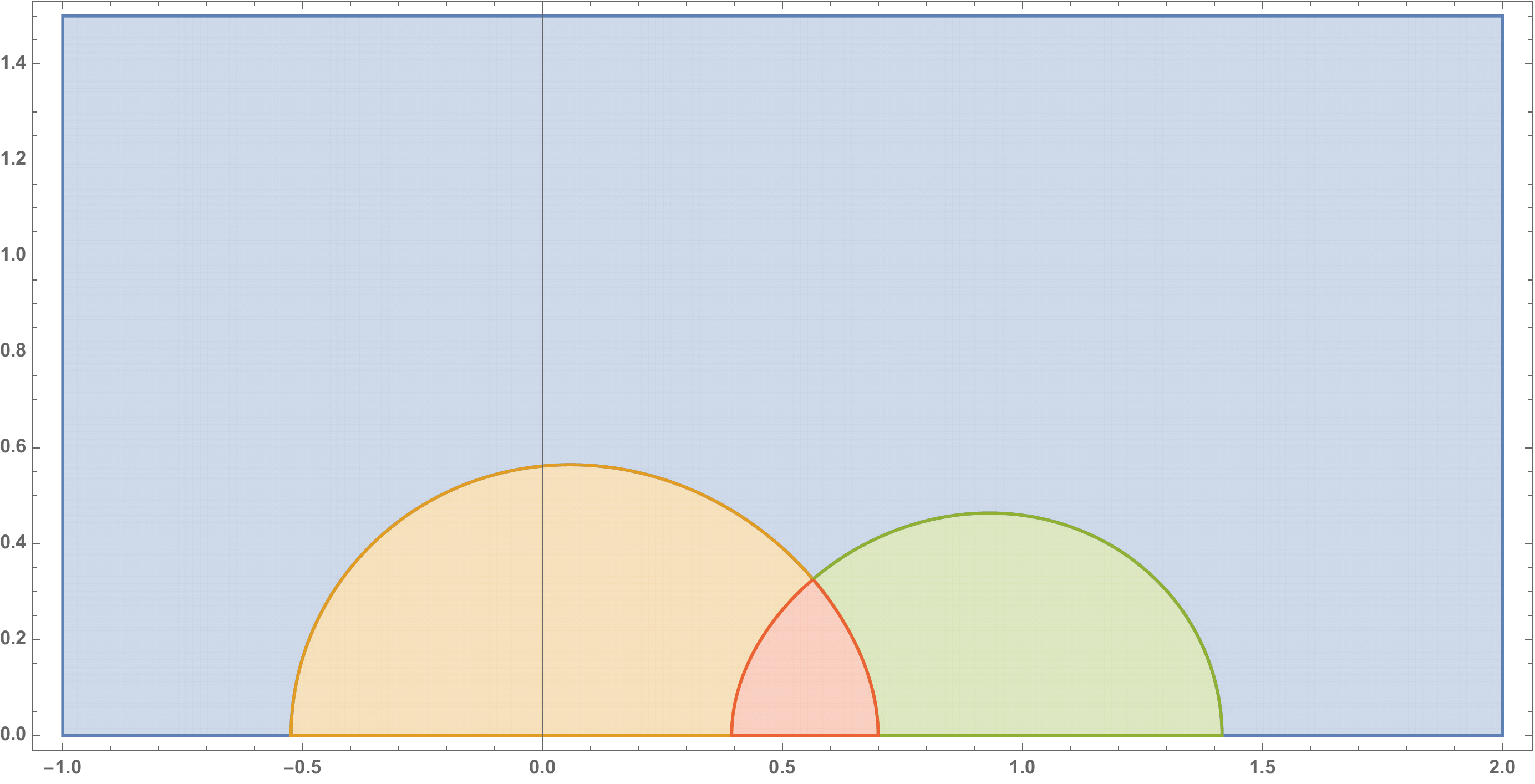}}
\end{center}
\caption{\label{fig:lightcone8} Local coordinate patches for the quartic vertex in the upper half plane. We have chosen $m=.6$ and $\lambda_1=.7,\lambda_2=.5$ and $\lambda_3=.8$.}
\end{figure}

\noindent Momentum conservation implies 
\begin{equation}\lambda_1+\lambda_3 =1+\lambda_2.\label{eq:quartcon}\end{equation} 
In addition we have the inequalities
\begin{equation}0<\lambda_2<\lambda_1<1,\ \ \ \ 0<\lambda_2<\lambda_3<1.\label{eq:quartin}\end{equation}
The inverse of the local coordinate maps are given by 
\begin{eqnarray}
(\flcm{1})^{-1}(u)\lineup = \frac{u-1}{\sqrt{\lambda_1(1-m)}}\left(\frac{\sqrt{\lambda_2m(1-m)}}{u-m}\right)^{\frac{\lambda_2}{\lambda_1}}\left(\frac{u}{\sqrt{\lambda_3 m}}\right)^{\frac{\lambda_3}{\lambda_1}},\label{eq:LCquartic1}\\
(\flcm{2})^{-1}(u)\lineup = \frac{u-m}{\sqrt{\lambda_2 m(1-m)}}\left(\frac{\sqrt{\lambda_3 m}}{u}\right)^{\frac{\lambda_3}{\lambda_2}}\left(\frac{\sqrt{\lambda_1(1-m)}}{1-u}\right)^{\frac{\lambda_1}{\lambda_2}},\\
(\flcm{3})^{-1}(u)\lineup = \frac{u}{\sqrt{\lambda_3 m}}\left(\frac{1-u}{\sqrt{\lambda_1(1-m)}}\right)^{\frac{\lambda_1}{\lambda_3}}\left(\frac{\sqrt{\lambda_2 m(1-m)}}{m-u}\right)^{\frac{\lambda_2}{\lambda_3}},\\
(\flcm{4})^{-1}(u)\lineup = - \left(\frac{\sqrt{\lambda_1(1-m)}}{u-1}\right)^{\lambda_1}\left(\frac{u-m}{\sqrt{\lambda_2 m(1-m)}}\right)^{\lambda_2}\left(\frac{\sqrt{\lambda_3 m}}{u}\right)^{\lambda_3}.\label{eq:LCquartic4}
\end{eqnarray}
For dealing with the fourth puncture at infinity it is useful to note
\begin{equation}
(I\circ\flcm{4})^{-1}(u) =  u(\sqrt{\lambda_3 m})^{\lambda_3}\left(\frac{\sqrt{\lambda_1(1-m)}}{1+u}\right)^{\lambda_1}\left(\frac{1+mu}{\sqrt{\lambda_2 m(1-m)}}\right)^{\lambda_2}.
\end{equation}
The derivatives of the local coordinate maps at the punctures are
\begin{eqnarray}
\flcmp{1}(0) \lineup =\sqrt{\lambda_1(1-m)}\Big(\!\sqrt{\lambda_3 m}\,\Big)^{\lambda_3/\lambda_1}\Bigg(\!\sqrt{\frac{1-m}{\lambda_2 m}}\,\Bigg)^{\lambda_2/\lambda_1},\\
\flcmp{2}(0)\lineup =\sqrt{\lambda_2  m(1-m)}\Bigg(\!\sqrt{\frac{1-m}{\lambda_1}}\,\Bigg)^{\lambda_1/\lambda_2}\Bigg(\!\sqrt{\frac{m}{\lambda_3}}\,\Bigg)^{\lambda_3/\lambda_2},\\
\flcmp{3}(0)\lineup =\sqrt{\lambda_3 m}\Big(\!\sqrt{\lambda_1(1-m)}\,\Big)^{\lambda_1/\lambda_3}\Bigg(\!\sqrt{\frac{m}{\lambda_2 (1-m)}}\,\Bigg)^{\lambda_2/\lambda_3},\\
(I\circ \flcm{4})'(0)\lineup =\Big(\!\sqrt{\lambda_1(1- m)}\,\Big)^{-\lambda_1}\Big(\!\sqrt{\lambda_3 m}\,\Big)^{-\lambda_3}\Big(\!\sqrt{\lambda_2 m(1-m)}\,\Big)^{\lambda_2}.
\end{eqnarray}
A picture of the local coordinate patches in the upper half plane is given in figure \ref{fig:lightcone8}. In the upper half plane coordinate, the interaction point is located where all four local coordinate patches touch, and is given by 
\begin{equation}
u_* = \frac{1}{2}\left(1+m\lambda_3-(1-m)\lambda_1+i\sqrt{\frac{4\lambda_1\lambda_2\lambda_3}{(1+\lambda_2)^2}-\left(\frac{\lambda_1-\lambda_3}{1+\lambda_2}+m\lambda_3-(1-m)\lambda_1\right)^2}\right).\label{eq:ustar}
\end{equation}
This must lie in the interior of the upper half plane, which in the coordinate $\rho$ means that the interaction point lies between the top and bottom edges of the strip of $B$. This implies that the object appearing under the square root in $u_*$ must be positive, which places a constraint on the moduli which can appear in the quartic vertex:
\begin{equation}
m_-<m<m_+,
\end{equation}
where 
\begin{equation}m_\pm = \frac{1}{(1+\lambda_2)^2}\Big(\lambda_3+\lambda_1\lambda_2\pm 2\sqrt{\lambda_1\lambda_2\lambda_3}\Big). \label{eq:mpm}\end{equation}
The quartic vertex region of the moduli space depends on the states which are interacting, as can be seen through the dependence on kinematic moduli.

The data so far define a surface state 
\begin{equation}\Sigmalcm_4(A,B,C,D) = \Big\langle\flcm{1}\circ A(0)\flcm{2}\circ B(0)\flcm{3}\circ C(0)\flcm{4}\circ D(0)\Big\rangle_\text{UHP},\label{eq:Sigmalc4}\end{equation}
which depends on a modulus $m\in[m_-,m_+]$. The quartic vertex is given by integrating from $m_+$ to $m_-$ with the appropriate $b$-ghost insertion for the measure. This is explained in subsection \ref{subsec:infinitesimal}. The result is 
\begin{eqnarray}
\Vlc_4(A,B,C,D)\lineup = \int_{m_-}^{m_+} dm\left\langle\!\left(\!\frac{u_*(u_*-m)(1-u_*)}{2m(1-m)}b(u_*)\! +\! \frac{\overline{u}_*(\overline{u}_*-m)(1-\overline{u}_*)}{2m(1-m)}b(\overline{u}_*)\!\right)\!\right.\nonumber\\
\lineup \ \ \ \ \ \ \ \ \ \ \ \ \ \ \ \ \ \ \ \left.\phantom{\bigg)}\times\flcm{1}\circ A(0)\flcm{2}\circ B(0)\flcm{3}\circ C(0)\flcm{4}\circ D(0)\right\rangle_\text{UHP}.\ \ \ \ \ \ \ \ \ \ \ \ \label{eq:V4complete}
\end{eqnarray}
A single $b$-ghost operator is inserted precisely at the interaction point, and, using the doubling trick, at its conjugate position in the lower half plane

If the  momenta are not arranged in the  standard configuration, all of the kinematic moduli $\lambda_1,\lambda_2,\lambda_3$ will be positive but at least one of them will be greater than 1, and $\lambda_2$ will not be the smallest. However, it turns out that all of the above formulas are still valid in this case.  This can be shown by cyclically permuting the quartic vertex to achieve the standard configuration,  and then performing an $SL(2,\mathbb{R})$ transformation on the upper half plane which permutes the punctures backwards until the vertex operator at infinity corresponds to the state $D$. 

\section{Some details of the lightcone SFT action}
\label{app:long}

In this paper we express the lightcone SFT action in the form
\begin{equation}
\Slc = -\frac{1}{2}\omega(\Psilc,c_0L_0\Psilc)-\frac{1}{3}\omega(\Psilc,\mlc_2(\Psilc,\Psilc))-\frac{1}{4}\omega(\Psilc,\mlc_3(\Psilc,\Psilc,\Psilc)),\label{eq:Slccov}
\end{equation}
where $\Psilc$ lives in the full matter/ghost BCFT vector space but is subject to the  constraint
\begin{equation} L_0^\parallel \Psilc = 0,\end{equation}
as described in section \ref{sec:setup}. In this appendix we translate this into an expression for the action which connects more directly to other discussions of lightcone SFT. 

The first issue concerns the description of the lightcone string field. Traditionally the lightcone string field is described as an element of the  vector space $\H_{X^i}$ of transverse free bosons which at the same time is a function of the longitudinal coordinates $x^+,x^-$:
\begin{equation}\Psi_\alpha(x^+)\in C^\infty(\mathbb{R}^{1,1})\otimes \H_{X^i}.\end{equation}
Typically the  dependence on $x^-$ is described in momentum space through a string length parameter~$\alpha$, related to the momentum $k_-$ through a conventional normalization
\begin{equation}\alpha  =  2k_-.\label{eq:alpha}\end{equation}
We write $\Psi_\alpha(k_+)$ when the dependence on $x^+$ is described in momentum space. The lightcone string field $\Psi_\alpha(x^+)$ is related to $\Psilc$ as follows. We expand $\Psilc$ in the oscillator basis as
\begin{equation}
\Psilc = \int\frac{d^{26}k}{(2\pi)^{26}}\sum_{N=0}^\infty \sum_{0\leq n_1\leq ...\leq n_N}\psi_{i_1...i_N}^{n_1,...n_N}(k)\alpha_{-n_1}^{i_1}...\alpha_{-n_N}^{i_N}c e^{ik\cdot X(0,0)}|0\rangle,\label{eq:Psilccov}
\end{equation}
where the $\psi$s are component fields which  depend on all 26 momenta.  We factorize the $SL(2,\mathbb{R})$ vacuum of the total BCFT into a tensor product of $SL(2,\mathbb{R})$ vacua of the transverse and longitudinal BCFTs,
\begin{equation}|0\rangle = |0\rangle_{X^i}\otimes |0\rangle_{X^\pm,b,c},\end{equation}
and further decompose the plane wave vertex operator into factors containing purely transverse and longitudinal momenta: 
\begin{equation}e^{ik\cdot X} = e^{ik_\perp\cdot X(0,0)}e^{ik_\parallel\cdot X(0,0)}.\end{equation}
Then we write 
\begin{eqnarray}
\Psilc = \int\frac{d^2k_\parallel}{(2\pi)^2}\left(\int\frac{d^{24}k_\perp}{(2\pi)^{24}}\sum_{N=0}^\infty \sum_{0\leq n_1\leq ...\leq n_N}\psi_{i_1...i_N}^{n_1,...n_N}(k_\parallel,k_\perp)\alpha_{-n_1}^{i_1}...\alpha_{-n_N}^{i_N}e^{ik_\perp\cdot X(0,0)}|0\rangle_{X^i}\right)\lineup\nonumber\\
\lineup \!\!\!\!\!\!\!\!\!\!\!\!\!\!\!\!\!\!\!\! \!\!\!\!\!\!\!\!\!\!\!\!\!\!\!\!\!\!\!\!\otimes\, c e^{ik_\parallel \cdot X(0,0)}|0\rangle_{X^\pm,b,c}.\  \ \ \ \ \ \ \  \ \  
\end{eqnarray}
The factor in parentheses is the lightcone string field $\Psi_\alpha(k^+)$. Therefore $\Psi_\alpha(x^+)$ and $\Psilc$ are related as
\begin{eqnarray}
\Psilc \lineup = \int\frac{d^2k_\parallel}{(2\pi)^2}\Psi_\alpha(k_+)\otimes  c e^{ik_\parallel \cdot X(0,0)}|0\rangle_{X^\pm,b,c}\phantom{\Bigg)}\label{eq:Psilcredmom}\\
\lineup = \int dx^+\frac{d\alpha}{4\pi}\Psi_\alpha(x^+)\otimes \int \frac{dk_+}{2\pi}e^{-ix^+k_+} c e^{ik_\parallel \cdot X(0,0)}|0\rangle_{X^\pm,b,c}\phantom{\Bigg)}.\label{eq:Psilcred}
\end{eqnarray}
The lightcone string field satisfies the reality condition
\begin{equation}\Psilc^\ddag =  \Psilc,\end{equation}
where $\ddag$ denotes the composition of Hermitian and BPZ conjugation. This implies 
\begin{equation}\Psi_\alpha(x^+)^\ddag = \Psi_{-\alpha}(x^+).\end{equation}
The sign of $\alpha$ changes because the momentum operator is BPZ odd.

Our  task is to rewrite \eq{Slccov} as an action for the  lightcone string field $\Psi_\alpha(x^+)$. The kinetic term is easily found to be 
\begin{equation}
-\frac{1}{2}\omega(\Psilc,c_0L_0\Psilc)\, =\, -\frac{1}{2}\int dx^+\frac{d\alpha}{4\pi}\Big\langle\Psi_{-\alpha}(x^+),\Big(-i\alpha \d_++L_0^\perp-1\Big)\Psi_\alpha(x^+)\Big\rangle^{X^i},\phantom{\Bigg{)}}
\end{equation}
where $\langle\cdot,\cdot\rangle^{X^i}$ is the  BPZ inner product of the transverse factor of the BCFT. We describe the evaluation of the cubic vertex in somewhat more detail.  Substituting \eq{Psilcredmom} into the cubic lightcone vertex as described in appendix \ref{app:lc3_vertex} gives
\begin{eqnarray}
\lineup -\frac{1}{3}\omega(\Psilc,\mlc_2(\Psilc,\Psilc)) = -\frac{1}{3}\Vlc_3(\Psilc,\Psilc,\Psilc)\phantom{\Bigg)}\nonumber\\
\lineup\ \ \ \ \ \  = -\frac{1}{3}\int \frac{d^2k_\parallel^1 d^2k_\parallel^2 d^2k_\parallel^3 }{(2\pi)^3}\Vlc_3\Big(\Psi_{\alpha_1}(k_+^1)\otimes c e^{ik_\parallel^1\cdot X},\,\Psi_{\alpha_2}(k_+^2)\otimes c e^{ik_\parallel^2\cdot X},\,\Psi_{\alpha_3}(k_+^3)\otimes c e^{ik_\parallel^3\cdot X}\Big).\ \ \ \ \ \ \ \ \ \ \ \ \label{eq:V3lctrans0}
\end{eqnarray}
Focus on the integrand:
\begin{eqnarray}
\lineup\!\!\!\!\!\!\!\!\!\Vlc_3\Big(\Psi_{\alpha_1}(k_+^1)\otimes c e^{ik_\parallel^1\cdot X},\,\Psi_{\alpha_2}(k_+^2)\otimes c e^{ik_\parallel^2\cdot X},\,\Psi_{\alpha_3}(k_+^3)\otimes c e^{ik_\parallel^3\cdot X}\Big)\nonumber\\
\lineup\!\!\!\! = \Big\langle \flc{1}\circ\Big(\Psi_{\alpha_1}(k_+^1)c e^{i k_\parallel^1\cdot X(0.0)}\Big)\flc{2}\circ\Big(\Psi_{\alpha_2}(k_+^2)c e^{i k_\parallel^2\cdot X(0,0)}\Big)\flc{3}\circ\Big(\Psi_{\alpha_3}(k_+^3)c e^{i k_\parallel^3\cdot X(0,0)}\Big)\Big\rangle_{\mathrm{UHP}}
\nonumber\\
\lineup\!\!\!\! = \big(\flcp{1}(0)\big)^{(k_\parallel^1)^2-1}\big(\flcp{2}(0)\big)^{(k_\parallel^2)^2-1}\big((I\circ\flc{3})'(0)\big)^{(k_\parallel^3)^2-1}\nonumber\\
\lineup\ \times
\Big\langle \!\Big(c e^{i k_\parallel^1\cdot X(1,1)} c e^{i k_\parallel^2\cdot X(0,0)} I\circ\big(c e^{i k_\parallel^3\cdot X(0,0)}\big)\Big)\Big(\flc{1}\circ\Psi_{\alpha_1}(k_+^1)\flc{2}\circ\Psi_{\alpha_2}(k_+^2) \flc{3}\circ\Psi_{\alpha_3}(k_+^3)\Big)\!\Big\rangle_{\mathrm{UHP}}.
\nonumber\\
\end{eqnarray}
In the last step we evaluated the conformal transformations of the longitudinal part of the vertex operators. Further evaluating the correlator in the longitudinal part of the BCFT produces a momentum conserving delta function. The correlator in the  transverse factor of the BCFT will be written as a ``transverse" lightcone cubic vertex
\begin{equation}
V_3^{\text{lc},X^i}(A,B,C) = \Big\langle \flc{1}\circ A(0)\flc{2}\circ B(0) \flc{3}\circ C(0)\Big\rangle_{\mathrm{UHP}}^{X^i},
\label{eq:V3lctrans}\end{equation}
for states $A,B,C\in\H_{X^i}$. The transverse lightcone cubic vertex depends on the string length parameters 
$\alpha_1,\alpha_2,\alpha_3$ through the implicit dependence on these parameters in the local coordinate maps. Note that we {\it define} the transverse lightcone vertex as a  correlation function in the upper half plane, and not, for example, as a correlation function in the interaction  coordinate of the Mandelstam diagram representing the cubic vertex. The distinction is important since the transverse free boson CFT has nonvanishing central charge. With this the integrand of \eq{V3lctrans} is expressed
\begin{eqnarray}
\lineup\!\!\!\!\!\! \Vlc_3\Big(\Psi_{\alpha_1}(k_+^1)\otimes c e^{ik_\parallel^1\cdot X},\,\Psi_{\alpha_2}(k_+^2)\otimes c e^{ik_\parallel^2\cdot X},\,\Psi_{\alpha_3}(k_+^3)\otimes c e^{ik_\parallel^3\cdot X}\Big)\nonumber\\
\lineup = \big(\flcp{1}(0)\big)^{(k_\parallel^1)^2-1}\big(\flcp{2}(0)\big)^{(k_\parallel^2)^2-1}\big((I\circ\flc{3})'(0)\big)^{(k_\parallel^3)^2-1}\nonumber\\
\lineup\ \ \ \ \ \ \times 2\pi\delta(k_+^1+k_+^2+k_+^3)4\pi\delta(\alpha_1+\alpha_2+\alpha_3)V_3^{\text{lc},X^i}\Big(\Psi_{\alpha_1}(k_+^1),\Psi_{\alpha_2}(k_+^2),\Psi_{\alpha_3}(k_+^3)\Big).\ \ \ \ \ \ \label{eq:V3lctransint}
\end{eqnarray}
Now we evaluate the conformal factors. Using \eq{dflc31}-\eq{dflc33} the conformal factors are
\begin{eqnarray}
\lineup\!\!\!\!\!\!  \big(\flcp{1}(0)\big)^{(k_\parallel^1)^2-1}\big(\flcp{2}(0)\big)^{(k_\parallel^2)^2-1}\big((I\circ\flc{3})'(0)\big)^{(k_\parallel^3)^2-1}\nonumber\\
\lineup \ \ \ \ \ \ \ \ \ \ \ \ \ \ \ \ \ \ \ \  = \big(|\lambda_1||\lambda_2|^{\lambda_2/\lambda_1}\big)^{(k_\parallel^1)^2-1}\big(|\lambda_2||\lambda_1|^{\lambda_1/\lambda_2}\big)^{(k_\parallel^2)^2-1}\big(|\lambda_1|^{-\lambda_1}|\lambda_2|^{-\lambda_2}\big)^{(k_\parallel^3)^2-1}.
\end{eqnarray}
We further express this in terms of the length parameters $\alpha_1,\alpha_2,\alpha_3$ and the plus momenta $k_+^1,k_+^2,k_+^3$. Noting $(k_\parallel)^2=\alpha k_+$ this gives
\begin{eqnarray}
\lineup\!\!\!\!\!\! \big(\flcp{1}(0)\big)^{(k_\parallel^1)^2-1}\big(\flcp{2}(0)\big)^{(k_\parallel^2)^2-1}\big((I\circ\flc{3})'(0)\big)^{(k_\parallel^3)^2-1}\nonumber\\
\lineup= \left(\left|\frac{\alpha_1}{\alpha_3}\right|\left|\frac{\alpha_2}{\alpha_3}\right|^{\alpha_2/\alpha_1}\right)^{\alpha_1k_+^1-1}
\left(\left|\frac{\alpha_2}{\alpha_3}\right|\left|\frac{\alpha_1}{\alpha_3}\right|^{\alpha_1/\alpha_2}\right)^{\alpha_2k_+^2-1}
\left(\left|\frac{\alpha_1}{\alpha_3}\right|^{\alpha_1/\alpha_3}\left|\frac{\alpha_2}{\alpha_3}\right|^{\alpha_2/\alpha_3}\right)^{\alpha_3k_+^3-1}\nonumber\\
\lineup = \left(\left|\frac{\alpha_1}{\alpha_3}\right|^{-\alpha_1\left(\frac{1}{\alpha_1}+\frac{1}{\alpha_2}+\frac{1}{\alpha_3}\right)}\left|\frac{\alpha_2}{\alpha_3}\right|^{-\alpha_2\left(\frac{1}{\alpha_1}+\frac{1}{\alpha_2}+\frac{1}{\alpha_3}\right)}\right)\left(  \left|\frac{\alpha_1}{\alpha_3}\right|^{\alpha_1(k_+^1+k_+^2+k_+^3)}\left|\frac{\alpha_2}{\alpha_3}\right|^{\alpha_2(k_+^1+k_+^2+k_+^3)}\right).\label{eq:lstfct}
\end{eqnarray}
Note that the last factor evaluates to unity due to momentum conservation. This means that the cubic vertex contains no explicit dependence on $k_+$, and therefore no lightcone time derivatives. What remains can be  simplified using momentum conservation to give 
\begin{equation}
\big(\flcp{1}(0)\big)^{(k_\parallel^1)^2-1}\big(\flcp{2}(0)\big)^{(k_\parallel^2)^2-1}\big((I\circ\flc{3})'(0)\big)^{(k_\parallel^3)^2-1}= Z_3(\alpha_1,\alpha_2,\alpha_3),
\end{equation}
where
\begin{equation}
 Z_3(\alpha_1,\alpha_2,\alpha_3) =\exp\left[-\left(\frac{1}{\alpha_1}+\frac{1}{\alpha_2}+\frac{1}{\alpha_3}\right)\Big(\alpha_1\ln|\alpha_1|+\alpha_2\ln|\alpha_2|+\alpha_3\ln|\alpha_3|\Big)\right].
\end{equation}
This is the well-known momentum-dependent ``form factor" of the cubic lightcone vertex. From our point of view the factor originates from evaluating the longitudinal BCFT correlator, but it can also be understood as the partition function of transverse free bosons on  the Mandelstam diagram representing the cubic vertex.  Equivalently, it can be understood to arise from the conformal anomaly in the process of transforming the transverse free boson correlator from the interaction coordinate on the  Mandelstam diagram into the upper half plane. Plugging this result into \eq{V3lctransint}, further into \eq{V3lctrans0}, and taking the Fourier transform of $k_+$ we obtain
\begin{eqnarray}
\lineup\!\!\!\!\!\!\!\!\!\!\!\!\!\!\!\! -\frac{1}{3}\omega(\Psilc,\mlc_2(\Psilc,\Psilc)) \nonumber\\
\lineup \!\!\!\!\!\!\!\!= -\frac{1}{3}\int dx^+ \frac{d\alpha_1d\alpha_2 d\alpha_3}{(4\pi)^3}4\pi\delta(\alpha_1+\alpha_2+\alpha_3)Z_3(\alpha_1,\alpha_2,\alpha_3)V_3^{\text{lc},X^i}\Big(\Psi_{\alpha_1}(x_+),\Psi_{\alpha_2}(x_+),\Psi_{\alpha_3}(x_+)\Big).\nonumber\\
\end{eqnarray}
One thing this expression makes manifest is that the cubic vertex contains no lightcone time derivatives. This is also true for the quartic vertex. Therefore the theory can be given a Hamiltonian description in lightcone time. The lightcone SFT action is in fact precisely the first order action derived from the lightcone SFT Hamiltonian. The existence of a Hamiltonian formulation of lightcone SFT has been a traditional route to argue for the unitarity of string perturbation theory. Now such arguments have also been given on the basis of covariant string field theories \cite{SenUnitary}.

It is worth mentioning that the absence of time derivatives is not a property shared by the lightcone vertices in the Kugo-Zwiebach theory. The longitudinal momenta explicitly appear in the squeezed state oscillator vertex in the combinations 
\begin{equation}
k_+^r k_-^s \bar{N}^{rs}_{00},\ \ \ \ \ \ \ \  k_-^r\alpha^{-,s}_n\bar{N}_{0n}^{rs},\ \ \ \ \ \ \ \ k_+^r\alpha^{+,s}_n\bar{N}_{0n}^{rs}.
\end{equation}
In a generic covariant SFT, the analogue of the first combination is the main origin of nonlocality in the interaction. In the lightcone vertex, this combination vanishes by the argument described below \eq{lstfct}. The second combination also vanishes as explained in subsection \ref{subsec:transfer}. The  third combination, however, does not vanish, and therefore the vertex contains lightcone time derivatives of arbitrarily high order, though only finite order time derivatives appear if the theory is truncated to a finite mass level. The time derivatives do not appear in lightcone SFT since the string field does not contain minus oscillator excitations. The only known covariant SFT which appears to allow a conventional Hamiltonian formulation is the Witten theory with time identified with the lightcone coordinate of the open string midpoint \cite{Maeno,GrossErler}. This formulation however is formal due to subtleties with the midpoint, and its significance is not fully clear. 

We can proceed in a similar way to evaluate the quartic vertex. Substituting \eq{Psilcred} into the quartic lightcone vertex as described in appendix  \ref{app:lc4_vertex} and evaluating the correlators in the  longitudinal BCFT gives 
\begin{eqnarray}
\lineup \!\! -\frac{1}{4}\omega(\Psilc,\mlc_4(\Psilc,\Psilc,\Psilc))\nonumber\\
\lineup =-\frac{1}{4}\int \! dx^+\frac{d\alpha_1d\alpha_2d\alpha_3 d\alpha_4}{(4\pi)^4} 4\pi\delta(\alpha_1+\alpha_2+\alpha_3+\alpha_4)\!\int_{m_-}^{m_+} \left(dm\frac{-2\alpha_4 \mathrm{Im}(u_*)}{m(1-m)}\right) \! Z_4(\alpha_1,\alpha_2,\alpha_3,\alpha_4,m)\nonumber\\
\lineup\ \ \ \ \ \ \times \Sigma_4^{\text{lc},m,X^i}\Big(\Psi_{\alpha_1}(x^+),\Psi_{\alpha_2}(x^+),\Psi_{\alpha_3}(x^+),\Psi_{\alpha_4}(x^+)\Big).
\end{eqnarray}
Here we introduce a ``transverse" surface state 
\begin{equation}
\Sigma_4^{\text{lc},m,X^i}(A,B,C,D) = \Big\langle \flcm{1}\circ A(0)\flcm{2}\circ B(0) \flcm{3}\circ C(0)\flcm{4}\circ D(0)\Big\rangle_{\mathrm{UHP}}^{X^i}.
\label{eq:V4lctrans}\end{equation}
The surface state is defined as a  transverse  free boson correlator in the upper half plane, and depends on the modulus $m$ and the  string length  parameters $\alpha_1,\alpha_2,\alpha_3,\alpha_3$ through the dependence on these variables in the conformal maps. We also assume that the surface state vanishes unless the  signs of $\alpha_1,\alpha_2,\alpha_3,\alpha_4$ alternate when listed in this order. The combination
\begin{equation}dm\frac{-2\alpha_4 \mathrm{Im}(u_*)}{m(1-m)} = d\theta\end{equation}
is the measure for integration over the modulus
\begin{equation}\theta = \frac{1}{i}\Big(\rho(u_*)-\rho(\overline{u}_*)\Big),\end{equation}
where $\rho(u)$ is the  Mandelstam mapping \eq{MandelstamV4} and $u_*$ is the interaction point \eq{ustar} in the upper half plane. The modulus $\theta$ is proportional to the distance between the interaction point on the Mandelstam diagram and  the open string boundary. For this reason it is a more natural coordinate on  the moduli space from the point of view of lightcone interactions than $m$. Finally, we obtain  
\begin{eqnarray}
\lineup\!\!\!\!\!\! Z_4(\alpha_1,\alpha_2,\alpha_3,\alpha_4,m) \nonumber\\
\lineup  =\frac{m(1-m)}{2 \alpha_4 \text{Im}(u_*)}\exp\left[\frac{1}{2}\left(\frac{\alpha_2}{\alpha_3}+\frac{\alpha_3}{\alpha_2}+\frac{\alpha_1}{\alpha_4}+\frac{\alpha_4}{\alpha_1}\right)\ln(m) + \frac{1}{2}\left(\frac{\alpha_1}{\alpha_2}+\frac{\alpha_2}{\alpha_1}+\frac{\alpha_3}{\alpha_4}+\frac{\alpha_4}{\alpha_3}\right)\ln(1-m)\right]\nonumber\\
\lineup\ \ \ \times\exp\left[-\frac{1}{2}\left(\frac{1}{\alpha_1}+\frac{1}{\alpha_2}+\frac{1}{\alpha_3}+\frac{1}{\alpha_4}\right)\Big(\alpha_1\ln|\alpha_1|+\alpha_2\ln|\alpha_2|+\alpha_3\ln|\alpha_3|+\alpha_4\ln|\alpha_4|\Big)\right].
\end{eqnarray}
This can be confirmed to agree with the general formula for the partition function of transverse free bosons on a tree-level Mandelstam diagram. See appendix C of \cite{BabaIshibashi} for a relatively recent and accessible derivation,\footnote{The relation to the result of \cite{BabaIshibashi} is given by
\begin{equation}
e^{-\Gamma[\phi]} = \frac{1}{\alpha_1\alpha_2\alpha_3\alpha_4}Z_4(\alpha_1,\alpha_2,\alpha_3,\alpha_4,m)^2.\end{equation}
The inverse product of $\alpha$s is a conventional normalization, and the square  appears since we consider open strings, rather than closed strings.} as well as earlier references \cite{MandelstamAnom,GSW}.

Below we display the lightcone SFT action truncated to the tachyon field. This gives a good impression of what the action looks like when expanded explicitly in Fock space component fields. Also the truncation to the tachyon is often of interest for studying rolling tachyon solutions and as a toy model for nonlocality in string theory. For recent work in this direction, see \cite{EKZ}. We write the tachyon field in momentum space as 
\begin{equation}T_\alpha(k_+,k_\perp) = \Tlc(k),\end{equation}
where $\Tlc(k)$ is the tachyon field as written in section \ref{sec:transformation}. We further transform $k_+,k_\perp$ into position space to arrive at the tachyon field $T_\alpha(x^+,x^\perp)$. The truncated action is a little too unwieldy to display as a single equation, so we write it as a sum of three terms
\begin{equation}
\Slc(T_\alpha) = \text{(kinetic term)}+\text{(cubic term)}+\text{(quartic term)}.
\end{equation}
The kinetic term is
\begin{equation}
\text{(kinetic term)} \ =\  -\frac{1}{2}\int dx^+ dx^\perp \frac{d\alpha}{4\pi}T_{-\alpha}(x^+,x^\perp)\Big(-i\alpha  \d_+ -\d_\perp^2-1\Big)T_\alpha(x^+,x^\perp).
\end{equation}
The cubic term is
\begin{eqnarray}
\text{(cubic term)} \lineup= \int dx^+dx^\perp \frac{d\alpha_1d\alpha_2d\alpha_3}{(4\pi)^3}4\pi\delta(\alpha_1+\alpha_2+\alpha_3)Z_3(\alpha_1,\alpha_2,\alpha_3)\nonumber\\
\lineup\ \ \ \ \ \times \Big(K_3(\alpha_1,\alpha_2,\alpha_3)^{-\frac{1}{2\alpha_1}\d_\perp^2} T_{\alpha_1}(x^+,x^\perp)\Big)\Big(K_3(\alpha_1,\alpha_2,\alpha_3)^{-\frac{1}{2\alpha_2}\d_\perp^2} T_{\alpha_2}(x^+,x^\perp)\Big)\nonumber\\
\lineup\ \ \ \ \ \times\Big(K_3(\alpha_1,\alpha_2,\alpha_3)^{-\frac{1}{2\alpha_3}\d_\perp^2} T_{\alpha_3}(x^+,x^\perp)\Big),
\end{eqnarray}
where
\begin{equation}K_3(\alpha_1,\alpha_2,\alpha_3) = |\alpha_1|^{\alpha_1}|\alpha_2|^{\alpha_2}|\alpha_3|^{\alpha_3}.\end{equation}
Finally the quartic term is
\begin{eqnarray}
\lineup \!\text{(quartic term)} \nonumber\\
\lineup = \int dx^+dx^\perp\frac{d\alpha_1d\alpha_2d\alpha_3d\alpha_4}{(4\pi)^4}4\pi\delta(\alpha_1+\alpha_2+\alpha_3+\alpha_4)\delta_{\alpha_1\alpha_3>0}\delta_{\alpha_2\alpha_4>0}\nonumber\\
\lineup\ \ \ \times \int_{m_-}^{m_+}\left(dm \frac{-2\mathrm{Im}(u_*)}{m(1-m)}\right)Z_4(\alpha_1,\alpha_2,\alpha_3,\alpha_4,m)\nonumber\\
\lineup\ \ \ \times  m^{-(\d_\perp^1+\d_\perp^2)^2}(1-m)^{-(\d_\perp^2+\d_\perp^3)^2}\Bigg[\Big(K_{(4,1)}(\alpha_1,\alpha_2,\alpha_3,\alpha_4,m)^{-\frac{1}{2\alpha_1}(\d_\perp^1)^2} T_{\alpha_1}(x^+,x^\perp_1)\Big)\nonumber\\
\lineup\ \ \ \times\Big(K_{(4,2)}(\alpha_1,\alpha_2,\alpha_3,\alpha_4,m)^{-\frac{1}{2\alpha_2}(\d_\perp^2)^2} T_{\alpha_2}(x^+,x^\perp_2)\Big)\Big(K_{(4,3)}(\alpha_1,\alpha_2,\alpha_3,\alpha_4,m)^{-\frac{1}{2\alpha_3}(\d_\perp^3)^2} T_{\alpha_3}(x^+,x^\perp_3)\Big)\phantom{\Bigg]}\nonumber\\
\lineup\ \ \ \left.\times \Big(K_{(4,4)}(\alpha_1,\alpha_2,\alpha_3,\alpha_4,m)^{-\frac{1}{2\alpha_4}(\d_\perp^4)^2} T_{\alpha_4}(x^+,x^\perp_4)\Big)\Bigg]\right|_{x^\perp_1=x^\perp_2=x^\perp_3=x^\perp_4=x^\perp},
\end{eqnarray}
where
\begin{eqnarray}
K_{(4,1)}(\alpha_1,\alpha_2,\alpha_3,\alpha_4,m)\lineup = |\alpha_1|^{\alpha_1}|\alpha_2|^{\alpha_2}|\alpha_3|^{\alpha_3}|\alpha_4|^{\alpha_4}\frac{m^{\alpha_2+\alpha_3}}{(1-m)^{\alpha_1+\alpha_2}},\\
K_{(4,2)}(\alpha_1,\alpha_2,\alpha_3,\alpha_4,m)\lineup = |\alpha_1|^{\alpha_1}|\alpha_2|^{\alpha_2}|\alpha_3|^{\alpha_3}|\alpha_4|^{\alpha_4}\frac{1}{m^{\alpha_2+\alpha_3}(1-m)^{\alpha_1+\alpha_3}},\\
K_{(4,3)}(\alpha_1,\alpha_2,\alpha_3,\alpha_4,m)\lineup = |\alpha_1|^{\alpha_1}|\alpha_2|^{\alpha_2}|\alpha_3|^{\alpha_3}|\alpha_4|^{\alpha_4}\frac{(1-m)^{\alpha_1+\alpha_2}}{m^{\alpha_2+\alpha_3}},\\
K_{(4,4)}(\alpha_1,\alpha_2,\alpha_3,\alpha_4,m)\lineup = |\alpha_1|^{\alpha_1}|\alpha_2|^{\alpha_2}|\alpha_3|^{\alpha_3}|\alpha_4|^{\alpha_4}m^{\alpha_2+\alpha_3}(1-m)^{\alpha_1+\alpha_2}.\phantom{\bigg)}
\end{eqnarray}
For comparison, the Witten theory truncated to the tachyon has the action
\begin{equation}
\SW(\TW) = \int d^{26}x\left( -\frac{1}{2}\TW(x)(-\d^2-1)\TW(x)-\frac{1}{3}K^{-3}\Big(K^{-\d^2}\TW(x)\Big)^3\right),
\end{equation}
where $K = \frac{4}{3\sqrt{3}}$.

\section{Proof of Claim \ref{claim:1}}
\label{app:claim1}

In this appendix we prove that 
\begin{equation}S\mathcal{O}S^{-1} = F^{-1}\circ_\text{lc}\mathcal{O},\label{eq:app1}\end{equation}
where $\mathcal{O}$ does not produce contractions with $X^+(z,\overline{z})$, the operator $S$ is given by
\begin{equation}S =  e^{-R},\ \ \ \ R = \frac{1}{\sqrt{2}p_-}\sum_{n\in \mathbb{Z}}\delta_{n\neq 0}\frac{1}{n}\alpha^+_{-n}L_n^\text{lc},\end{equation}
the function $F^{-1}$ is the inverse of the conformal map,
\begin{equation}F(\xi) = e^{-\frac{ix^+}{2p_-}}\exp\left[\frac{i}{p_-}X^+(\xi)\right],\end{equation}
and $\circ_\text{lc}$ is the conformal transformation computed with respect to the  lightcone energy-momentum tensor. Before proceeding to the main argument we make a few comments about this result:
\begin{itemize}
\item The operator $S$ looks similar to an operator implementing a finite conformal transformation expressed in the form
\begin{equation}U = \exp\left[\sum_{n\in \mathbb{Z}}v_n L_n\right],\end{equation}
where $v_n$ are numbers which define Laurent coefficients of a holomorphic vector field. It is known that the conformal transformation implemented by $U$ is given by the solution of Julia's equation \cite{Sch_wedge}. This does not apply here since the plus oscillators, unlike the $v_n$s, do not commute with the Virasoros. It is in fact not obvious that the transformation by $S$ can be interpreted as a conformal transformation in the appropriate sense.
\item We have not given an expression for~$F^{-1}(\xi)$. This requires defining a compositional inverse of a worldsheet operator, which is an unusual and obscure notion. To see how to deal with this, write $\xi'=F(\xi)$ and transform to the strip coordinate $w=\ln \xi$ to find
\begin{equation}w' = w + \frac{i}{p_-}\widetilde{X}^+(e^w).\end{equation}
Here we can apply Lagrange's reversion theorem \cite{Lagrange} to determine $w$ as a function of $w'$ in the form of a power series in $1/p_-$:
\begin{equation}w = w' +\sum_{n=1}^\infty \frac{1}{n!}\left(\frac{i}{p_-}\right)^n\frac{d^{n-1}}{dw'^{n-1}}\Big(\widetilde{X}^+(e^{w'})^n\Big).\label{eq:Finvw}\end{equation}
For future use, it is helpful to introduce scaling differential operators in the weight $h$ representation
\begin{equation}\ell_0^{(h)} = h + \xi\frac{d}{d\xi}.\end{equation}
Writing \eq{Finvw} back in the coordinate $\xi$, we find
\begin{equation}
F^{-1}(\xi) = \xi \prod_{n=1}^\infty\exp\left[\frac{1}{n!}\left(\frac{i}{p_-}\right)^n(\ell_0^{(0)})^{n-1}\Big(\widetilde{X}^+(\xi)^n\Big)\right].
\end{equation}
The expression is unwieldy, but can be expanded explicitly in terms of plus oscillators.
\item The similarity transformation maps operators on $\Hlc$ into operators on $\Hcov$. One might expect that the inverse transformation, from $\Hcov$ into $\Hlc$, might be given by inverting the conformal map: 
\begin{equation}F\circ_\text{lc} \mathcal{O} \stackrel{?}{=}  S^{-1}\mathcal{O} S. \end{equation}
when $\mathcal{O}$ has no contractions with $X^+(z,\overline{z})$. However this is incorrect. To see why, note that we have the equality
\begin{equation}F(F^{-1}(\xi))=\xi,\end{equation}
but this does not imply 
\begin{equation}F\circ_\text{lc} F^{-1}\circ_\text{lc} \mathcal{O} \stackrel{?}{=}\mathcal{O},\end{equation}
because lightcone conformal transformations act nontrivially on $\widetilde{X}^+$. 
\end{itemize}

Now we turn to the proof. For simplicity, we will consider $\mathcal{O}$ as a local operator $\phi(\xi)$ which is primary with respect to the lightcone energy-momentum tensor with weight $h$.  From this we can build more general operators. By assumption $\phi(\xi)$ does not have contractions with $X^+$. We introduce an expansion parameter 
\begin{equation}\lambda = -\frac{i}{p_-}.\end{equation}
The first step is computing the left hand side of \eq{app1}. Since $S=e^{-R}$ this requires computing repeated commutators
\begin{equation}\lambda^n S_n^{(h)}[\phi(\xi)] = [[...[\phi(\xi),\underbrace{R]....R],R]}_{n\text{ times}},\end{equation}
where $S_n^{(h)}$ is a linear differential operator on $\phi(\xi)$. The left hand side of \eq{app1} will then be expressed
\begin{equation}S^{-1}\phi(\xi) S = \sum_{n=0}^\infty \frac{\lambda^n}{n!}S_n^{(h)}[\phi(\xi)].\end{equation}
To compute $S_n^{(h)}$ we need the commutators
\begin{eqnarray}
[\phi(\xi),R] \lineup =\lambda \Big(\ell_0^{(h)}(\widetilde{X}^+\phi(\xi))+(h-1)\xi \d \widetilde{X}^+ \phi(\xi)\Big),\\
\ [\widetilde{X}^+(\xi),R ] \lineup = \lambda\xi \widetilde{X}^+\d \widetilde{X}^+(\xi).
\end{eqnarray}
We can work out the first few orders
\begin{eqnarray}
S_0^{(h)}[\phi(\xi)]\lineup  = \phi(\xi),\phantom{\bigg)}\\
S_1^{(h)}[\phi(\xi)] \lineup = \ell_0^{(h)}\Big(\widetilde{X}^+\phi(\xi)\Big)+(h-1)\xi \d \widetilde{X}^+\phi(\xi),\phantom{\bigg)}\\
S_2^{(h)}[\phi(\xi)] \lineup = (\ell_0^{(h)})^2\Big((\widetilde{X}^+)^2\phi(\xi)\Big)+2(h-1)\xi \ell_0^{(h+1)}\Big(\widetilde{X}^+\d \widetilde{X}^+\phi(\xi)\Big)+h(h-1)\xi^2(\d \widetilde{X}^+)^2\phi(\xi),\phantom{\bigg)}\ \ \ \ \ \ \  \\
S_3^{(h)}[\phi(\xi)] \lineup = (\ell_0^{(h)})^3\Big((\widetilde{X}^+)^3\phi(\xi)\Big)+3(h-1)\xi (\ell_0^{(h+1)})^2\Big((\widetilde{X}^+)^2\d \widetilde{X}^+\phi(\xi)\Big)\phantom{\bigg)}\nonumber\\
\lineup\ \ \ \ \ +3h(h-1)\xi^2 \ell_0^{(h+2)}\Big(\widetilde{X}^+(\d \widetilde{X}^+)^2\phi(\xi)\Big)+(h+1)h(h-1)\xi^3(\d \widetilde{X}^+)^3\phi(\xi),\phantom{\bigg)}\phantom{\bigg)}\\
\lineup \vdots .\nonumber
\end{eqnarray}
From this we can see a general pattern 
\begin{equation}
S_n^{(h)}[\phi(\xi)] = \sum_{k=0}^n\left({n \atop k}\right)\frac{(h-2+k)!}{(h-2)!}\xi^k (\ell_0^{(h+k)})^{n-k}\Big((\widetilde{X}^+)^{n-k}(\d \widetilde{X}^+)^k\phi(\xi)\Big),\label{eq:Snh}
\end{equation}
which can be confirmed by showing that $\lambda S_{n+1}^{(h)}[\phi(\xi)] =[S_n^{(h)}[\phi(\xi)],R]$.

Next we compute the right hand side. We write
\begin{equation}F^{-1}\circ_\text{lc}\phi(\xi) = S^{(h)}[\phi(\xi)],\end{equation}
where $S^{(h)}$ is a linear functional defined by
\begin{equation}
S^{(h)}[\phi(\xi)] =\left(\frac{d F^{-1}(\xi)}{d\xi}\right)^h \phi(F^{-1}(\xi)).
\end{equation}
The goal is to show that the coefficients operators in the expansion of $S^{(h)}$ match what we have found in \eq{Snh}. To do this we derive a differential equation which $S^{(h)}$ must satisfy. Start by taking a derivative with respect to $\lambda$:
\begin{equation}
\frac{d}{d\lambda}S^{(h)}[\phi(\xi)] = h\left(\frac{d F^{-1}}{d\xi}\right)^{h-1}\frac{d}{d\xi}\left(\frac{dF^{-1}}{d\lambda}\right)\phi(F^{-1}(\xi))+\left(\frac{d F^{-1}}{d\xi}\right)^{h}\left.\frac{d \phi(z)}{dz}\right|_{z=F^{-1}(\xi)}\frac{d F^{-1}}{d\lambda}.
\end{equation}
To go further we need the derivative of $F^{-1}(\xi)$ with respect to $\lambda$. The derivative of $F(\xi)$ with respect to $\lambda$ is 
\begin{equation}\frac{d F(\xi)}{d\lambda} = -\widetilde{X}^+(\xi)F(\xi),\end{equation}
and this determines the corresponding derivative of $F^{-1}(\xi)$ as follows. We note
\begin{equation}\frac{d}{d\lambda}F^{-1}(F(\xi)) = \frac{d}{d\lambda}\xi = 0.\end{equation}
Computing the derivative leads to 
\begin{equation}
\left.\frac{dF^{-1}(z)}{d\lambda}\right|_{z=F(\xi)}+\left.\frac{d F^{-1}(z)}{dz}\right|_{z=F(\xi)}\frac{d F(\xi)}{d\lambda}=0,
\end{equation}
and therefore
\begin{equation}
\left.\frac{dF^{-1}(z)}{d\lambda}\right|_{z=F(\xi)} = \widetilde{X}^+(\xi)F(\xi) \left.\frac{d F^{-1}(z)}{dz}\right|_{z=F(\xi)}.
\end{equation}
Now replacing $F(\xi)$ with $\xi$ we find that
\begin{equation}
\frac{d F^{-1}(\xi)}{d\lambda} = \xi \widetilde{X}^+(F^{-1}(\xi))\frac{d F^{-1}(\xi)}{d\xi}.
\end{equation}
Now we can plug this back in and continue computing the derivative of $S^{(h)}$ with respect to $\lambda$:
\begin{eqnarray}
\frac{d}{d\lambda}S^{(h)}[\phi(\xi)]\lineup  = h\left(\frac{d F^{-1}}{d\xi}\right)^{h-1}\frac{d}{d\xi}\left(\xi \widetilde{X}^+(F^{-1}(\xi))\frac{d F^{-1}}{d\xi}\right)\phi(F^{-1}(\xi))\nonumber\\
\lineup \ \ \ \ \ +\left(\frac{d F^{-1}}{d\xi}\right)^{h}\xi \widetilde{X}^+(F^{-1}(\xi))\frac{d}{d\xi}\phi( F^{-1}(\xi)).
\end{eqnarray}
With some rearrangement this reads
\begin{equation}
\frac{d}{d\lambda}S^{(h)}[\phi(\xi)] = \ell_0^{(h)}S^{(h)}[\widetilde{X}^+(\xi)\phi(\xi)]+(h-1)\xi S^{(h+1)}[\d \widetilde{X}^+(\xi) \phi(\xi)].
\end{equation}
This is a differential equation which determines the right hand side.

Expanding $S^{(h)}$ in powers of $\lambda$,
\begin{equation}S^{(h)}[\phi(\xi)] = \sum_{n=0}^\infty \frac{\lambda^n}{n!}S_n^{(h)}[\phi(\xi)],\end{equation}
the differential equation implies a recursive formula
\begin{equation}
S_{n+1}^{(h)}[\phi(\xi)] = \ell_0^{(h)}S_n^{(h)}[\widetilde{X}^+(\xi)\phi(\xi)] +(h-1)\xi S_n^{(h+1)}[\d \widetilde{X}^+(\xi)\phi(\xi)].
\end{equation}
The solution of this recursion is uniquely determined by the initial value
\begin{equation}S^{(h)}_0[\phi(\xi)] = \phi(\xi),\end{equation}
which follows since $S^{(h)}$ is the identity transformation at $\lambda = 0$. Now it is simply a matter of algebra to check that \eq{Snh} solves this recursion. This completes the proof.

\end{appendix}

\end{document}